\documentclass{elsarticle}
\usepackage{lineno,hyperref}
\usepackage{color}
\modulolinenumbers[5]

\usepackage{multicol}
\usepackage{multirow}
\usepackage{subfigure}

\usepackage{xcolor}

\usepackage[normalem]{ulem}

\journal{}

\providecommand{\bo}{\mathbf}
\providecommand{\bs}{\boldsymbol}
\newcommand{\V}{\bs V}
\newcommand{\X}{\bs X}
\newcommand{\x}{\bs x}
\newcommand{\m}{\bs m}
\newcommand{\Z}{\bs Z}
\providecommand{\cov}{\mathrm{COV}}
\providecommand{\lcov}{\mathrm{LCOV}}
\providecommand{\scov}{\mathrm{SCOV}}
\providecommand{\tcov}{\mathrm{TCOV}}
\providecommand{\ucov}{\mathrm{UCOV}}
\def\covF{\mathop{\mathrm{COV}_4}\nolimits}
\providecommand{\mlt}{\mathrm{MLC}}
\providecommand{\mcd}{\mathrm{MCD}}
\providecommand{\rmcd}{\mathrm{RMCD}}
\providecommand{\diag}{\mathrm{diag}}

\newtheorem{prop}{Proposition}

\newcommand{\mat}[1]{\boldsymbol{#1}}   
\newcommand{\vect}[1]{\boldsymbol{#1}}  
\newcommand{\obs}[1]{\mathbf{#1}}		    

\newcommand{\pkg}[1]{\texttt{#1}}
\newcommand{\proglang}[1]{\textsf{#1}}

\usepackage{amsmath,amssymb,amsfonts}
\usepackage{graphicx}
\usepackage[toc,page]{appendix}

\biboptions{authoryear}

\begin{document}

\begin{frontmatter}

\title{Tandem clustering with invariant coordinate selection  
}

\author[eur]{Andreas Alfons\corref{mycorrespondingauthor}}
\cortext[mycorrespondingauthor]{Corresponding author}
\ead{alfons@ese.eur.nl}
\author[eur,tbs]{Aurore Archimbaud}
\author[jyu]{Klaus Nordhausen}
\author[tse]{Anne Ruiz-Gazen}

\address[eur]{Erasmus School of Economics, Erasmus University Rotterdam, Netherlands}
\address[tbs]{Department of Information, Operations and Management Sciences, \\ TBS Business School, France}
\address[jyu]{Department of Mathematics and Statistics, University of Jyv\"askyl\"a, Finland}
\address[tse]{Toulouse School of Economics, Universit\'e Toulouse Capitole, France}





\begin{abstract}
For multivariate data, tandem clustering is a well-known technique aiming to improve cluster identification through initial dimension reduction. Nevertheless, the usual approach using principal component analysis (PCA) has been criticized for focusing solely on inertia so that the first components do not necessarily retain the structure of interest for clustering. To address this limitation, a new tandem clustering approach based on invariant coordinate selection (ICS) is proposed. By jointly diagonalizing two scatter matrices, ICS is designed to find structure in the data while providing affine invariant components. Certain theoretical results have been previously derived and guarantee that under some elliptical mixture models, the group structure can be highlighted on a subset of the first and/or last components. However, ICS has garnered minimal attention within the context of clustering. Two challenges associated with ICS include choosing the pair of scatter matrices and selecting the components to retain. For effective clustering purposes, it is demonstrated that the best scatter pairs consist of one scatter matrix capturing the within-cluster structure and another capturing the global structure. For the former, local shape or pairwise scatters are of great interest, as is the minimum covariance determinant (MCD) estimator based on a carefully chosen subset size that is smaller than usual. The performance of ICS as a dimension reduction method is evaluated in terms of preserving the cluster structure in the data. In an extensive simulation study and empirical applications with benchmark data sets, various combinations of scatter matrices as well as component selection criteria are compared in situations with and without outliers. Overall, the new approach of tandem clustering with ICS shows promising results and clearly outperforms the PCA-based approach.
\end{abstract}


\end{frontmatter}


\section{Introduction}
Clustering or cluster analysis is the task of finding groups of observations in a data set.  Clustering problems arise in various application fields such as marketing for customer segmentation or in biostatistics for analysing gene expression data. In the present paper, our primary focus is on hard clustering, where observations are divided into subsets referred to as clusters. These clusters form a partition of the data set with each observation belonging exclusively to one cluster. As highlighted, for instance, in \citet[p.~2]{hennig2015handbook}, the definition of a cluster is neither unique nor precise. Consequently, numerous clustering algorithms have been developed and studied.  Their usefulness and relevance depend on factors such as the context, the data type, and the objectives of the data analyst. Several books and surveys delve into this subject, e.g., \citet{kaufman2009finding}, \citet{xu2005survey}, \citet{aggarwal2013}, and \citet{hennig2015handbook}.

The commonly accepted objective in clustering is to group similar observations together within the same cluster, while dissimilar ones should be assigned to different clusters. Among the many existing approaches, common clustering methods are either distance-based, model-based, or density-based. In the present paper, we focus on centroid-based clustering techniques which belong to the distance-based category, as well as clustering through Gaussian mixture models, which belong to the model-based category. The data in consideration consist of observations characterized by numerical variables or features.

When the cluster structure lies in a low-dimensional latent subspace, or when some variables are irrelevant for identifying the cluster structure, a preprocessing step like feature selection or dimension reduction can be beneficial. Tandem analysis, as termed by \cite{ArabieHubert1994}, is a well-known method involving principal component analysis (PCA) followed by clustering on the selected principal components. However, several authors including \citet{chang1983using}, \citet{soete1994k}, \citet{vichi2001factorial}, and \citet{radojivcic2021large} warn against this approach, as it may mask the cluster structure of the data. The explanation is that PCA focuses on conserving inertia, linked to data variability, and does not necessarily extract the relevant information for clustering. Hence, many papers propose alternatives to tandem analysis (see, e.g., \citealp{ding2007adaptive,vichi2009clustering,timmerman2013subspace,van2019special}; and references therein). 

In the present paper, we propose a new tandem clustering approach in which we substitute PCA with invariant coordinate selection (ICS), as proposed by \citet{tyler2009invariant}. ICS serves as a dimension reduction method akin to PCA, involving the computation and selection of linear combinations of the original features. While PCA is derived from the eigendecomposition of a single scatter matrix, ICS relies on the simultaneous diagonalization of two scatter matrices \citep[see, e.g.,][]{nordhausen2022usage}. Unlike PCA, ICS can extract components that are pertinent for clustering. More precisely, \citet{tyler2009invariant} show that for a location mixture of elliptical distributions and any pair of scatter matrices, ICS can identify Fisher's linear discriminant subspace without knowledge of the class labels. Additionally, ICS is affine invariant while PCA is merely orthogonally invariant.

The use of ICS as a preprocessing step for clustering is not new. ICS and its earlier versions have been recognized as a method for dimension reduction preceding group identification in \citet{art1982data}, \citet{caussinus1990interesting},\citet{caussinus1993ascona,Caussinus1995metrics,caussinus2007classification}, \citet{gnanadesikan1993mahalanobis}, \citet{tyler2009invariant}, \citet{Hennig2009}, \citet{pena2010eigenvectors}, \citet{Fekri2015}, \citet{AlashwaliKent2016}, and \citet{FischerHonkatukiaTuiskulaHaavistoNordhausenCaveroPreisingerVilkki:2017, FISCHER2020e05732}. However, there is no in-depth study of ICS in combination with clustering, and the present paper aims to fill this gap. Following the study by \citet{archimbaud2018CSDA} on ICS for outlier detection, we first highlight the relevance of ICS when the cluster structure lies in a low-dimensional subspace. We then compare several pairs of scatter matrices and  discuss the selection of invariant coordinates. It is important to note that, akin to the standard PCA approach, ICS is not well-suited to high-dimensional data. In this article, we assume a scenario where the number of observations is large compared to the number of variables, and the number of variables is moderate.

In the context of clustering, ICS can be compared to linear discriminant analysis (LDA) but in an unsupervised situation where the groups are unknown. When the groups are known, and assuming similar cluster shapes, it becomes possible to compute within-group and between-group scatter matrices. LDA \citep[see][]{mardia1982multivariate} is essentially a joint diagonalization of these specific scatter matrices, equivalent to the joint diagonalization of within-group and total covariance matrices. Keeping the analogy with LDA, in situations where the groups are unknown, we anticipate ICS yielding valuable results when preprocessing data before clustering if we jointly diagonalize a scatter matrix close to the within-group covariance matrix and another scatter close to the total covariance matrix. Regarding the within-group matrix, this paper further explores local scatter matrices \citep{Hennig2009}, weighted pairwise scatter matrices \citep{caussinus1990interesting,tyler2009invariant}, and minimum covariance determinant \citep{rousseeuw1985multivariate} scatter matrices based on a small percentage of observations. Interestingly, \citet{kettenring2006practice} claims that cluster analysis and LDA are ``at the opposite ends of a spectrum''. While no information is available about the number or the types of clusters in clustering, the number of groups and their content is known in LDA. Following this idea, ICS is positioned in the middle of the spectrum, making assumptions about the type of clusters (ideally Gaussian distributions with different means and the same covariance matrices), but lacking information about the number of groups and how the observations are distributed among the groups. Similar to LDA, some theoretical properties of ICS can be derived for mixtures of Gaussian distributions with different means and equal covariance matrices \citep[see][]{tyler2009invariant, archimbaud2018CSDA}, and we predominantly focus on this model throughout the paper. However, theoretical properties of ICS extend to other types of mixture types \citep[see][]{archimbaud2018statistical}, and in practice, ICS proves useful beyond this specific framework, as demonstrated in artificial and empirical examples.

The paper is organized as follows. In Section~\ref{sec:highdim}, we examine a mixture of Gaussian distributions with different means and equal covariance matrix. We derive a property at the population level when the number of variables increases while the number of groups is kept fixed. In this scenario where the dimension of the affine subspace containing the cluster means decreases relative to the total dimension, we highlight the difficulty of identifying clusters when using the total or the within-group Mahalanobis distances. In contrast, the dimension reduction induced by ICS allows for solving the problem. Section~\ref{sec:ics} provides a detailed presentation of ICS, along with the definition of several scatter matrices relevant in a clustering context, as well as a description of component selection criteria. Section~\ref{sec:sim} presents results from an extensive simulation study for Gaussian mixtures with two, three, and five groups. We compare a variety of dimension reduction strategies using ICS or PCA followed by some centroid-based and model-based clustering algorithms. Section~\ref{sec:applications} illustrates the use of tandem clustering with ICS in three empirical examples, and Section~\ref{sec:conclusion} concludes the paper and discusses further perspectives.

\section{Highlighting the relevance of ICS in case of low-dimensional cluster structure}
\label{sec:highdim}

Consider the following probability model introduced in \citet{caussinus1993ascona}. Let $\bo X$ be a $d$-variate real random vector distributed according to the following mixture distribution:
\begin{equation}\label{eq:modelP}
    \int {\cal N}_d(\x,\bs\Sigma_W) \ dP(\x),
\end{equation}
where the probability measure $P$ is concentrated on a unknown $l$-dimensional linear manifold 
with $l<d$, and $ {\cal N}(\x,\bs\Sigma_W)$ denotes the $d$-dimensional normal distribution with mean $\x$ and covariance $\bs\Sigma_W$. That is, the mixing probability $P$ corresponds to the unknown structure of interest. An example of such a  model is the case where $P$ is a mixture of Dirac probability measures at points $\bs\mu_1$,~\ldots,~$\bs\mu_q$. This example is a finite mixture of Gaussian distributions with different means and equal covariance matrices, with $l$ equal to the dimension of the affine subspace which contains the mean vectors $\bs\mu_1, \dots, \bs\mu_q$. In this paper, we focus on the case where $l$ is small compared to $d$. Considering the Mahalanobis distances, we present a result for the case of a Gaussian mixture distribution with different means and equal covariance matrices. The result is analogous to Proposition 1 in \cite{archimbaud2018CSDA} but tailored to the clustering context rather than outlier detection. The number of groups $q$ remains constant as the overall dimension $d$ increases to infinity, which means that the intrinsic dimension $l$ is small compared to the original dimension $d$ in model~\eqref{eq:modelP}. In such a context, ICS is particularly suitable for data reduction before clustering, as it allows to transform the data from the $d$-dimensional space into the $l$-dimensional subspace associated with the structure of interest.

The question of generalizing the $k$means algorithm to the use of the Mahalanobis distance instead of the Euclidean distance has been studied in the literature for a long time. \citet{gnanadesikan1993mahalanobis} provide an example featuring three groups in three dimensions where the Euclidean distance proves inappropriate, in contrast to the Mahalanobis distance based on the within-group dispersion. The latter accounts for the elliptical nature of the groups (i.e., correlations). Another advantage of using the Mahalanobis distance is its invariance under affine transformations of the data, eliminating the need for variable standardization before clustering \citep[see][for more details]{cerioli2005k}. However, the use of the Mahalanobis distance is not the most common choice when applying a generalized variant of $k$means. The primary challenge lies in adapting the $k$means algorithm to incorporate the within-group covariance matrix (see \citealp{cerioli2005k, lapidot2018convergence}; and references therein). In our proposed tandem approach, Euclidean distances computed on the ICS components represent Mahalanobis distances with respect to one of the scatter matrices on the original variables. Opting for a scatter matrix close to the within-group scatter matrix allows the consideration of elliptical clusters without modifying the $k$means algorithm. 

Let $\bo X$ be a $d$-variate real random vector distributed according to a mixture of $q$ Gaussian distributions with $q < d$ distinct location parameters $\bs\mu_1, \dots, \bs\mu_q$, and the same positive definite covariance matrix $\bs\Sigma_W$:
\begin{equation}\label{mixt}
{\bo X} \sim \sum_{h=1}^{q} \epsilon_h \, {\cal N}(\bs\mu_h,\bs\Sigma_W),
\end{equation}
where $\epsilon_{1}, \dots, \epsilon_{q}$ are mixture weights with $\epsilon_1 + \cdots + \epsilon_q = 1$. Let $\bs \Sigma$ denote the total covariance matrix of $\bo X$. We now examine the squared pairwise Mahalanobis distances between two random vectors $\bo X$ and $\bo Y$ with the Gaussian mixture distribution defined in~\eqref{mixt}:
\begin{equation*}
r^2(\bo X, \bo Y) = (\bo X-\bo Y)^\top{\bs\Sigma}^{-1}(\bo X-\bo Y)
\end{equation*}
and
\begin{equation*}
r_{\bs W}^2(\bo X, \bo Y) = (\bo X-\bo Y)^\top{\bs\Sigma_W}^{-1}(\bo X-\bo Y).
\end{equation*}
These distances are affine invariant, meaning that $r^2(\bo A \bo X + \bo b, \bo A \bo Y + \bo b)=r^2(\bo X, \bo Y)$ holds for any full rank $d\times d$ matrix $\bo A$ and any $d$-dimensional vector $\bo b$.

To discern groups, it is crucial that the distance between two independent random vectors $\bo X_{h}$ and $\bo Y_{h'}$ from two different groups $h\neq h'$ is stochastically larger than the distance between two independent random vectors $\bo X_{h}$ and $\bo Y_{h}$  from the same group~$h$. However, we can prove that asymptotically, the variance of the difference between these distances increases with the dimension $d$ when the number of groups $q$ is held constant, making cluster identification increasingly challenging.

\begin{prop} \label{propMD}
Let $\bo X_h,\bo Y_h \sim {\cal N}(\bs\mu_h,\bs\Sigma_W)$ and  $\bo Y_{h'}\sim {\cal N}(\bs\mu_{h'},\bs\Sigma_W)$ be independent random vectors with $h,h' \in \{1,\ldots,q\}$ and $h\neq h'$. 
Assume that $q>1$ is fixed and that 
\[
\sum_{h=1}^q \epsilon_h (\bs\mu_h-\bs \mu)' \bs\Sigma_W^{-1} (\bs\mu_h-\bs\mu) =o(d).
\]
Then under model~\eqref{mixt}, the expressions
$$
\frac{c}{\sqrt{d}} 
\Big[ \left( r^2(\bo X_{h},\bo Y_{h'}) - r^2(\bo X_{h}, {\bo Y}_h) \right) - 
\mathbb{E} \left( r^2(\bo X_{h},\bo Y_{h'}) -r^2(\bo X_{h}, {\bo Y}_h)\right)  \Big]
$$
and
$$
\frac{c_{\bs W}}{\sqrt{d}} 
\Big[ \left( r^2_{\bs W}(\bo X_{h},\bo Y_{h'}) -r^2_{\bs W}(\bo X_{h}, {\bo Y}_h) \right) - 
\mathbb{E}\left( r^2_{\bs W}(\bo X_{h},\bo Y_{h'}) -r^2_{\bs W}(\bo X_{h}, {\bo Y}_h)\right) \Big]
$$
converge in distribution to standard Gaussian distributions as $d$ goes to infinity with some constants $c$ and $c_{\bs W}>0$. Moreover, the expectations of the differences of the Mahalanobis distances remain bounded as $d$ goes to infinity.
\end{prop}

The proof is similar to the proof of Proposition 1 in \cite{archimbaud2018CSDA} and is not detailed further. The key element of the proof lies in the fixed dimension $l \leq q-1$ of the subspace of interest (where $q$ denotes the number of groups), while the total number of dimensions $d$ grows to infinity. This implies that the number of dimensions $d-l \geq d-q+1$ in which there is no cluster structure is also increasing to infinity. In these dimensions, the expectation of the difference of distances remains bounded as $d$ goes to infinity, whereas its variance inflates and masks the difference in expectation in the subspace of interest. If the relevant subspace of dimension $l$ is known, the issue arising from the $d-l$ irrelevant dimensions can be addressed by projecting the data set onto that subspace and computing distances based only on the relevant  $l$ dimensions. This is exactly what ICS is all about, offering the data-analyst the capability to identify a subspace revealing the clusters (in an unsupervised way), and to project the data onto it. Note that Proposition~\ref{propMD} describes a property at the distribution level for $d \rightarrow \infty$, while this paper primarily focuses on the sample setting with $d < n$. At the sample level, it is still possible to satisfy $d < n$ if $d$ increases to infinity as long as $n$ increases at a faster rate.

\begin{figure}[t!]
\centering
\includegraphics[width=0.275\textwidth]{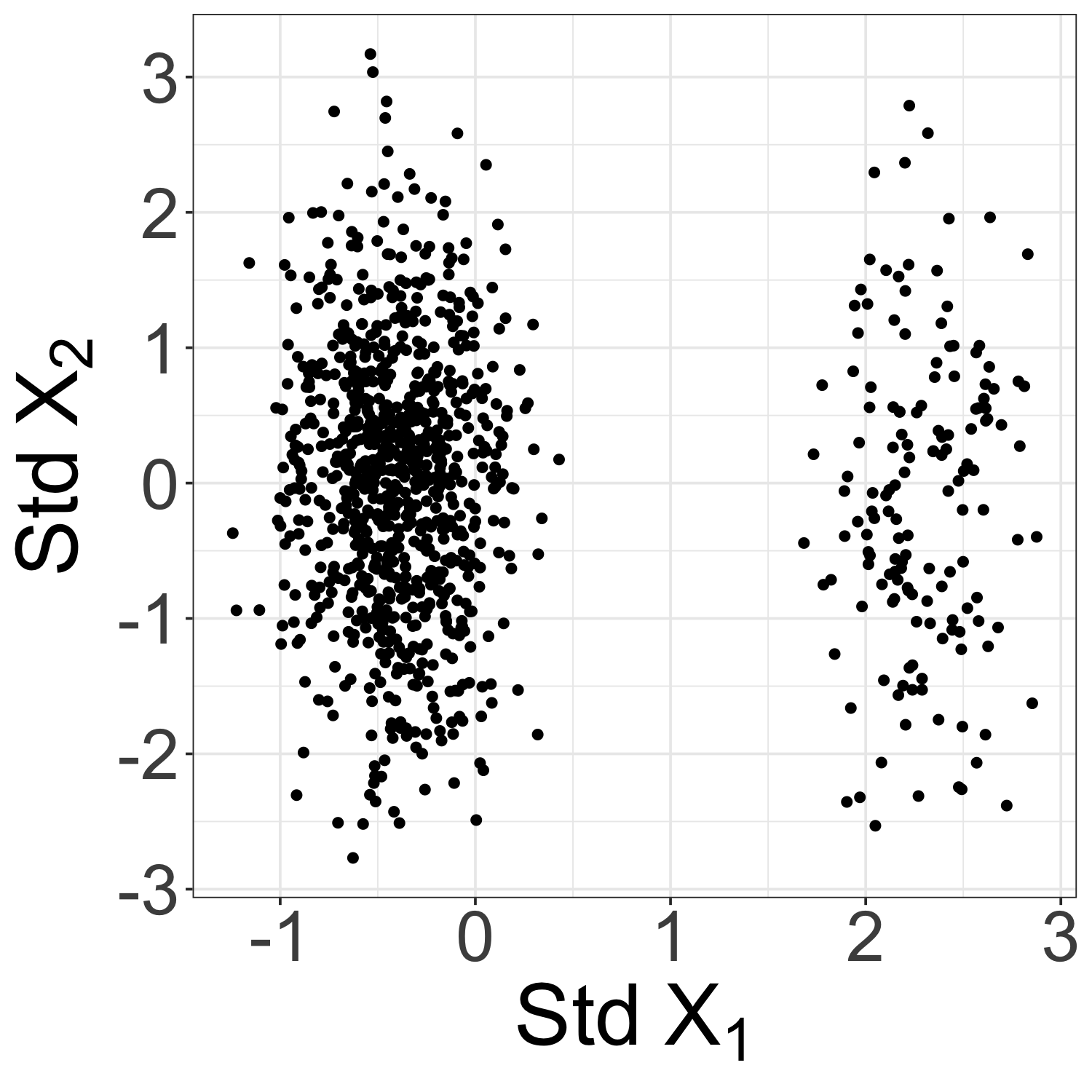}
\includegraphics[width=0.275\textwidth]{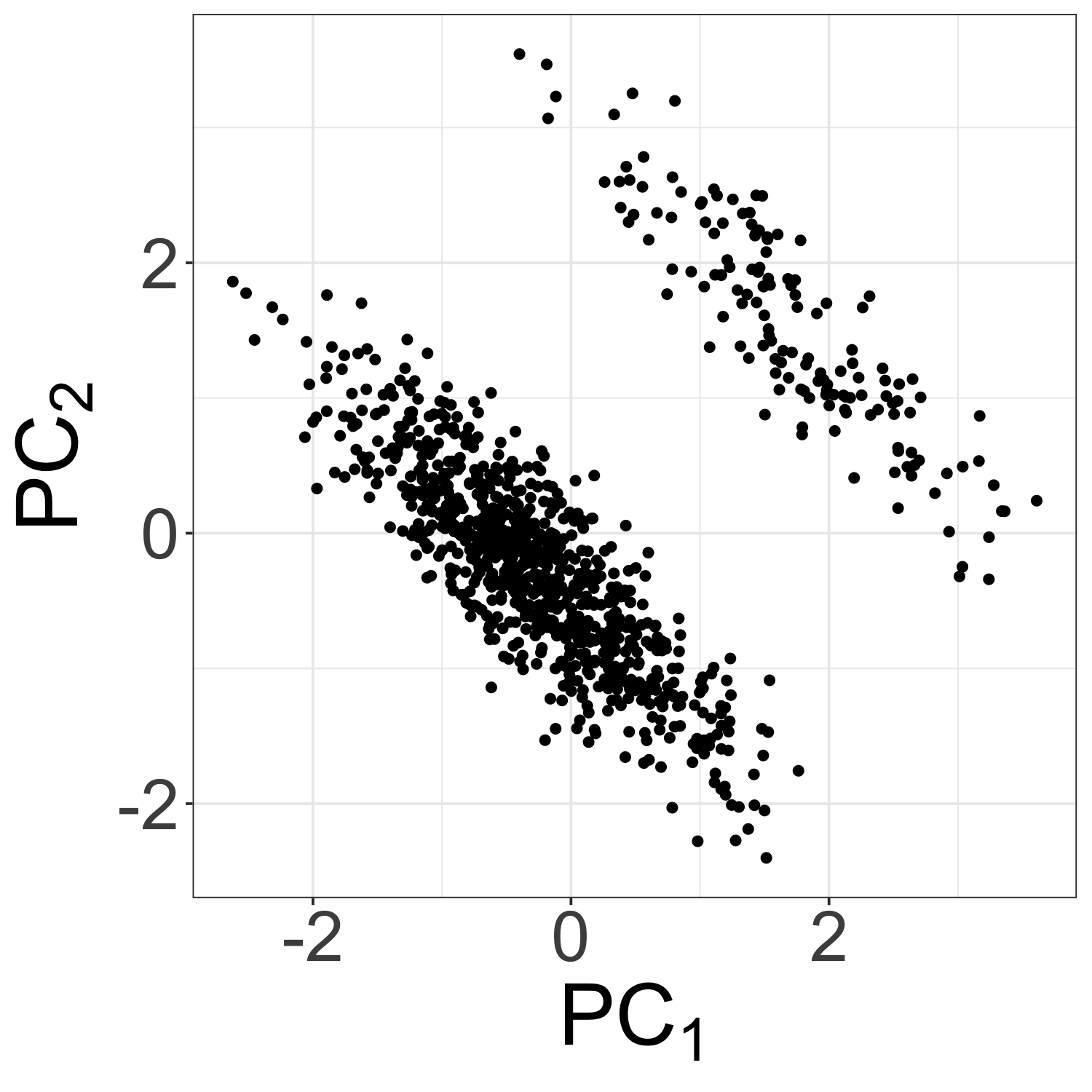}
\includegraphics[width=0.275\textwidth]{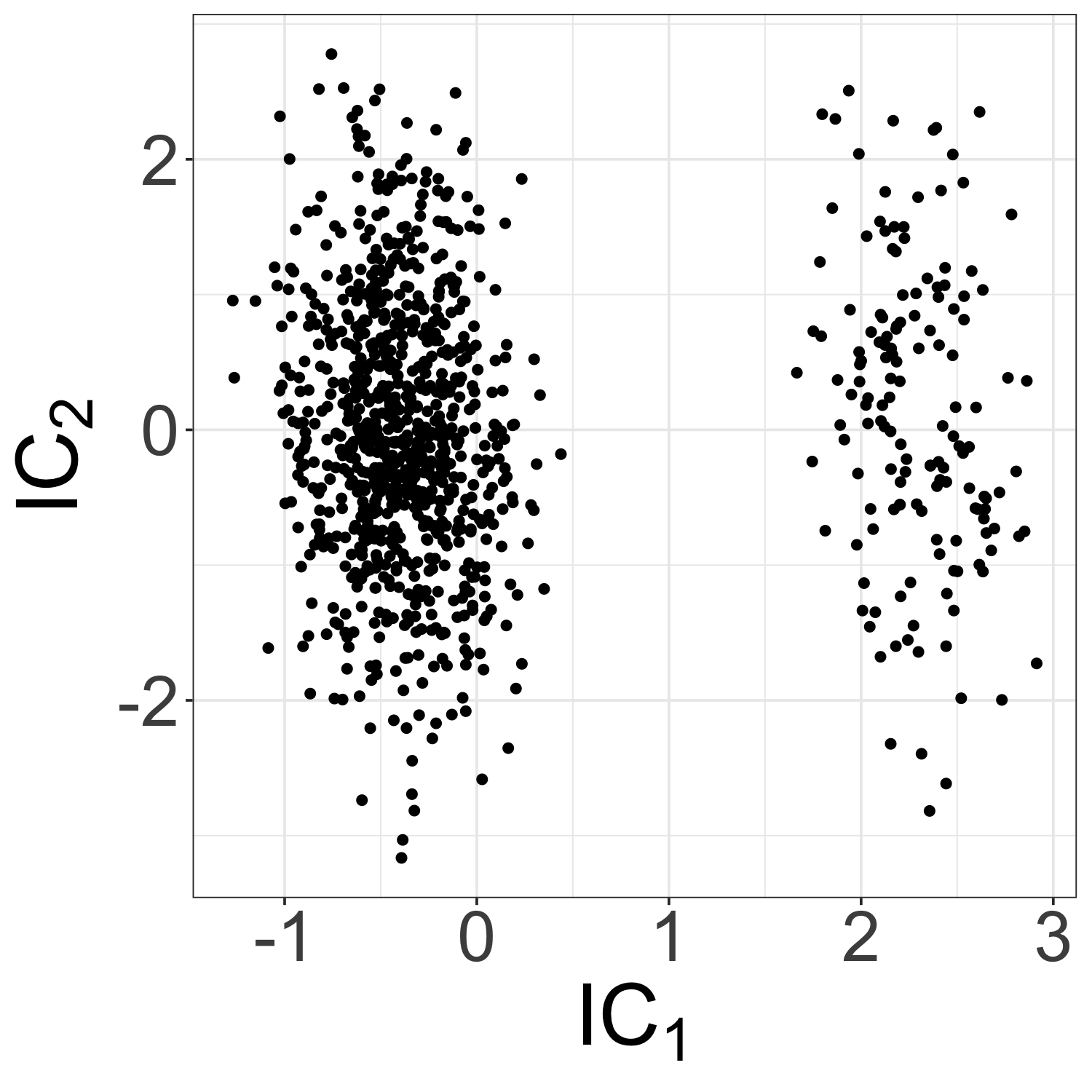} \\
\includegraphics[width=0.275\textwidth]{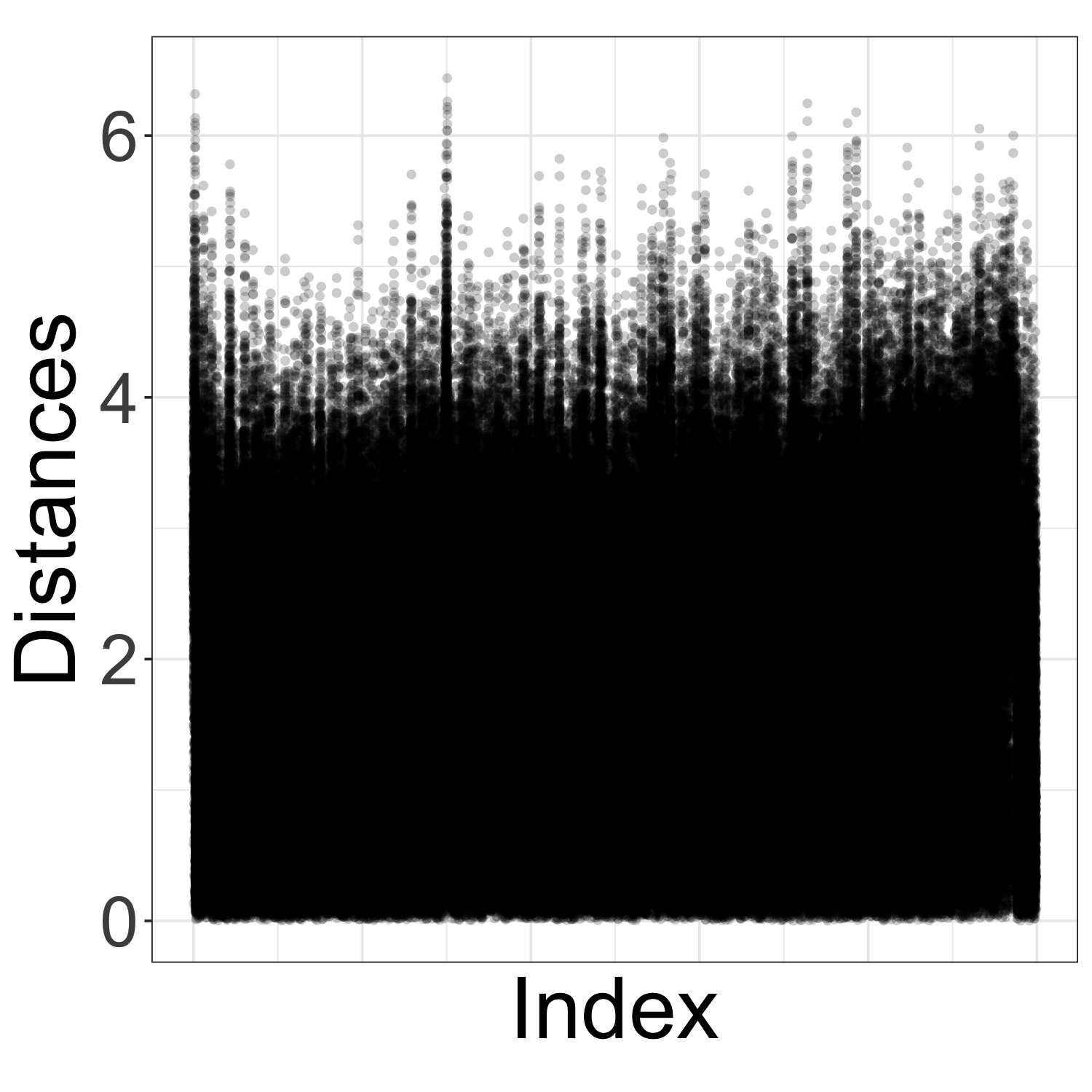}
\includegraphics[width=0.275\textwidth]{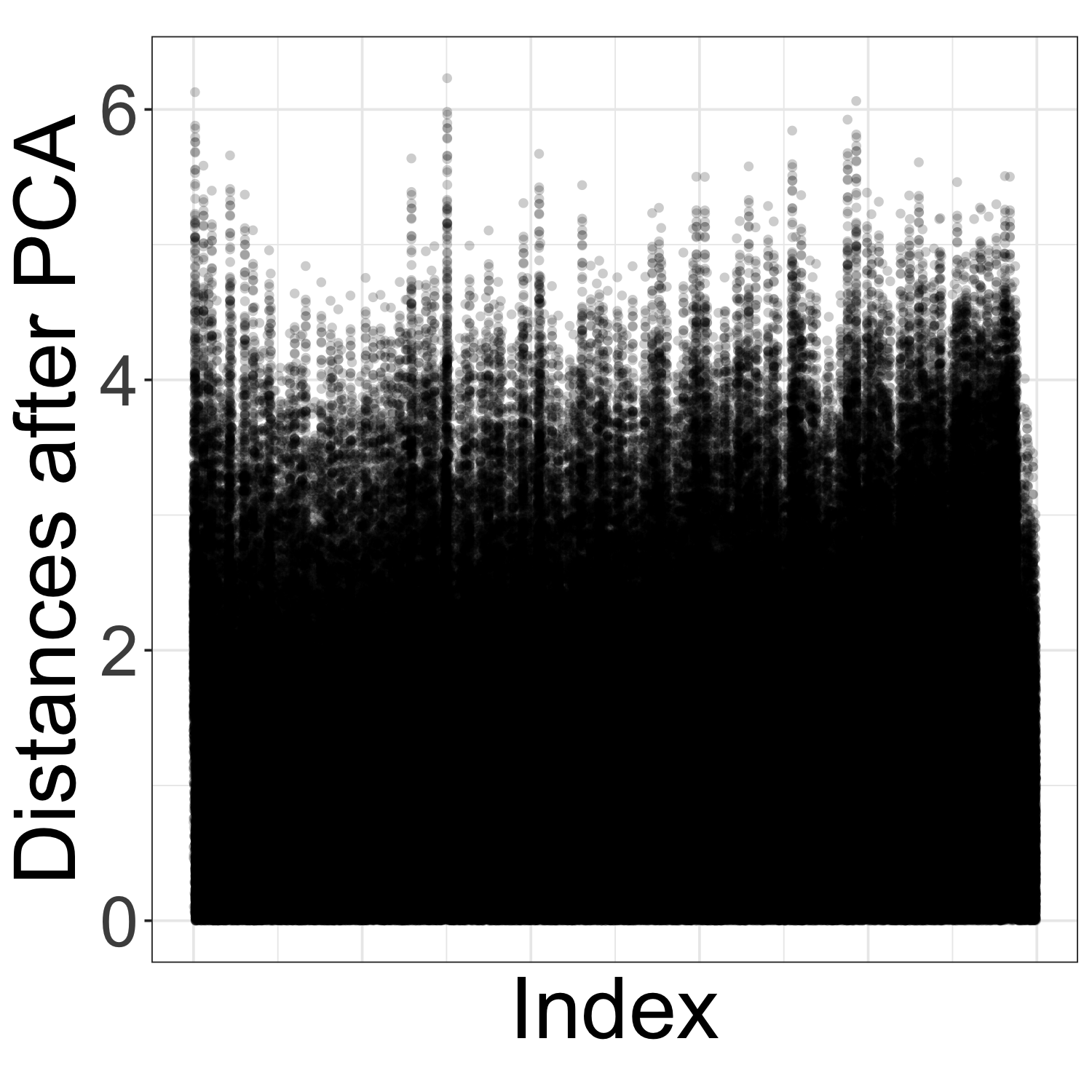}
\includegraphics[width=0.275\textwidth]{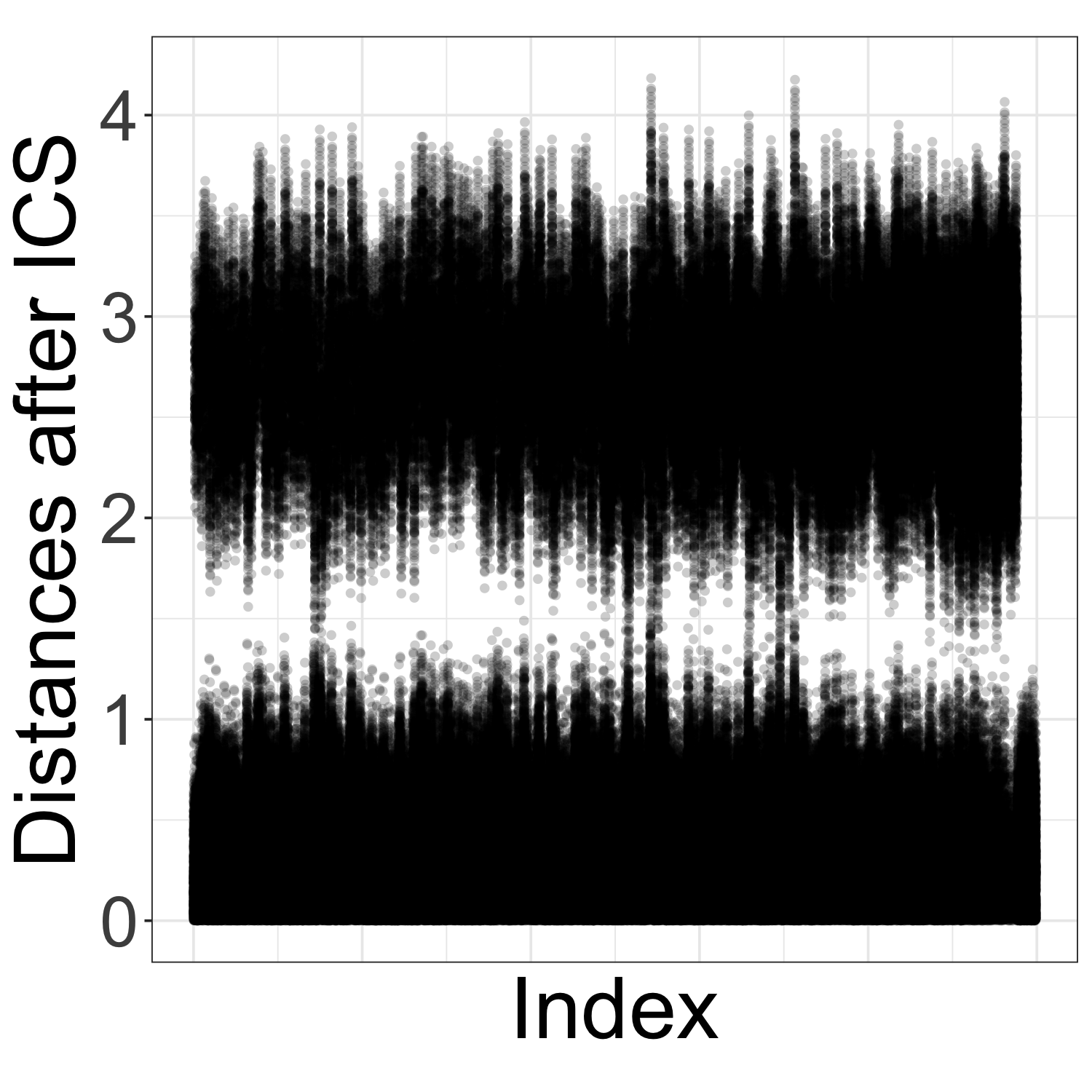}\\
\includegraphics[width=0.275\textwidth]{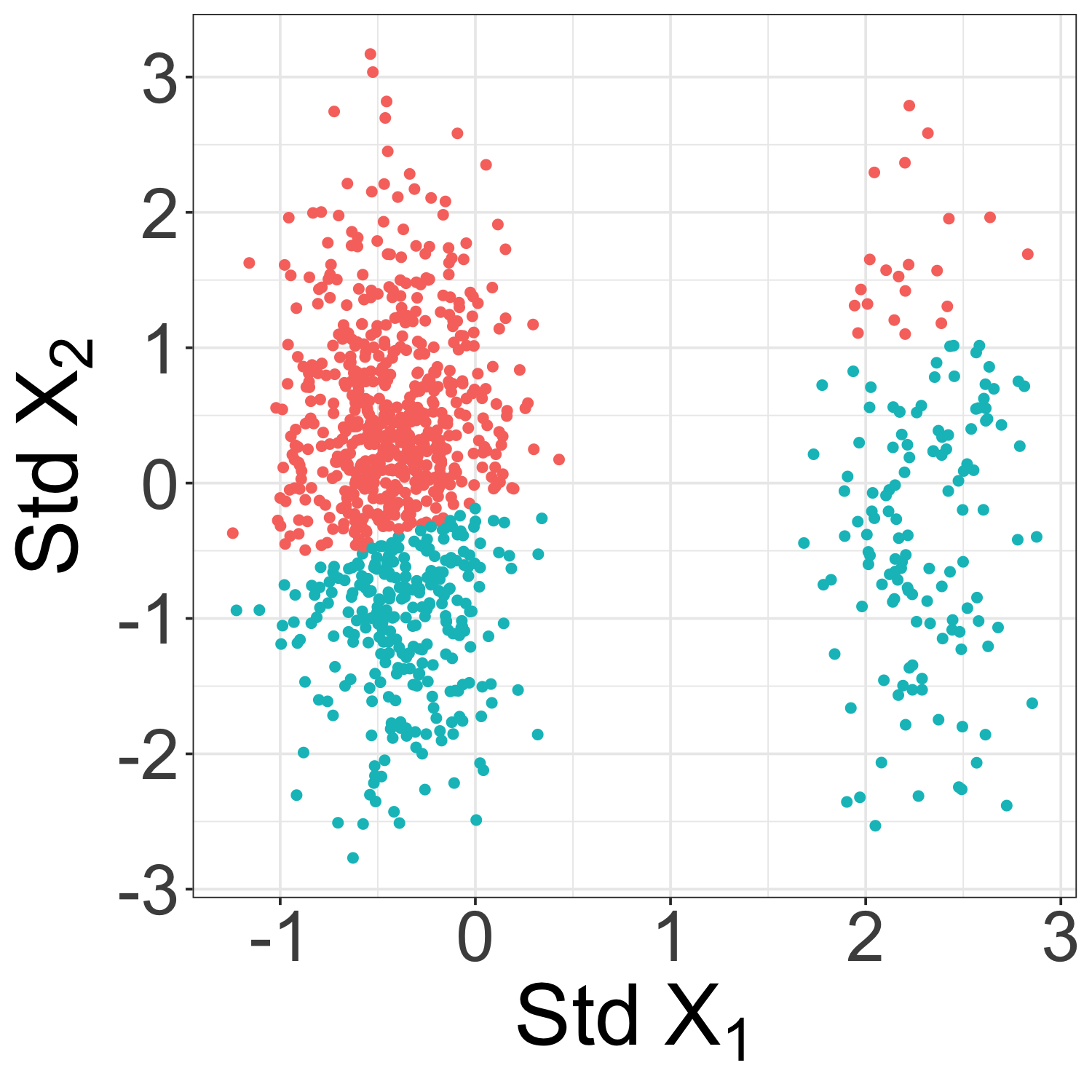}
\includegraphics[width=0.275\textwidth]{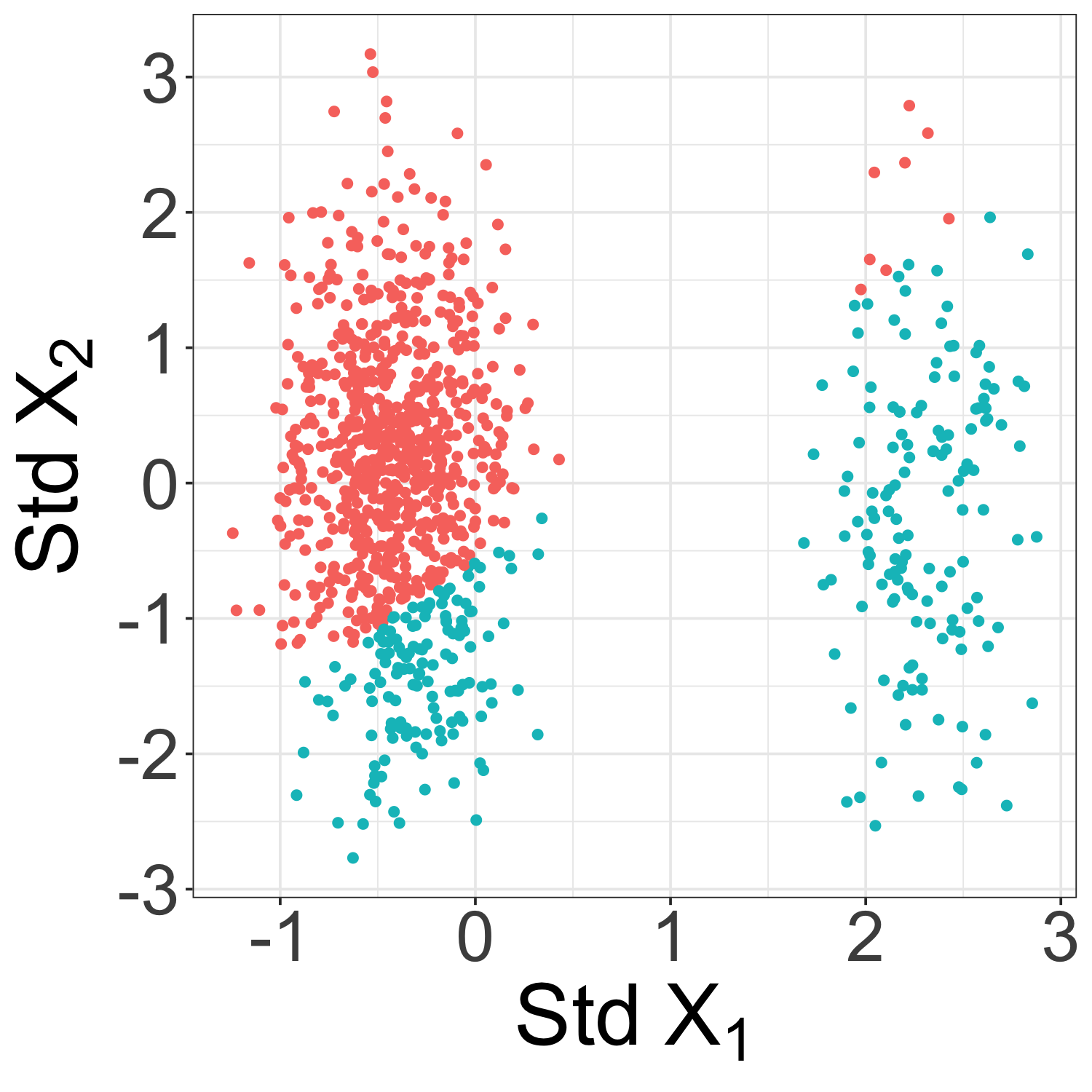}
\includegraphics[width=0.275\textwidth]{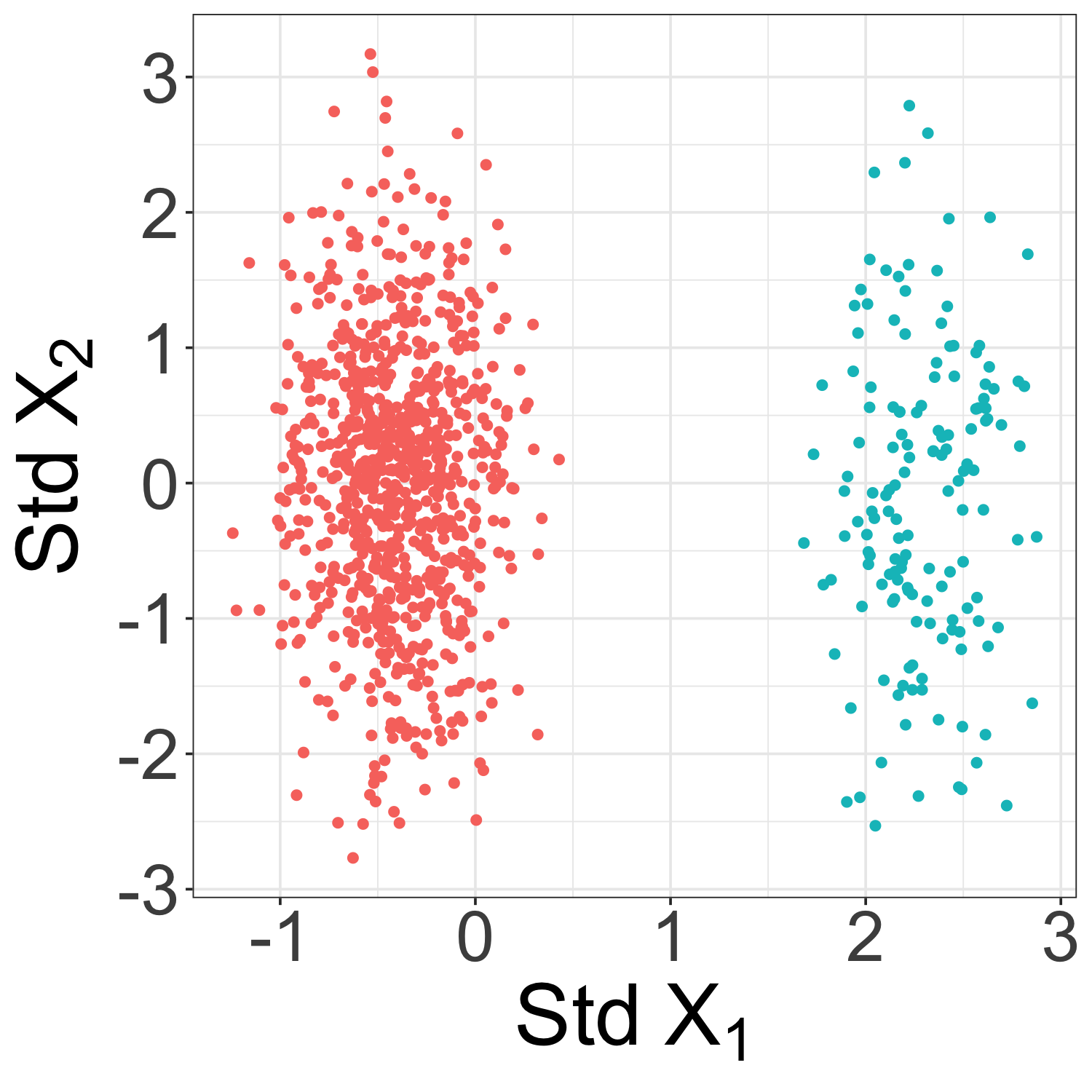}
\caption{Illustration of $k$means clustering on simulated data. The columns correspond to the standardized data (left), the PCA components (center), and the ICS components (right). 
The top row displays the corresponding scatterplots, the middle row visualizes pairwise Euclidean distances, and the bottom row shows scatterplots of the standardized data, with colors indicating the clusters obtained by $k$-means on the original data and on the reduced data from PCA and ICS.}
\label{fig:intro}
\end{figure}

To illustrate the above discussion, Figure~\ref{fig:intro} highlights the utility of dimension reduction for $k$means clustering in two dimensions. The example features two groups: 850 observations generated from the distribution $\mathcal{N}_{d=2}(0,\bo I_d)$ and 150 observations from $\mathcal{N}_{d=2}(\bo (10,0)^\top,\bo I_d)$. The distinction between the groups is apparent in the first dimension, while the second dimension carries no structure.

We compare the outcomes of the standard $k$means algorithm applied to (i)~the standardized data (variables Std $X_1$ and Std $X_2$), (ii)~the first component of PCA ($\text{PC}_{1}$) using the sample correlation matrix, and (iii) the first component of ICS ($\text{IC}_{1}$) using the $\cov-\covF$ pair of scatter matrices (see Section~\ref{sec:ics_scatters}). Figure~\ref{fig:intro} comprises 9 plots. The first row presents scatterplots of the standardized data (left), the PCA components (center), and the ICS components (right). While ICS successfully captures the dimension of interest on the first component, this is not the case for PCA. The middle row of Figure~\ref{fig:intro} visualizes pairwise Euclidean distances between the observations. For the standardized data (left), distinguishing between distances of observations within the same group and those from different groups is not feasible, which is consistent with the result of Proposition~\ref{propMD}. For PCA and ICS, pairwise distances are computed on $\text{PC}_{1}$ (center) and $\text{IC}_{1}$ (right), respectively. Only for $\text{IC}_{1}$ do we observe a discernible gap between the two types of distances. The bottom row of Figure~\ref{fig:intro} presents scatterplots of the standardized data with observations colored based on different $k$means clusterings. The two groups are not correctly identified when $k$means is applied to the original standardized data (left) or $\text{PC}_{1}$ (center), while they are accurately recovered for $\text{IC}_{1}$ (right).

\section{Invariant coordinate selection}\label{sec:ics}

In Section~\ref{sec:ics_scatters}, the principle of ICS and various scatter matrices are detailed. Section~\ref{sec:ics_prepro} discusses ICS in a clustering context, while Section~\ref{sec:ics_selection} introduces criteria for selecting which invariant coordinates to keep.

\subsection{Scatter matrices and the principle of ICS}\label{sec:ics_scatters}

ICS was initially introduced as generalized PCA in \citet{caussinus1990interesting}, \citet{caussinus1993ascona,Caussinus1995metrics,caussinus2007classification}, \citet{caussinus2003monitoring},  and received its current name in \citet{tyler2009invariant}, where many of its properties were also derived. The main ingredients of ICS are two scatter matrices. In this context, a scatter matrix is any $d \times d$ matrix-valued  functional $\V(F_{\X})=\V(\X)$ that is affine equivariant, meaning that
$$
\V(F_{\bs A\X + \bs b}) = \bs A \V(F_{\X}) \bs A^\top
$$
holds for all $d$-dimensional vectors $\bs b$ and all full rank $d \times d$ matrices $\bs A$. Here, $\X$ denotes a $d$-variate random vector with cumulative distribution function (CDF) $F_{\X}$. For a sample $\X_n =(\x_1,\ldots,\x_n)^\top$, the scatter functional is applied to the empirical CDF $F_{\X_n}$ and referred to as a scatter matrix.

The literature is abundant with suggestions for scatter matrices, with the most popular being the sample covariance matrix
\[
\cov(\X_n)= \frac{1}{n-1} \sum_{i=1}^n (\x_i-\bar{\x}_n)(\x_i-\bar{\x}_n)^\top,
\]
where $\bar{\x}_n= \frac{1}{n} \sum_{i=1}^n \x_i$ is the sample mean.
Other popular proposals originate from the robustness literature, such as the M-estimator of scatter based on the likelihood of the Cauchy distribution \citep{KentTyler1991}, denoted by $\mlt(\X_n)$. It can be computed using the simultaneous estimation equations
\begin{equation*}
\m_C(\X_n) = \sum_{i=1}^n \frac{d+1}{r^2_C(\x_i)+1} \x_i \bigg/ \sum_{i=1}^n \frac{d+1}{r^2_C(\x_i)+1}
\end{equation*}
and
\begin{equation*}
\mlt(\X_n) = \frac{1}{n} \sum_{i=1}^n \frac{d+1}{r^2_C(\x_i)+1}(\x_i - \m_C(\X_n))(\X_i - \m_C(\X_n))^\top,
\end{equation*}
where $r^2_C(\x_i)=(\x_i - \m_C(\X_n))^\top \mlt(\X_n) ^{-1} (\x_i-\m_C(\X_n))$ and $\m_C(\X_n)$ is the corresponding location estimator.

One of the most widely-used robust scatter matrices is the minimum covariance determinant estimator ($\mcd$) \citep{rousseeuw1985multivariate}. For a tuning parameter $\alpha$, the $\mcd$ selects out of the $n$ observations those $n_\alpha = \lceil \alpha n \rceil$ observations $\x_{i_1}, \ldots, \x_{i_{n_\alpha}}$ for which the sample covariance matrix has the smallest determinant, i.e., 
$$
\mcd_{\alpha}(\X_n) = c_\alpha \frac{1}{n_\alpha} \sum_{j=1}^{n_\alpha} (\x_{i_j}-\bar{\x}_{\alpha,n})(\x_{i_j}-\bar{\x}_{\alpha,n})^\top, 
$$ 
where $\bar{\x}_{\alpha,n}$ is the sample mean of the selected set of observations and $c_\alpha$ is a consistency factor. In the robustness literature, $\alpha$ is typically chosen from the interval $[0.5,1]$. For a value of $\alpha$ close to 0.5, the $\mcd_{\alpha}$ has a high breakdown point but low efficiency. Therefore, it is often combined with a reweighting step to increase efficiency, in which case the the $\mcd_{\alpha}$ is sometimes referred to as the raw $\mcd$. Here, we denote the reweighted version by $\rmcd_\alpha$.
For more details about $\mcd_\alpha$ and $\rmcd_\alpha$, see for example \citet{HubertDebruyne2010}, \citet{CatorLopuhaa2012}, and \citet{HubertDebruyneRousseeuw2018}.

On the other hand, there are scatter matrices for which robustness is not the primary focus, such as the matrix of fourth moments
$$
\covF(\X_n) =\frac{1}{n} \frac{1}{d+2} \sum_{i=1}^n r^2(\x_i)(\x_i - \bar{\x}_n)(\x_i - \bar{\x}_n)^\top,
$$
where $r^2(\x_i)=(\x_i - \bar{\bo x}_n)^\top \cov(\X_n) ^{-1} (\x_i - \bar{\x}_n)$ is the squared Mahalanobis distance. 

Several other scatter matrices have been proposed specifically in the context of ICS. For a positive, decreasing weight function $w$ and a tuning constant $\beta > 0$, a one-step M-estimator of scatter can be computed as follows:
$$
\scov_\beta(\X_n) =  \frac{\sum_{i=1}^n 
w(\beta\,r^2(\x_i)) (\x_i-\bar{\x}_n)(\x_i-\bar{\x}_n)^\top}
{\sum_{i=1}^n  w(\beta\,r^2(\x_i))},
$$
where $r^2(\x_i)$ is again the squared Mahalanobis distance. Closely related  to $\scov_\beta(\X_n)$ is the estimator 
$$
\tcov_\beta(\X_n) =  \frac{\sum_{i=1}^{n-1} \sum_{j=i+1}^n 
w(\beta\,r^2(\x_i,\x_j)) (\x_i-\x_j)(\x_i-\x_j)^\top}
{\sum_{i=1}^{n-1}\sum_{j=i+1}^n w(\beta r^2(\x_i,\x_j))},
$$
where $r^2(\x_i,\x_j) = (\x_i - \x_j)^\top \cov(\X_n) ^{-1} (\x_i - \x_j)$. Note that $\tcov_\beta(\X_n)$ is based on pairwise differences and therefore does not require a location estimator. Both estimators are studied in great detail in \citet{caussinus1993ascona,Caussinus1995metrics}. \citet{RuizGazen1996} suggests in this context to consider also
$$
\ucov_\beta(\X_n) = (\scov_\beta(\X_n)^{-1}-\beta\,\cov(\X_n)^{-1})^{-1}.
$$
For these scatter matrices, we use the weight function $w(x) = \exp(-x/2)$, and we set $\beta = 2$ for $\tcov$ and $\beta=0.2$ for $\ucov$.

The last scatter matrix we introduce is the local shape matrix $\lcov_{\V_0,\beta}$ suggested by \citet{Hennig2009}. For $\lcov_{\V_0,\beta}$, a scatter $\V_0$ is computed first, and the distance matrix of the corresponding pairwise Mahalanobis distances is calculated. Then, for each observation, the sample covariance matrix is computed based on the $n_\beta = \lceil \beta n \rceil$ nearest neighbors. All $n$ covariance matrices are subsequently averaged, for which they first need to be appropriately scaled to be comparable. We standardize them in such a way that they all have determinant one. Note that this is not a scatter in our original sense but only a shape matrix \citep{NordhausenTyler2015}, which suffices for ICS. We use $\lcov_{\cov,0.1}$, but one could also use $\lcov_{\mcd_{0.8},0.1}$ as in \citet{Hennig2009}. 

In a clustering framework, $\tcov$ and $\lcov$ can be seen as estimators of the within-group covariance matrix (cf.\ Figure~\ref{fig:scatter_shapes}). An alternative estimator proposed by \cite{art1982data} and \cite{gnanadesikan1993mahalanobis} is based on the smallest within-group pairwise differences of observations. However, it will not be considered further, as the algorithm is iterative and involves a graphical aid to determine the number of pairwise differences to consider.

A general result is that all scatter matrices are proportional to each other at the population level (if they exist) for exchangeable sign-symmetric distributed random vectors, which includes all elliptical distributed random vectors \citep{NordhausenTyler2015}. This implies that for such models, when whitening the data \citep[e.g.,][]{NordhausenOjaTyler:2008} or performing principal component analysis (PCA), it does not matter which scatter matrix and location estimator are used, as the results differ only in scale. Similarly, \citet{Tyler2010} showed that a comparison of two scatter matrices is only meaningful if the sample size $n$ is larger than the dimension $d$, making this a requirement for ICS.

Given a sufficient sample size, in a clustering framework assuming for example a Gaussian mixture model, different scatter matrices will measure different population quantities. The idea behind ICS is to exploit these differences. An ICS matrix $\bs W(\X_n)$ is a matrix that simultaneously diagonalizes two scatter matrices $\V_1(\X_n)$ and $\V_2(\X_n)$ in the following way:
\begin{equation*}
\bs W(\X_n) \V_1(\X_n)  \bs W(\X_n)^\top= \bs I_d
\end{equation*}
and
\begin{equation*}
\bs W(\X_n)\bo \V_2(\X_n)  \bs W(\X_n)^\top= \bs D(\X_n),
\end{equation*}
where $\bs D(\X_n)$ is a diagonal matrix with diagonal elements $\lambda_1 \geq \lambda_2 \geq \dots \geq \lambda_d$. ICS can be interpreted as a transformation where $\X_n$ is whitened with respect to $\bo V_1$, and then a subsequent PCA is performed with respect to $\V_2$ to see what structure is still left after the whitening with $\V_1$. ICS is usually computed via a generalized eigenvalue decomposition \citep[see][]{tyler2009invariant}. 

The so-called invariant coordinates are then obtained as
$$
\Z_n =(\X_n - \bo 1_n\bs T(\X_n)^\top)\bs W({\X_n})^\top, 
$$
where $\bo 1_n$ denotes an $n$-variate vector full of ones, and $\bs T(\X_n)$ denotes a location estimator, usually the one that goes along with $\V_1$. In case $\V_1$ has no natural accompanying location, any location estimator can be used for the purpose of centering, or centering can be omitted when computing the ICs. The name of ICS is motivated by a property of the (centered) invariant coordinates, which are often also called invariant components or simply ICs. These ICs are invariant in the sense that for any linear transformation of $\X_n$, i.e., $\X_n^* = \bo \X_n \bo A^\top + \bo 1_n \bo b^\top$, it holds that
$$
(\X_n^* - \bo 1_n \bs T(\X_n^*)^\top) \bs W(\X_n^*)^\top  =  (\X_n - \bo 1_n\bo T(\X_n)^\top) 
\bs W({\X_n})^\top \bs J
$$
if the diagonal elements of $\bs D(\X_n) = \bs D(\X_n^*)$ are all distinct, where $\bs A$ and $\bs b$ are as above and $\bs J$ is a $d \times d$ sign-change matrix, i.e., a diagonal matrix with $\pm 1$ on its diagonal. Thus, if ICS is performed on the same data but measured in different coordinates systems, the resulting ICs will differ at most by their signs. Note that in the following, we will drop the dependency of the matrices on $\X_n$ in the notation when the context is clear.

Not only do the values $\lambda_1,\ldots,\lambda_d$ fix the order of the components, but they also have a concrete interpretation as a measure of kurtosis for the corresponding components in the sense of $\V_1$ and $\V_2$, since they correspond to the ratio of two scale measures, i.e.,
\[
\kappa(\bs w^\top \X) = \frac{\bs w^\top \V_{2}(\X) \bs w}{\bs w^\top \V_{1}(\X) \bs w},
\]
where $\bs w$ is a $d$-dimensional vector. If $\V_1=\cov$ and $\V_2=\covF$, this measure relates to the classical kurtosis measure of Pearson \citep{NordhausenOjaOllila2011b}. \citet{tyler2009invariant} show then that the maximal and minimal kurtosis in the sense of $\V_1$ and $\V_2$ that can be obtained for $\X_n \bs w^\top$ is captured by $\lambda_1$ and $\lambda_d$, respectively.

\subsection{ICS as preprocessing step for clustering}\label{sec:ics_prepro}

While ICS has diverse applications, including its use in a transformation-retransformation framework, outlier detection, and as a method for independent component analysis \citep{tyler2009invariant,NordhausenOjaOllila2011b,archimbaud2018CSDA,NordhausenOjaTyler:2008}, our focus is on studying ICS within the context of clustering.

The most common framework of clustering are location mixture models. In scenarios where class labels are known, a natural approach would be to maximize the ratio of the between-group and within-group covariance matrices to obtain Fisher's linear discriminant subspace. However, ICS addresses situations where class labels are unknown. The objective is for two scatter matrices, which are in general not proportional to each other in this model, to capture this ratio. Indeed, \citet{tyler2009invariant} proved that in an elliptical mixture model where the $q$ populations have proportional scatter matrices and different means, ICS will generally produce at most $q-1$ values among $\lambda_1,\ldots,\lambda_d$ which differ from the remaining non-informative values, which are all equal and denoted  $\lambda_0$. In the context of ICS as a preprocessing step for clustering, the goal is to use only the invariant coordinates spanning Fisher's linear discriminant subspace and discard other coordinates. We seek the $l \times d$ matrix $\bs W_{DS}$ containing rows of $\bs W$ corresponding to values of interest in $\lambda_1\ldots,\lambda_d$, i.e., those with $\lambda_i \neq \lambda_0$, $i\in \{1,\ldots,d\}$. The problem is that the values of $\lambda_0$ and $l \leq q-1$ depend on the distributions of the groups, their mixture weights, and the selected scatters $\V_1$ and $\V_2$. 
Unlike PCA where the first few PCs are typically of interest, the interesting ICs can be the first, last, or a combination thereof. Assuming $\V_1$ is more robust than $\V_2$, it is expected, heuristically speaking and based on the generalized kurtosis interpretation, that groups with different sizes will exhibit large $\V_1$-$\V_2$ kurtosis values, while groups with similar group sizes will have small $\V_1$-$\V_2$ kurtosis values. This implies that large and small clusters will be found in the different extremes \citep[e.g.,][]{pena2010eigenvectors}.

It is well established that there is no general best scatter combination for ICS, but that some combinations are better suited for certain applications than others. For example, \citet{pena2010eigenvectors} consider the combination $\cov-\covF$, known as FOBI \citep{Cardoso1989,NordhausenVirta2019}, to reveal clusters. However, this combination is highly sensitive to extreme observations leading \citet{archimbaud2018CSDA} to recommend it for outlier detection. This combination is also of interest due to being completely moment-based and therefore easy to study. For example, one can show that in a Gaussian mixture framework, the value of the non-interesting (Gaussian) components is $\lambda_0=1$. For a mixture of two Gaussians with the same covariance matrix, it can be shown that if the mixture weight for the smaller cluster is $(3+\sqrt{3})^{-1} \approx 21$\%, then the kurtosis of the mixture is the same as for a pure Gaussian component, independent of the difference in locations. Therefore, the generalized eigenvalue of such a component equals 1 \citep[see, e.g.,][]{tyler2009invariant}.

When using ICS as a preprocessing step for clustering, the general motivation is that one scatter matrix should measure within-group scatter and the other between-group scatter (or equivalently, total scatter). Therefore, it seems natural to use one scatter matrix to estimate local structure and another to estimate global structure. For instance, \citet{Hennig2009} suggests $\lcov$ for the former. Robust scatter matrices could also be an interesting option to capture the dispersion of a single cluster. Specifically, for $\mcd_\alpha$, it might be worth considering values of $\alpha$ that are smaller than 0.5 for this purpose.

While different scatter combinations have been considered for ICS in a clustering framework, a thorough comparison is still lacking in the literature. To conclude this part, we emphasize that ICS is performed in an unsupervised manner without using class label information. If such information is available, one of the scatter matrices involved in ICS can use this information, converting it to a supervised scatter matrix, which leads to supervised invariant coordinate selection (SICS). Many supervised dimension reduction methods, such as linear discriminant analysis (LDA), sliced inverse regression (SIR), and sliced average variance estimation (SAVE) can be seen as special cases of SICS \citep{LiskiNordhausenOja2014}.

\subsection{Selection of the invariant coordinates to retain}\label{sec:ics_selection}

Another topic often ignored in the literature is the selection of the ICs that should be retained for clustering. When the number of groups $q$ is known, theoretical properties of ICS suggest that the number of invariant coordinates to retain is typically less than or equal to $q-1$. In what follows, we compare several criteria for the selection of invariant coordinates. Two of the criteria that we consider further are based on the assumption of a known $q$, while one criterion operates without such an assumption. In a specific data analysis, the most natural approach might be to look at the scatterplot matrix of the invariant coordinates and at the generalized eigenvalues to determine which $l<d$ invariant coordinates carry relevant information for clustering. However, for a more formal approach,  we discuss potential rules below.

Assume the data come from a Gaussian mixture model with equal covariance matrices and $q < d$ mixture components. In this case, non-interesting coordinates follow a Gaussian distribution whereas the informative ones are non-Gaussian. This setup aligns with the framework of non-Gaussian component analysis (NGCA), aiming to identify a non-Gaussian subspace. \citet{nordhausen2017asymptotic,nordhausen2022asymptotic} show that $\cov-\cov4$ serves as an NGCA method recovering this subspace. They derive asymptotic and bootstrap tests for the number of non-Gaussian coordinates, which can be used in a successive testing strategy to obtain an estimate for $l$. The key concept is that all Gaussian coordinates have a generalized eigenvalue of 1, resulting in small variance among these generalized eigenvalues. Resampling-based estimators of $l$ for this scatter combination are considered in \citet{LuoLi2016,luo2021order}. These tests, incorporating a joint estimation strategy, were extended to a resampling framework for all scatter combinations in an NGCA setting in \citet{radojivcic2019non}. The challenge lies in the unknown value of the non-interesting (Gaussian) generalized eigenvalues, except for their shared generalized eigenvalue. The idea in \citet{radojivcic2019non} is to go for a specific $l_0$ through all combinations of $l_0$ neighboring generalized eigenvalues to choose the ones with minimal variance, and to use bootstrapping to decide if assuming equality is reasonable. However, bootstrapping is quite computationally expensive in this context. 

If the number of clusters $q$ is assumed known, one rule could be to find the $d-q+1$ generalized eigenvalues with the smallest variance and keep the components belonging to the non-included generalized eigenvalues. This rule is referred to as the \emph{var criterion}. While initially outlined in an NGCA framework, this approach can also be applied in a more general framework of elliptical mixtures, where again the generalized eigenvalues of the purely elliptical part will be equal, and the generalized eigenvalues of components containing cluster information will differ.

A simpler criterion than the \emph{var criterion} within the framework of an elliptical mixture model is to assume $q \leq d/2$, which implies that the majority of the generalized eigenvalues should be equal. Then a rule could be to choose the $q-1$ components whose generalized eigenvalues deviate the most from the median of all generalized eigenvalues. This approach is referred to as the \emph{med criterion}. 

In a Gaussian mixture model, a simple approach is to apply marginal tests for normality and select all non-normal components, which we call accordingly the \emph{normal criterion}. This approach proved successful for outlier detection via ICS in \citet{archimbaud2018CSDA}. Note that since those authors work in an outlier detection framework, they look for non-Gaussian components only among the first components. In a clustering context, it is natural to consider the first and last components. 

To summarize the section on ICS, it is essential to highlight that the key characteristic of ICS as a preprocessing step for clustering is that the components are related to generalized kurtosis measures. Extreme generalized kurtosis values serve as indicators of potential interesting structures for clustering. This motivation does not strictly rely on the assumption of an underlying elliptical location-mixture model but can be argued for other models with group structures. Nevertheless, the theory and many of the component selection methods discussed above rely on a Gaussian mixture model, one of the most popular clustering models. Additionally, ICS is most meaningful when the clustering information is embedded in a lower dimensional subspace, and the resulting dimension reduction is independent of the subsequent clustering method applied. In the following simulation study, we apply some popular clustering methods after component selection using ICS, although, in general, any other clustering method could be used. This makes ICS a general tool for dimension reduction in a clustering context, distinguishing it from numerous subspace estimation methods tailored for specific clustering methods, see for instance \citet{soete1994k,vichi2001factorial,BouveyronCeleuxMurphyRaftery2019}.

\section{Simulation study}
\label{sec:sim}

In order to investigate the performance of tandem clustering with ICS and PCA, we perform simulations. The simulation design is described in Section~\ref{sec:sim-design}, and the main results are presented in Section~\ref{sec:sim-results}. Additional findings are discussed in Section~\ref{sec:sim-additional}, while Section~\ref{sec:sim-discussion} provides a brief summary and discusses some limitations of the simulation study.

\subsection{Simulation design}
\label{sec:sim-design}

Focusing on a data generating process in which the cluster structure lies in a low-dimensional subspace, we compare the performance of ICS and PCA with respect to reducing the dimensionality of the data while keeping the relevant structure for clustering. We study the impact of the scatter pair and the component selection criterion in different cluster settings and in the presence of outliers. For assessing the performance of the methods, we compute $\eta^{2} = 1 - \Lambda$, where $\Lambda$ denotes Wilks' lambda. That is, we compute
\begin{equation*}
\eta^{2} = 1 - \frac{\det(\mat{E})}{\det(\mat{T})},
\end{equation*}
where $\mat{E}$ is the within-group sum of squares and cross-products matrix and $\mat{T}$ is the total sum of squares and cross-products matrix. The $\eta^{2}$ is a measure of discriminatory power commonly used in discriminant analysis (see, e.g., \citealp{mclachlan1992discriminant}; as well as \citealp{todorov2007robust}, for a robust version of Wilks' lambda). It is in the interval $[0, 1]$, with a higher value indicating better discriminatory power. 

After dimension reduction, we apply clustering methods to the obtained components to confirm the results on the discriminatory power. Our aim is not to compare clustering methods, hence for the most part we limit our study to five well-known methods: partitioning around medoids (PAM) \citep{kaufman2009finding}, $k$means \citep{hartigan1979algorithm}, t$k$means (a trimmed version of $k$means that identifies a specified proportion of observations as outliers) \citep{cuesta1997trimmed}, as well as model-based clustering with Gaussian mixtures \citep{fraley2002model} without allowing for noise (mclust) and with allowing for noise (rmclust). These clustering methods take the number of clusters $k$ as an input, which we set equal to the true number of clusters $q$ to keep the simulations simple. As an evaluation measure, we compute the adjusted Rand index (ARI) \citep{hubert1985comparing}, which has a maximum value of 1 in the case of perfect clustering, while a value of 0 is expected for a random cluster assignment.
For comparison, we also apply the clustering methods to the simulated data without dimension reduction, in which case we standardize the data using the mean and standard deviation before $k$means, and we employ robust standardization using the median and median absolute deviation (MAD) before PAM and t$k$means.

We generate $n=1000$ observations on $d=10$ variables from the model in~\eqref{mixt} for different numbers of clusters $q \in \{2,3,5\}$ with $\mat{\Sigma}_W = \bs I_d$, $\vect{\mu}_1 = \obs{0}$, and $\vect{\mu}_{h+1} = \delta \vect{e}_h$, $h = 1, \dots, q-1$, where $\delta=10$ and $\vect{e}_h$ is a $d$-dimensional vector with one in the $h$-th coordinate and zero elsewhere. In this setting, the cluster structure lies in a low-dimensional subspace of dimension $l=q-1$. We thereby consider 22 different combinations of the mixture weights $\epsilon_1, \dots, \epsilon_q$ (see \ref{app:sim-details} for details).

In addition to a baseline setting without outliers, we consider settings in which 2\% and 5\% of the observations are replaced by outliers.  We draw the outliers from a uniform distribution on a large hyperrectangle with a central cutout in the shape of a smaller hyperrectangle around the original data. The length of each side of the smaller hyperrectangle is given by the range of the original data in the respective variable, while the corresponding side of the larger hyperrectangle is twice as long.  Note that we treat outliers as an additional group in the computation of the $\eta^{2}$ and the ARI.

Regarding scatter matrices, we apply ICS with the following scatter pairs:
$\lcov-\cov$, 
$\tcov-\cov$, 
$\tcov-\ucov$, 
$\mcd_{\alpha}-\cov$ and $\rmcd_{\alpha}-\cov$  with $\alpha \in \{0.1, 0.2, 0.25, 0.5, 0.75\}$,  
$\mcd_{0.25}-\mcd_{0.95}$ and  $\rmcd_{0.25}-\rmcd_{0.95}$, 
$\mlt-\cov$, and
$\cov-\covF$ 
(see Section~\ref{sec:ics_scatters} for details on those scatter matrices). Note that $\mcd_{\alpha}$ and $\rmcd_{\alpha}$ are computed using the FAST-MCD algorithm of \citep{rousseeuw1999fast}. 
Furthermore, we apply PCA using the correlation matrices derived from $\cov$ and $\rmcd_{0.75}$ \citep[the latter being advocated by, e.g.,][]{croux1999influence}. Other robust methods for PCA such as ROBPCA \citep{hubert2005robpca} or MacroPCA \citep{hubert2019} are beyond the scope of our analysis.

Concerning the selection of components, we use four criteria for ICS (see Section~\ref{sec:ics_selection} for details): the med criterion and the var criterion (which select $k-1$ components), the normal criterion using the D’Agostino test of skewness \citep{dagostino1970transformation} at a 5\% significance level (which is not restricted to selecting a certain number of components), and an oracle criterion selecting the $k-1$ components (among the first and last components) that maximize the discriminatory power for the true clusters (as measured by the $\eta^{2}$). For PCA, we use two criteria: the $80\%$ criterion (which keeps as many components as necessary to explain at least $80\%$ of the variability in the data), and the $k-1$ criterion (which keeps the first $k-1$ components).

We simulate 100 data sets for each of the 66 different cluster and outlier settings. Overall, we compare 72 different dimension reduction strategies: 68 for ICS (17 scatter pairs and 4 criteria) and 4 for PCA (2 scatters and 2 criteria).

\subsection{Main results}
\label{sec:sim-results}

Due to the large number of dimension reduction strategies, we only show a selection of the most relevant results here. Section~\ref{sec:best-eta2} presents the results for the best-performing methods in terms of discriminatory power of the selected components, while Section~\ref{sec:best-ARI} discusses clustering performance of selected clustering methods after dimension reduction. 

A detailed discussion on the selected dimension reduction methods and component selection criteria is given in \ref{app:selection}. For ICS, the best-performing scatter pairs are $\lcov-\cov$, $\tcov-\cov$, as well as the robust scatter pair $\tcov-\ucov$. Furthermore, we compare the med criterion with the oracle criterion for component selection. The former is a representative selection since there is not much difference in the selection criteria for those best-performing scatter pairs, while the latter is included for reference purposes as it uses information on the true clusters and can therefore not be computed in practice. For PCA, the best-performing method is $\rmcd_{0.75}$ combined with the $80\%$ criterion for selecting the components.

\subsubsection{Discriminatory power of selected components}
\label{sec:best-eta2}

\begin{figure}[t!]
\includegraphics[width=\textwidth]{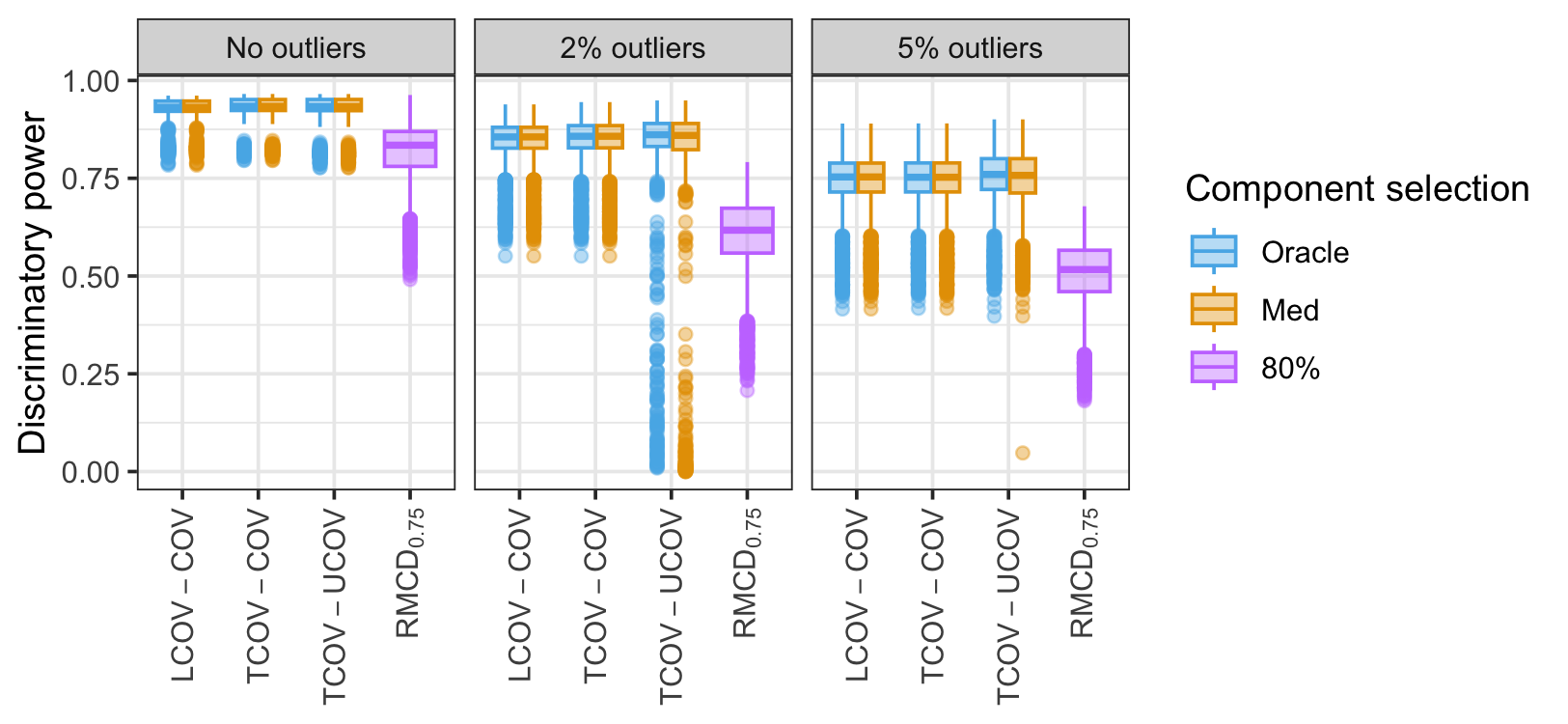}
\caption{Discriminatory power as measured by the $\eta^{2}$ of the best-performing dimension reduction methods for ICS and PCA, respectively, using the corresponding best component selection criteria. Boxplots display the results across the different cluster settings with $100$ simulation runs each. Results for different outlier settings are shown in separate panels.}
\label{fig:best-eta2}
\end{figure}

We first focus on the overall results across the investigated cluster settings. Figure~\ref{fig:best-eta2} shows boxplots of the discriminatory power of the selected methods for different outlier settings. 

With and without outliers, ICS with $\lcov-\cov$ and $\tcov-\cov$ clearly outperforms PCA with $\rmcd_{0.75}$ in terms of discriminatory power. This is not surprising, as PCA is not designed to pick up the relevant structure for clustering. Although ICS with the robust scatter pair $\tcov-\ucov$ also clearly outperforms PCA with $\rmcd_{0.75}$ when there are no outliers or 5\% outliers, a conclusion cannot be drawn in the setting with 2\% outliers. While the median $\eta^{2}$ is much higher for $\tcov-\ucov$ than for $\rmcd_{0.75}$, the former also exhibits many points with a very small $\eta^{2}$ in Figure~\ref{fig:best-eta2}. This behavior is investigated in more detail below in the context of different cluster settings. 

As the amount of outliers increases, we observe that the discriminatory power decreases for all methods, together with a slight increase in variability. Nevertheless, the discriminatory power of ICS with $\lcov-\cov$ and $\tcov-\cov$ remains good in all settings. For PCA, the decrease in discriminatory power is much more pronounced despite using the robust scatter matrix $\rmcd_{0.75}$. This indicates that PCA as a preprocessing technique for clustering is much more affected by outliers than ICS with a suitable scatter pair.

\begin{figure}[t!]
\includegraphics[width=\textwidth]{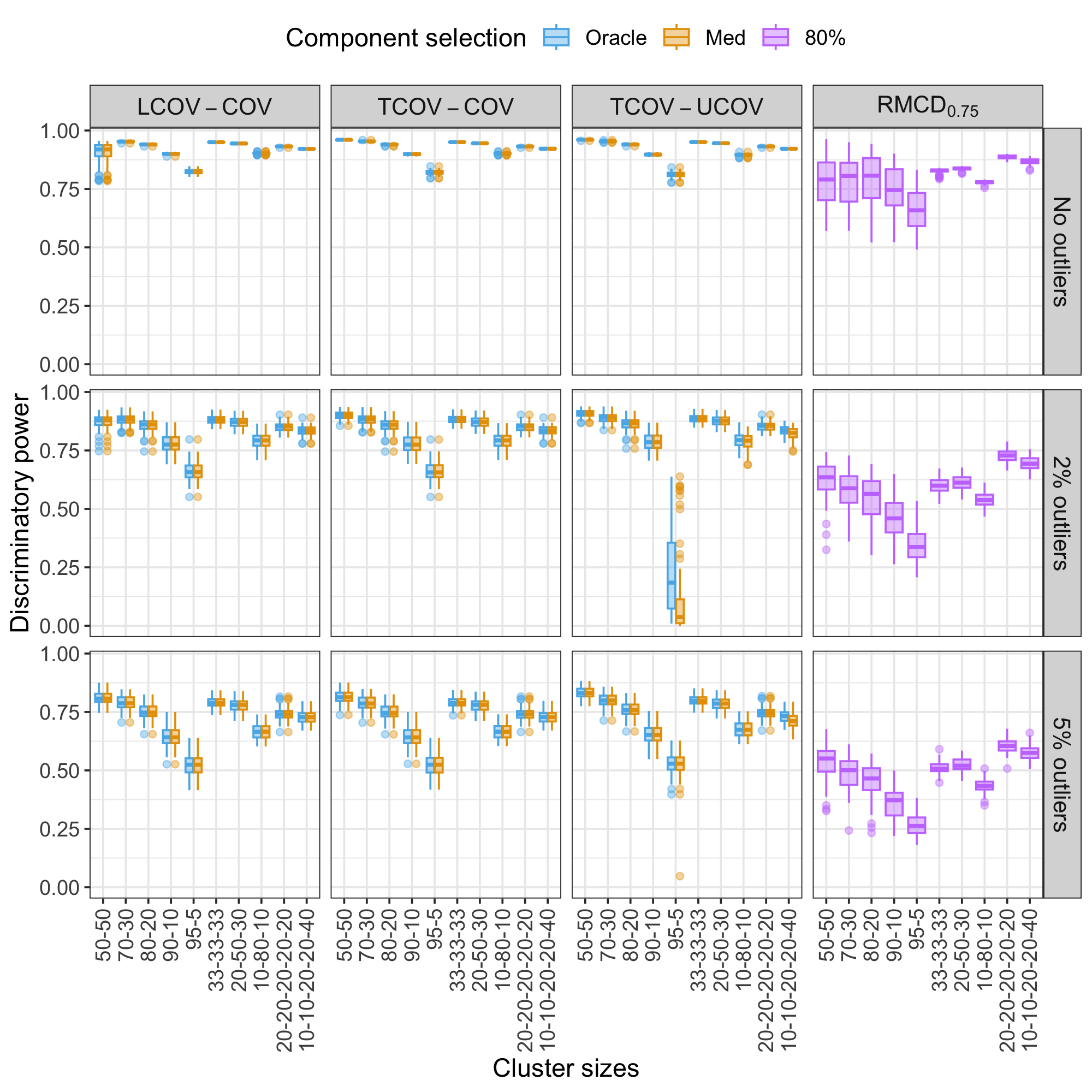}
\caption{Discriminatory power as measured by the $\eta^{2}$ of the best-performing dimension reduction methods for ICS and PCA, respectively, using the corresponding best component selection criteria. Boxplots display the results for selected cluster settings over $100$ simulation runs. Results for different outlier settings are shown in separate rows, and results for different methods are shown in separate columns.}
\label{fig:best-eta2-clusters}
\end{figure}

Next, we take a look at specific cluster settings to gain further insights. Since we investigate a large number of cluster settings, we present boxplots for a representative selection of cluster settings in Figure~\ref{fig:best-eta2-clusters}. The main finding is that $\tcov-\cov$ may be preferred over $\lcov-\cov$, as the former has lower variability for two perfectly balanced clusters (50--50). However, both scatter pairs yield very similar and excellent discriminatory power in the investigated cluster settings. Furthermore, note that the med criterion essentially yields the same discriminatory power as the oracle criterion for those two scatter pairs.

Figure~\ref{fig:best-eta2-clusters} also provides more insights into the behavior of the robust scatter pair $\tcov-\ucov$. Specifically, we observe that this scatter pair leads to similar discriminatory power as $\lcov-\cov$ and $\tcov-\cov$, except in the setting with $q=2$ highly unbalanced clusters (95--5) and with 2\% of outliers. In this specific situation, the discriminatory power considerably drops towards 0 and the variability increases. This is due to an artifact of our simulation design, where the outlyingness structure is likely to be constrained to a low-dimensional subspace if the number of outliers is small. Both $\tcov$ and $\ucov$ treat a small cluster as outliers, meaning that this scatter pair tends to ignore (some of) the relevant structure in such settings. There is of course a philosophical discussion as to where the line is between a group of outliers and a small cluster of observations, but such a discussion is beyond the scope of this paper.

For PCA with $\rmcd_{0.75}$, we find much larger variability in settings with two clusters compared to the other cluster settings. 
Moreover, in all investigated cluster settings, the discriminatory power of PCA with $\rmcd_{0.75}$ is lower or at best comparable to that of ICS with $\tcov-\cov$ and $\lcov-\cov$.

\subsubsection{Clustering performance}
\label{sec:best-ARI}

In order to validate our findings with respect to discriminatory power, we apply five popular and relatively simple clustering methods. To keep computation time low, we apply the clustering methods only for a subset of cluster settings (see \ref{app:sim-details} for details).

\begin{figure}[t!]
\includegraphics[width=\textwidth]{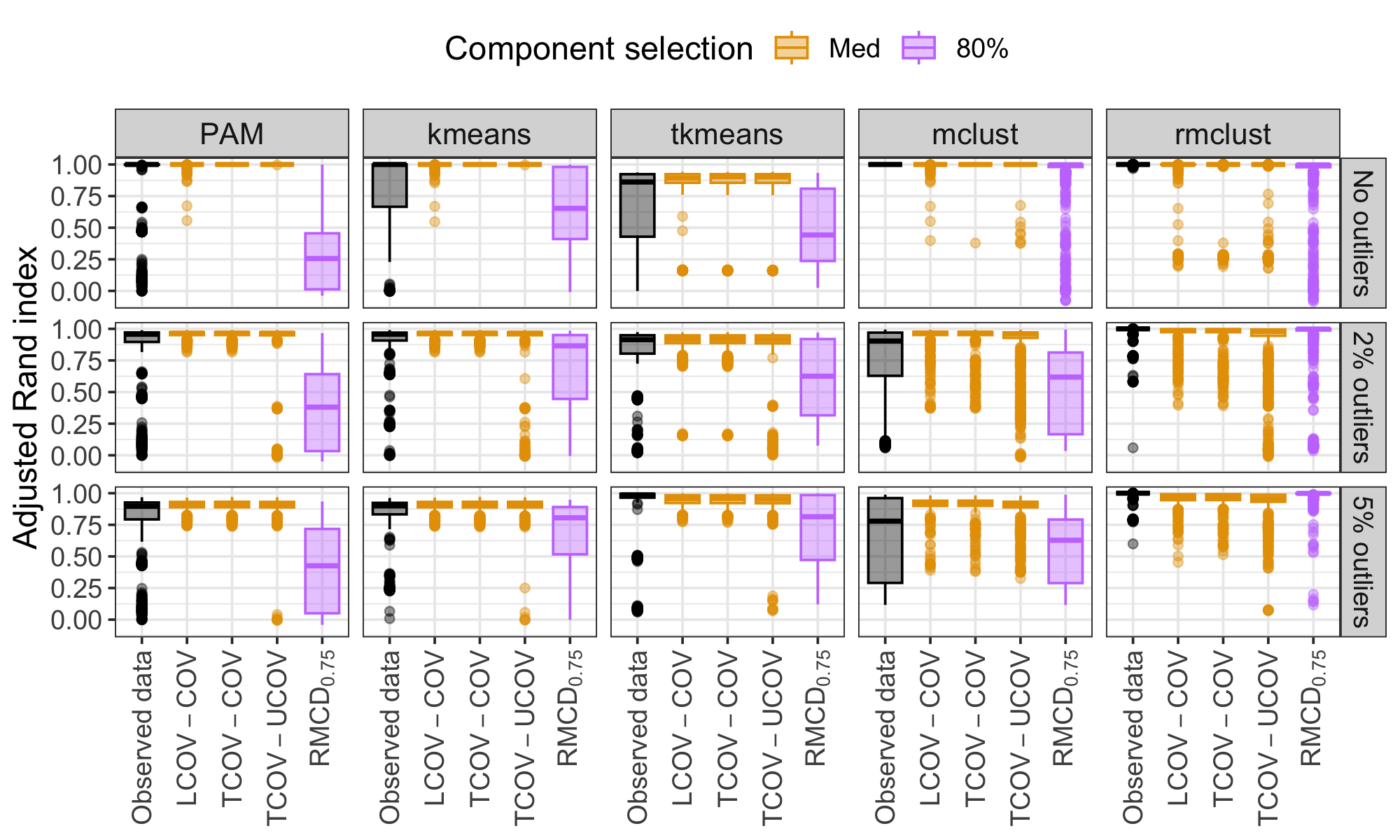}
\caption{Clustering performance as measured by the ARI after the best-performing dimension reduction methods for ICS and PCA, respectively, using the corresponding best component selection criteria. Boxplots display the results across the different cluster settings with $100$ simulation runs each. Results for different clustering methods are shown in separate columns, and results for different outlier settings are shown in separate rows.}
\label{fig:best-ARI}
\end{figure}

Figure~\ref{fig:best-ARI} displays overall results of the ARI of the selected methods. It contains boxplots across the selected cluster settings for the different clustering methods and outlier settings.  Compared to the ARI on the original data without dimension reduction, ICS with $\lcov-\cov$ and $\tcov-\cov$ improves the performance of PAM, $k$means, t$k$means, and mclust. For PAM and $k$means, there is not much change in the median ARI as it is already near perfect for the observed data, but variability is reduced or there are fewer instances where clustering fails. For t$k$means, we observe a similar effect, taking into account that this method trims away some observations so that the resulting outlier component is reflected accordingly in a somewhat lower ARI. For mclust, ICS yields a substantial improvement in ARI when outliers are present. For rmclust, on the other hand, the ARI is almost perfect for the observed data, as expected in this simulation design with Gaussian clusters and outliers. Nevertheless, there are some instances where rmclust fails. Note that ICS with $\lcov-\cov$ or $\tcov-\cov$ and a simple clustering method such as $k$means lead to an ARI as high as for rmclust for any situation and never completely breaks down.

As the amount of outliers increases, the median ARI of PAM and $k$means decreases while the median ARI of t$k$means increases accordingly. This behavior is expected and can be observed on the original data and after ICS. Also note that in the setting with 5\% outliers, t$k$means on the observed data performs almost perfectly in most instances (but fails in some instances). Keep in mind that for t$k$means, we used the default trimming proportion of 5\%. In practice, the trimming proportion needs to be carefully tuned, and the interplay of the trimming proportion and the number of clusters is quite complex \citep{garcia2011}. Hence, this result could be due to the trimming proportion being equal to the true amount of outliers.

Otherwise, PAM, $k$means, and t$k$means after ICS are not much affected by the outliers. This reflects that the discriminatory power of the selected components remains high, even though the aforementioned clustering methods fail in some instances for $\tcov-\ucov$. This phenomenon is more pronounced for mclust and rmclust, where all scatter pairs yield a considerable number of unsuccessful clustering instances. 

PCA with $\rmcd_{0.75}$ deteriorates the overall performance of all clustering methods compared to clustering the observed data, either by a substantially lower median or a higher variability in ARI. This demonstrates that PCA struggles to capture the relevant structure for clustering.

From the overall results across cluster settings, we can already see that the results for clustering performance reflect the results for discriminatory power very well. A detailed discussion of the clustering performance for the selected cluster settings is therefore moved to \ref{app:ARI-clusters}.

\subsection{Additional findings}
\label{sec:sim-additional}

As we have seen from the previous section that clustering performance nicely follows the results for the discriminatory power of the selected components, Sections~\ref{sec:MCD-COV} and~\ref{sec:COV-COV4} focus on discriminatory power using the $\mcd_{\alpha}-\cov$ and $\cov-\covF$ scatter pairs, respectively. Furthermore, Section~\ref{sec:ARI-GMM} provides a comparison of model-based clustering after ICS with methods that integrate dimension reduction into Gaussian mixture modeling.

\subsubsection[Discriminatory power with MCD-COV scatter pairs]{Discriminatory power with $\mcd_{\alpha}-\cov$ scatter pairs}
\label{sec:MCD-COV}

The MCD is typically applied in the context of robust statistics, where a subset size based on the parameter $\alpha \geq 0.5$ is used to capture the covariance structure of the majority of the data. However, in the context of ICS as a preprocessing step for clustering, $\alpha \leq 0.5$ is better suited to capture the local within-cluster structure. Hence, we focus on the scatter pairs $\mcd_{0.1}-\cov$, $\mcd_{0.25}-\cov$, and $\mcd_{0.5}-\cov$. Details on the selection of these scatter pairs can be found in \ref{app:selection}.

\begin{figure}[t!]
\includegraphics[width=\textwidth]{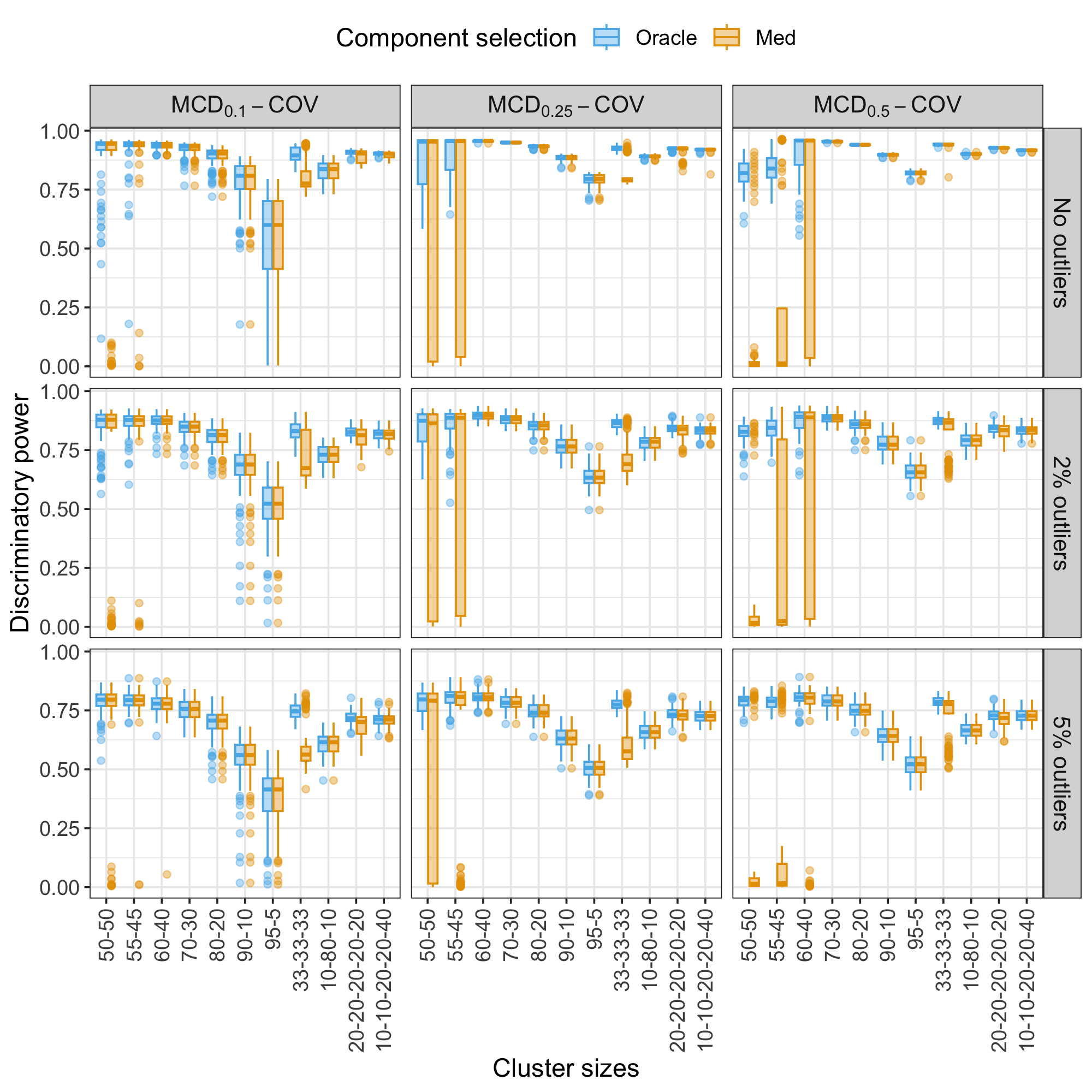}
\caption{Discriminatory power as measured by the $\eta^{2}$ of ICS with $\mcd_{\alpha}-\cov$ scatter pairs, using the best component selection criteria. Boxplots display the results for selected cluster settings over $100$ simulation runs. Results for different outlier settings are shown in separate rows, and results for different methods are shown in separate columns.}
\label{fig:MCD-eta2-clusters}
\end{figure}

Figure~\ref{fig:MCD-eta2-clusters} shows boxplots of the discriminatory power for the different outlier settings and a representative selection of cluster settings. The main finding is that the performance of $\mcd_{\alpha}-\cov$ scatter pairs in different cluster settings depends on the parameter $\alpha$. We first focus on the setting without outliers. For $q=2$ clusters, for instance, a lower $\alpha$ performs better for balanced clusters but yields larger variability for unbalanced clusters (90--10, 95--5), while a larger $\alpha$ is preferable for unbalanced clusters but yields lower discriminatory power or larger variability for relatively balanced clusters (50--50, 55--45, 60--40). However, as the number of clusters increases, this becomes less of an issue. 

Regarding the selection of components, the med criterion yields a performance similar to that of the oracle criterion, except that the aforementioned effect is more pronounced. Specifically, the med criterion fails or exhibits large variability for some cases (50-50 and 55-45 for $\alpha = 0.25$ and $\alpha = 0.5$, and additionally 60--40 for $\alpha = 0.5$). With outliers, the discriminatory power somewhat drops but remains good overall. Otherwise, results remain similar to the setting without outliers.

Overall, the results reveal that ICS with $\mcd_{\alpha}-\cov$ has potential in a clustering context, but further research is needed. First, since results in different cluster settings depend on the parameter $\alpha$, it could be interesting to consider $\alpha$ as a tuning parameter to be selected, e.g., together with the number of clusters. Second, the results for the oracle criterion indicate that the scatter pair $\mcd_{\alpha}-\cov$ succeeds in revealing the relevant structure for clustering, but component selection criteria that can be computed in practice fail in certain cluster settings.

\subsubsection[Discriminatory power with the COV-COV4 scatter pair]{Discriminatory power with the $\cov-\covF$ scatter pair}
\label{sec:COV-COV4}

\begin{figure}[b!]
\includegraphics[width=\textwidth]{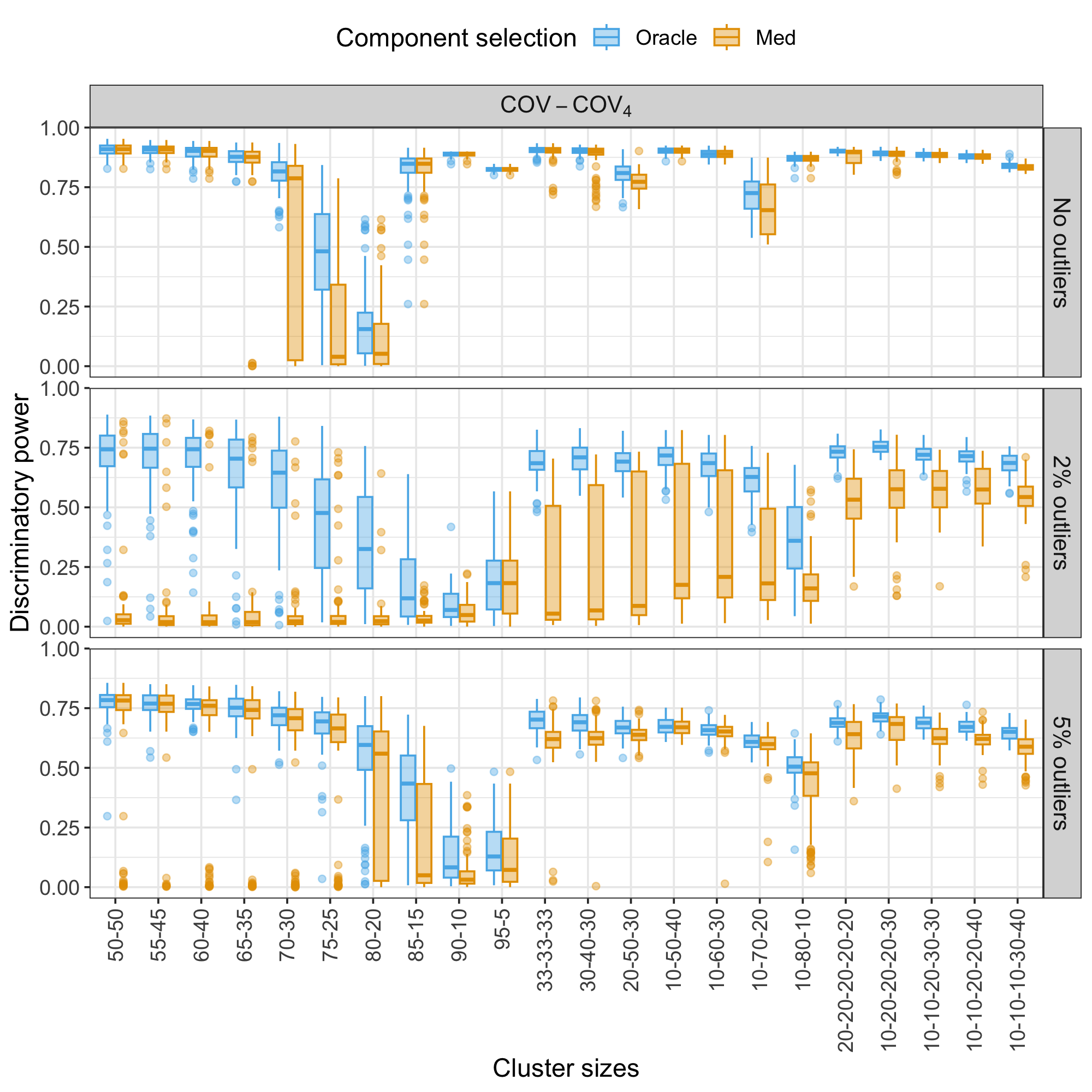}
\caption{Discriminatory power as measured by the $\eta^{2}$ of ICS with the $\cov-\covF$ scatter pair, using the best component selection criteria. Boxplots display the results for all investigated cluster settings over $100$ simulation runs. Results for different outlier settings are shown in separate rows.}
\label{fig:COV-COV4-eta2}
\end{figure}

As one of the most widely used scatter pairs in ICS, we take a closer look at the performance of $\cov-\covF$ in a clustering context \citep[see also][]{pena2010eigenvectors}. Boxplots of the discriminatory power for different cluster and outlier settings are displayed in Figure~\ref{fig:COV-COV4-eta2}.

We first discuss the setting without outliers. For $q=2$ clusters, the results are as expected from the theory explained in Section~\ref{sec:ics_prepro}. In that case, it is known that the relevant structure is contained in the first component for very unbalanced clusters and in the last component for (somewhat) balanced clusters, while neither the first nor the last component can separate the clusters if the smaller cluster contains about 21\% of observations. Indeed, we find that discriminatory power is very high except for the cluster settings 75--25 and 80--20, where the method fails. For $q=3$ clusters, the method still cannot recover the cluster with 20\% of observations (20--50--30, 10--70--20) but the discriminatory power is quite good since the two remaining clusters are highlighted. For $q=5$ clusters, discriminatory power is very high throughout even though there are various cluster settings with clusters of size 20\%. This is a promising result that needs to be verified from a theoretical point of view in further research. Furthermore, the med criterion is close to the oracle criterion in terms of performance, except in settings where the method struggles either way.

With 2\% of outliers, there is a considerable drop in performance for the oracle criterion. But more importantly, the med criterion fails almost completely. In particular, the discriminatory power is close to 0 for $q=2$ clusters, while the discriminatory power is very low and variability is large for $q=3$ clusters. In these settings, the med criterion fails to pick the components that carry the relevant structure. Surprisingly, the results of the med criterion improve again for 5\% of outliers, being more closely aligned again with the results of the oracle criterion. In fact, with 2\% of outliers and our sample size, the outlyingness structure is artificially contained in a low-dimensional space. $\cov-\covF$ is designed to discover such a structure and therefore highlights it on the first component. However, the clusters are visible on the last component, which is not kept with the med criteria based on only $k-1$ components. When the percentage of outliers increases, the outlying observations span the entire space so that the selected component(s) again capture the low-dimensional cluster structure, explaining the good results with 5\%.

\subsubsection{Clustering performance for Gaussian mixture modeling with dimension reduction}
\label{sec:ARI-GMM}

As Sections~\ref{sec:best-ARI} and~\ref{app:ARI-clusters} already provide a detailed comparison of mclust and rmclust (among other methods) on the observed data as well as after ICS and PCA, we now focus on comparing mclust and rmclust after ICS to methods that integrate dimension reduction into Gaussian mixture modeling. Specifically, we consider the approach of \citet{mclachlan2003modelling} that employs mixtures of factor analyzers (denoted by \emph{MFA}) and the approach of \citet{raftery2006variable} that incorporates variable selection into Gaussian mixture modeling (denoted by \emph{clustvarsel}). Note that \citet{scrucca2010} also introduces a dimension reduction method for Gaussian mixture modeling, but this approach applies dimension reduction after clustering for visualization purposes. That is, clustering is performed on the observed data rather than in a lower-dimensional subspace, hence this approach is not considered further here.

\begin{figure}[t!]
\includegraphics[width=\textwidth]{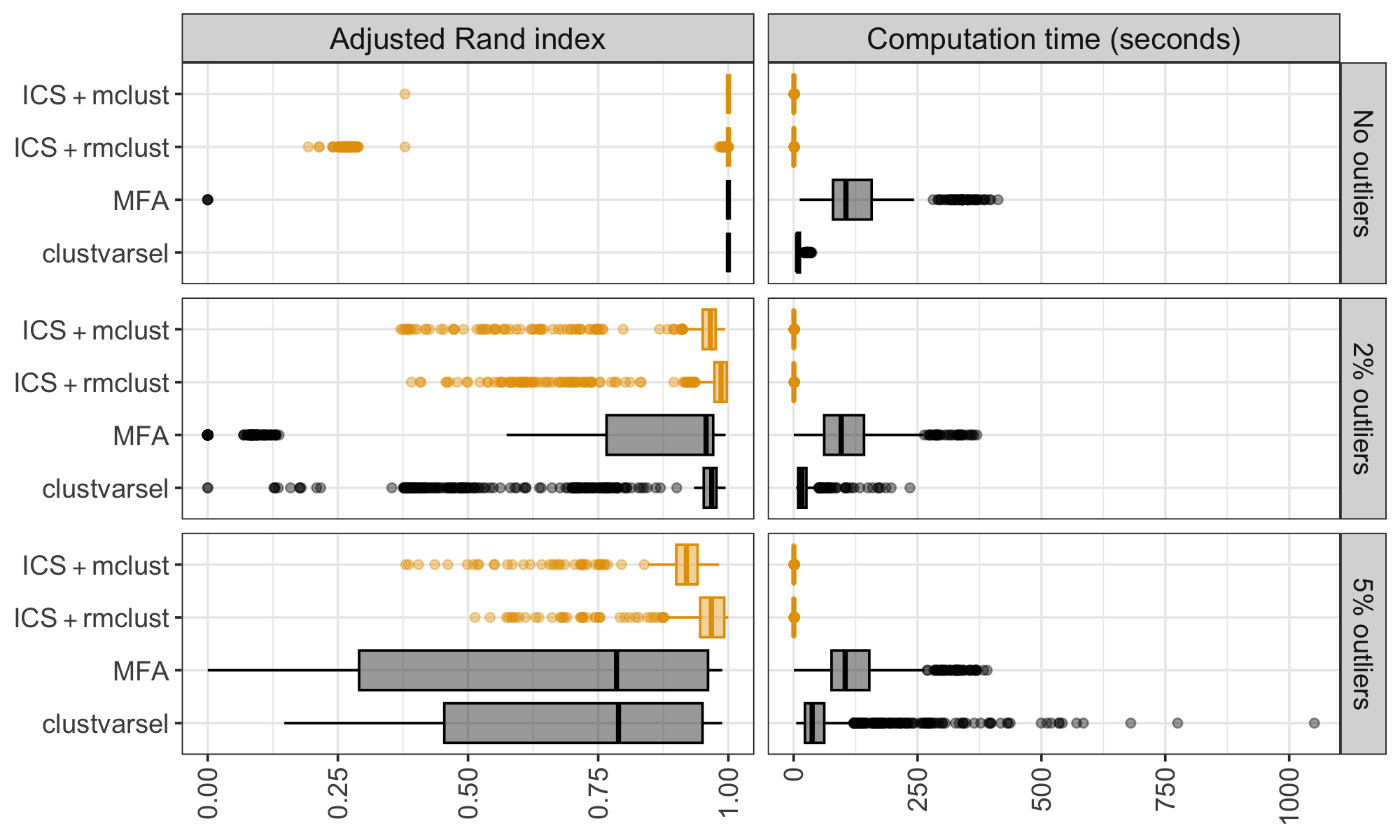}
\caption{Clustering performance as measured by the ARI (left column) and computation time in seconds (right column) for Gaussian mixture modeling after ICS with $\tcov-\cov$ and the med criterion, as well as for methods that integrate dimension reduction into Gaussian mixture modeling. Boxplots display the results across the different cluster settings with 100 simulation runs each. Results for different outlier settings are shown in separate rows.}
\label{fig:GMM_reduc_dim}
\end{figure}

Figure~\ref{fig:GMM_reduc_dim} compares MFA and clustvarsel with mclust and rmclust after ICS with $\tcov-\cov$. The left column displays the ARI across the 10 investigated cluster settings. Without outliers, MFA and clustvarsel perform (almost) perfectly, but also mclust and rmclust after ICS with $\tcov-\cov$ yield excellent results. With increasing percentage of outliers, however, the performance of MFA and (to a slightly lesser extent) clustvarsel deteriorates dramatically, most notably manifesting in a large variability in the ARI. On the other hand, rmclust but also mclust after ICS with $\tcov-\cov$ remain much more stable in the presence of outliers. In addition to this increase in stability, mclust and rmclust after ICS with $\tcov-\cov$ are also orders of magnitude faster to compute, as shown in the right column of Figure~\ref{fig:GMM_reduc_dim}.

On a final note, we emphasize that our simulation design places the relevant structure for clustering on a subset of the observed variables, which is in line with the assumptions of clustvarsel. When transforming the generated data so that the relevant structure lies in a latent lower-dimensional space, clustvarsel can no longer be expected to succeed in dimension reduction, whereas ICS is invariant to affine transformations. Nevertheless, we did not investigate this further to keep the paper at a reasonable length.

\subsection{Discussion of the simulation study}
\label{sec:sim-discussion}

The main findings from our simulations are that (i) ICS with $\tcov-\cov$ and $\lcov-\cov$ performs best for tandem clustering, (ii) the scatter pair $\mcd_{\alpha}-\cov$ with a value of $\alpha \leq 0.5$ could be an interesting alternative when $\alpha$ is treated as a tuning parameter but further research is needed, and (iii) PCA is not a suitable preprocessing step for clustering purposes and yields large variability in clustering performance.

However, there are several limitations of our simulations. First, we only consider one value each for the number of observations $n$ and the number of variables $d$. We compare a large number of scatter pairs for ICS with different component selection criteria, and our main aim is to study those for different numbers of clusters with various cluster sizes. As such, we had to make some concessions to other parameters to keep the paper concise. Second, we only consider a simple mean shift model for $q$ clusters that are defined in a subspace of dimension $q-1$. An important consideration for this choice is that it allows us to use simple clustering methods such as $k$means to validate our findings for the discriminatory power of the selected components. More complex models with, e.g., different covariance structures in different clusters or non-normal distributions are beyond the scope of this paper and are left for further research.

As a short glimpse into the performance of tandem ICS in a non-normal setting, we provide an illustrative example in \ref{app:barrowwheel} with a simulated data set coming from the so-called barrow wheel distribution. This distribution goes beyond Gaussian mixtures, and it is mentioned in the discussion of \cite{tyler2009invariant}, where the authors noticed that the theoretical properties of ICS are also valid for this distribution \citep[see also][for more theoretical details]{archimbaud2018statistical}.

\section{Empirical applications}
\label{sec:applications}

To further study the performance of tandem clustering with ICS, we consider three publicly available benchmark datasets. First, the \emph{crabs} data set \citep{campbell1974multivariate} contains morphological measurements on crabs of both sexes and from two color-different species.  We take log-transformations of all variables, which is common for this data set \citep[e.g.][]{archimbaud2022numerical}. Second, the well-known \emph{iris} data set \citep{anderson1936, fisher1936} gathers measurements of flowers from three different species.  Third, the \emph{Philips} data set contains measurements of a certain part for television sets, collected in 1997 by Philips Mecoma (The Netherlands) and studied by \cite{rousseeuw1999fast} and \citet{hubert2008high} for anomaly detection purposes. In this specific case, the engineers agreed on two clear abnormal phenomena regarding 15\% and 11\% of the data, which we treat as small clusters. An overview of the dimensions and cluster sizes of the three data sets is given in Table~\ref{tab:datasets}.

\begin{table}[b!]
\centering
\begin{tabular}{lrrrcp{4.75cm}}
  \hline\noalign{\smallskip}
  Name & $n$ & $d$ & $q$ & Cluster sizes (\%) & Source (\proglang{R} package) \\ 
  \noalign{\smallskip}\hline\noalign{\smallskip}
  crabs & 200 &   5 &   4 & 25--25--25--25 & \pkg{MASS} \citep{mass} \\ 
  \noalign{\smallskip}
  iris & 150 &   4 &   3 & 33--33--33 &  \pkg{datasets} \citep{R} \\ 
  \noalign{\smallskip}
  Philips & 677 &   9 &   3 & 74--15--11 &  \pkg{cellWise} \citep{cellWise} \\ 
  \noalign{\smallskip}\hline
\end{tabular}
\caption{Overview of the three benchmark data sets, where $n$ denotes the number of observations, $d$ the number of variables, and $q$ the number of true clusters.}
\label{tab:datasets}
\end{table}

\begin{figure}[t!]
\includegraphics[width=\textwidth]{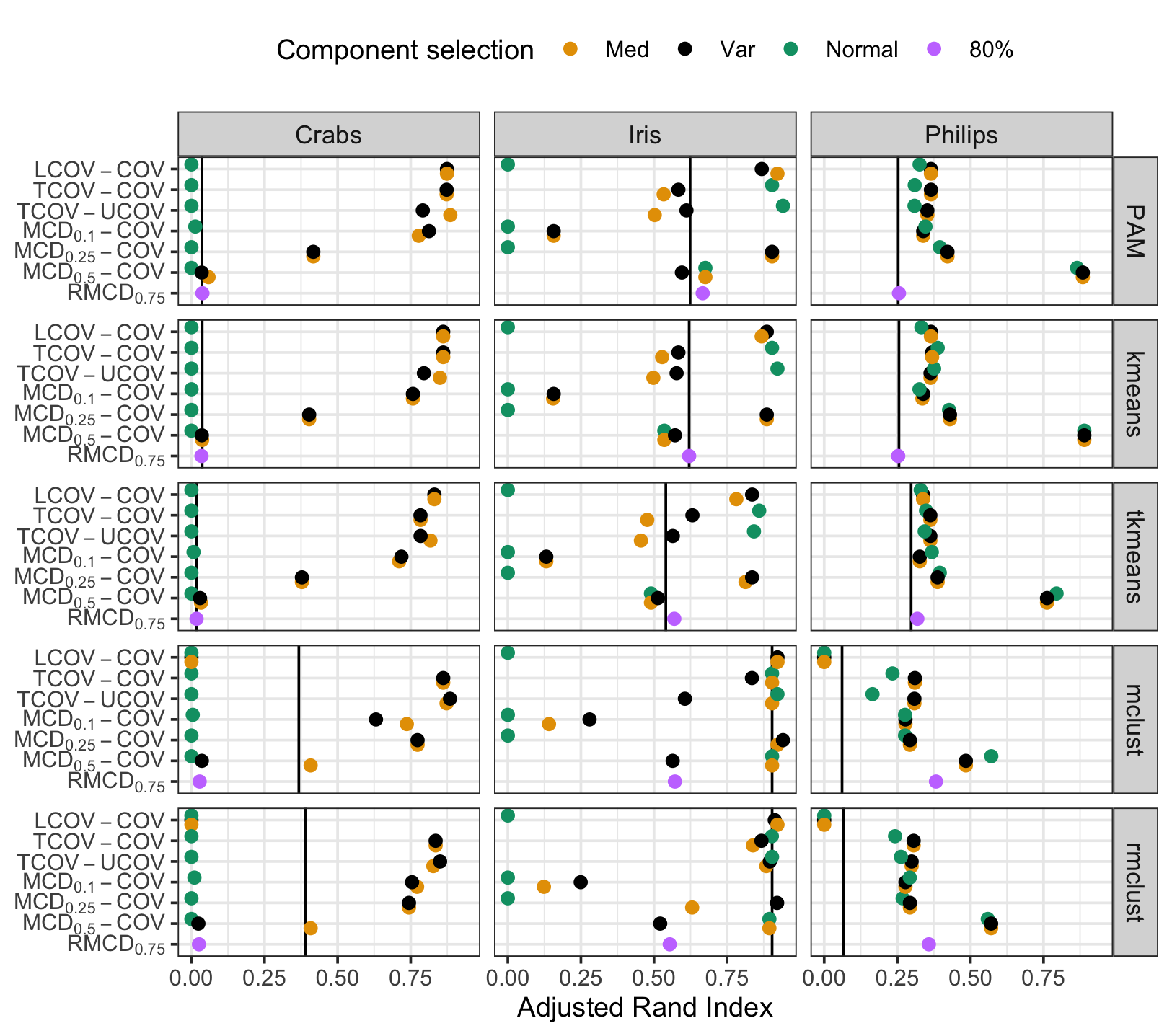}
\caption{Adjusted Rand Index (ARI) in the empirical applications for different clustering methods after selected dimension reduction methods for ICS and PCA, respectively, using different component selection criteria.  Results for different clustering methods are shown in separate rows, and results for different data sets are shown in separate columns. Black reference lines indicate the ARI obtained on initial data without dimension reduction.}
\label{fig:ARI_datasets}
\end{figure}

We focus on the clustering performance on these data sets using the same five clustering methods and the best-performing dimension reduction methods from Section~\ref{sec:sim}, thereby selecting the components using the med, var, and normal criteria for ICS, as well as the $80\%$ criterion for PCA. The clustering performance is once more evaluated by the Adjusted Rand Index (ARI) \citep{hubert1985comparing}. The results are shown in Figure~\ref{fig:ARI_datasets}, where reference lines indicate the ARI on the initial data without dimension reduction. Even though the data sets are relatively low-dimensional, ICS improves the identification of clusters in most cases, as reflected by an increase in the ARI. As in the simulation results, the med and var criteria lead to similar performance.  By contrast, PCA yields roughly the same or a lower ARI in all cases except for mclust and rmclust for the Philips data. Furthermore, the Philips data set is an interesting case where mclust and rmclust yield a lower ARI on the observed data than simple methods such as PAM, $k$means, and t$k$means. Note that the ARI slightly deteriorates in some instances also for ICS, hence the selection of the scatter pair and the component selection criterion is crucial. For a better understanding, we take a deeper look at the components obtained via ICS in order to relate them to the clustering performance. We thereby focus on the performance of $k$means and mclust, as the results are qualitatively similar for $k$means, t$k$means and PAM, as well as for mclust and rmclust.

\begin{figure}[t!]
\centering
\includegraphics[width=\textwidth]{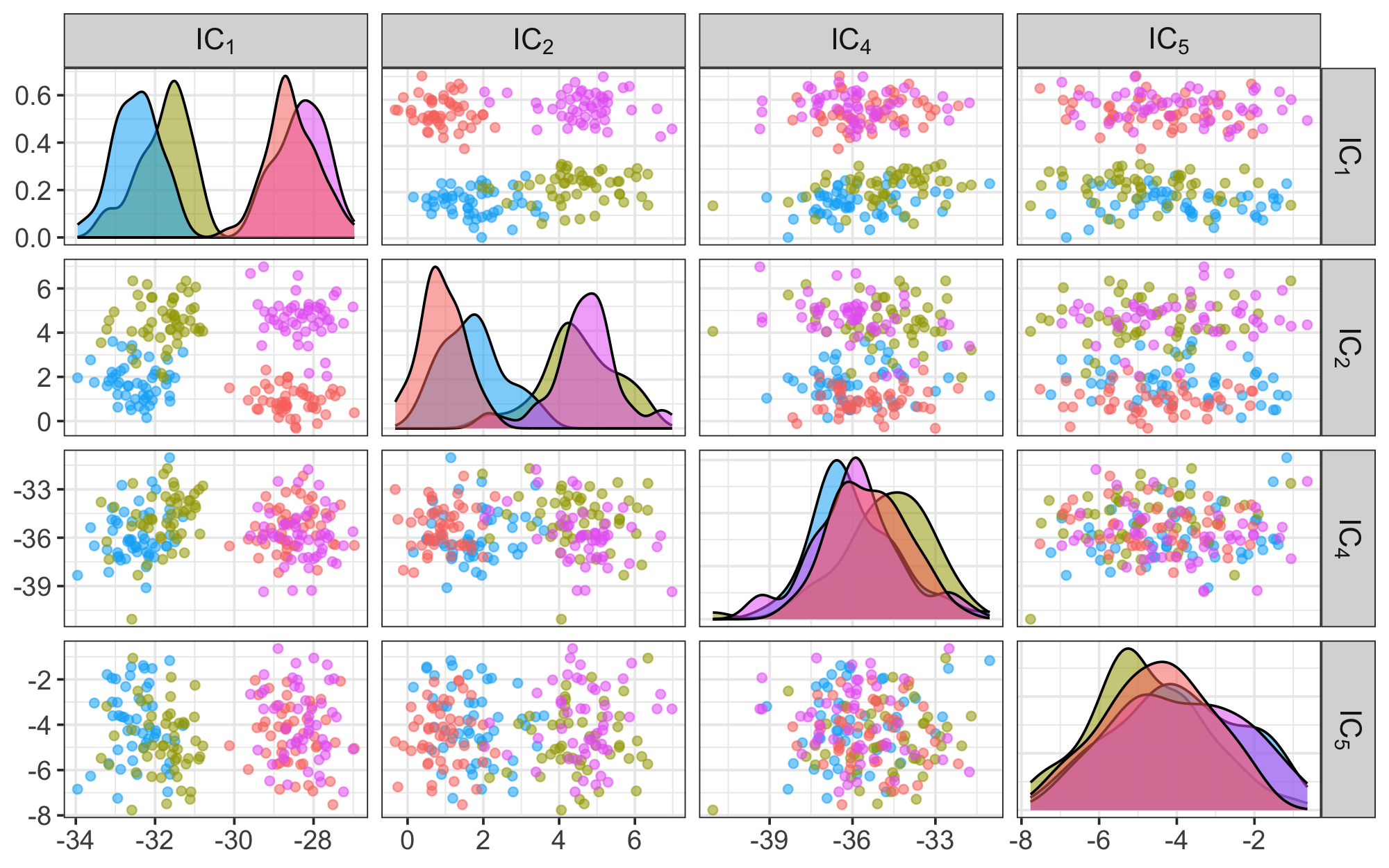}
\caption{First two and last two components from ICS with $\tcov-\cov$ on the crabs data set: scatterplot matrix with density estimates on the diagonal. The true clusters are indicated by different colors.}
\label{fig:ICS_crabs}
\end{figure}

For the crabs data set, the ARI on the initial data is 0.04 for $k$means and 0.37 for mclust. The improvement is substantial for ICS with the scatter pairs $\tcov-\cov$ and $\tcov-\ucov$, each yielding an ARI between 0.78 and 0.89 depending on the criterion used and the clustering method. The $\mcd_{\alpha}-\cov$ scatter pairs, on the other hand, achieve almost the same performance when $\alpha$ is small but the ARI decreases when $\alpha$ increases. Note that $\lcov-\cov$ works well for $k$means but not for mclust. Another interesting fact is that the normal criterion fails to select any components for any of the scatter pairs. To investigate this further, Figure~\ref{fig:ICS_crabs} displays the first two and last two components from ICS with the scatter pair $\tcov-\cov$, which clearly shows that the first two components reveal the relevant structure for clustering. The plot also visualizes the nested nature of the $q=4$ clusters in the crabs data: there are two species (blue and orange) and two sexes (male and female). ICS highlights the two balanced groups regarding the species on one component (here, the first) and the two balanced clusters regarding the sex on the another component (here, the second). However, for two balanced groups, the D'Agostino test of skewness \citep{dagostino1970transformation} does not reject the null hypothesis of normality, explaining why the normal criterion results in the selection of no components. The med criterion, on the other hand, is based on the selection of $k-1=3$ components (where $k=4$ denotes the number of clusters specified in the clustering algorithm). It selects the first two components, which carry the relevant structure, but also the last component. It is possible that clustering can be further improved by selecting only the first two components.

\begin{figure}[t!]
\centering
\includegraphics[width=\textwidth]{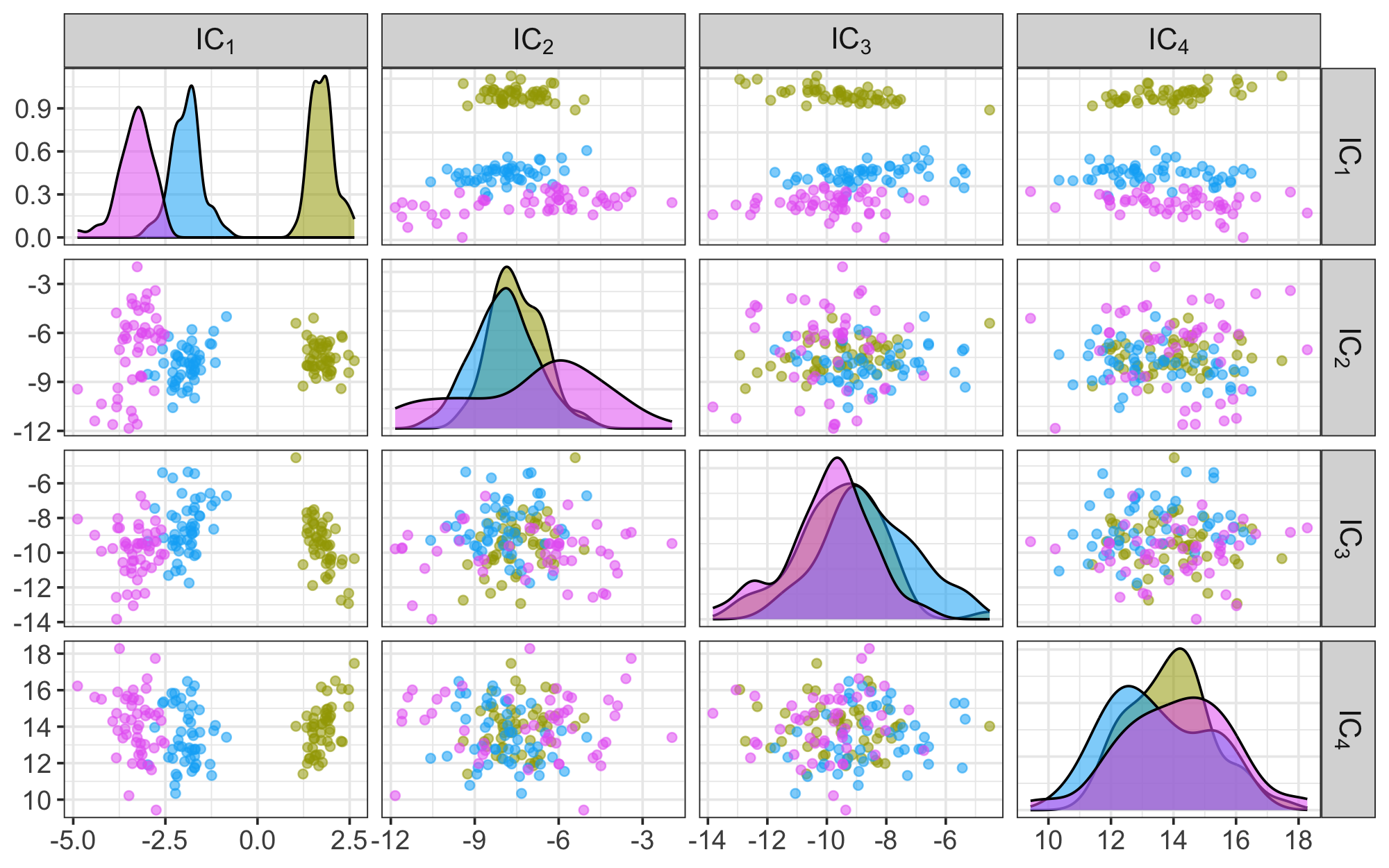}
\caption{All four components from ICS with $\tcov-\cov$ on the iris data set: scatterplot matrix with density estimates on the diagonal. The true clusters are indicated by different colors.}
\label{fig:ICS_iris}
\end{figure}

For the iris data set, the ARI on the initial data is already excellent for mclust with a value of 0.90. However, it is noteworthy that for $k$means, the ARI of 0.62 on the initial data is improved upon with the scatter pairs $\lcov-\cov$, $\tcov-\cov$,  $\tcov-\ucov$, and $\mcd_{0.25}-\cov$ to achieve competitive ARI values between 0.87 and 0.92. For the scatter pairs involving $\tcov$, the normal criterion selects only the first component, whereas the var criterion keeps the first two components and the med criterion selects the first and the last components, with the former resulting in a much higher ARI. Figure~\ref{fig:ICS_iris} displays the four components from ICS with the scatter pair $\tcov-\cov$, confirming that the relevant structure is indeed contained only in the first component. For $\lcov-\cov$, the var criterion also selects the first two components and med criterion the first and last component, here resulting in an excellent ARI, while the normal criterion does not select any component. It should be noted that the iris data set was used as an example in \citet{tyler2009invariant}, too. Those authors also find that the relevant structure is captured by the first component, but they conjecture that the results are similar ``for almost any pair of scatter matrices that we may choose''. Our results are an indication that the choice of scatter pair matters more for this data set than initially thought.

\begin{figure}[t!]
\centering
\includegraphics[width=\textwidth]{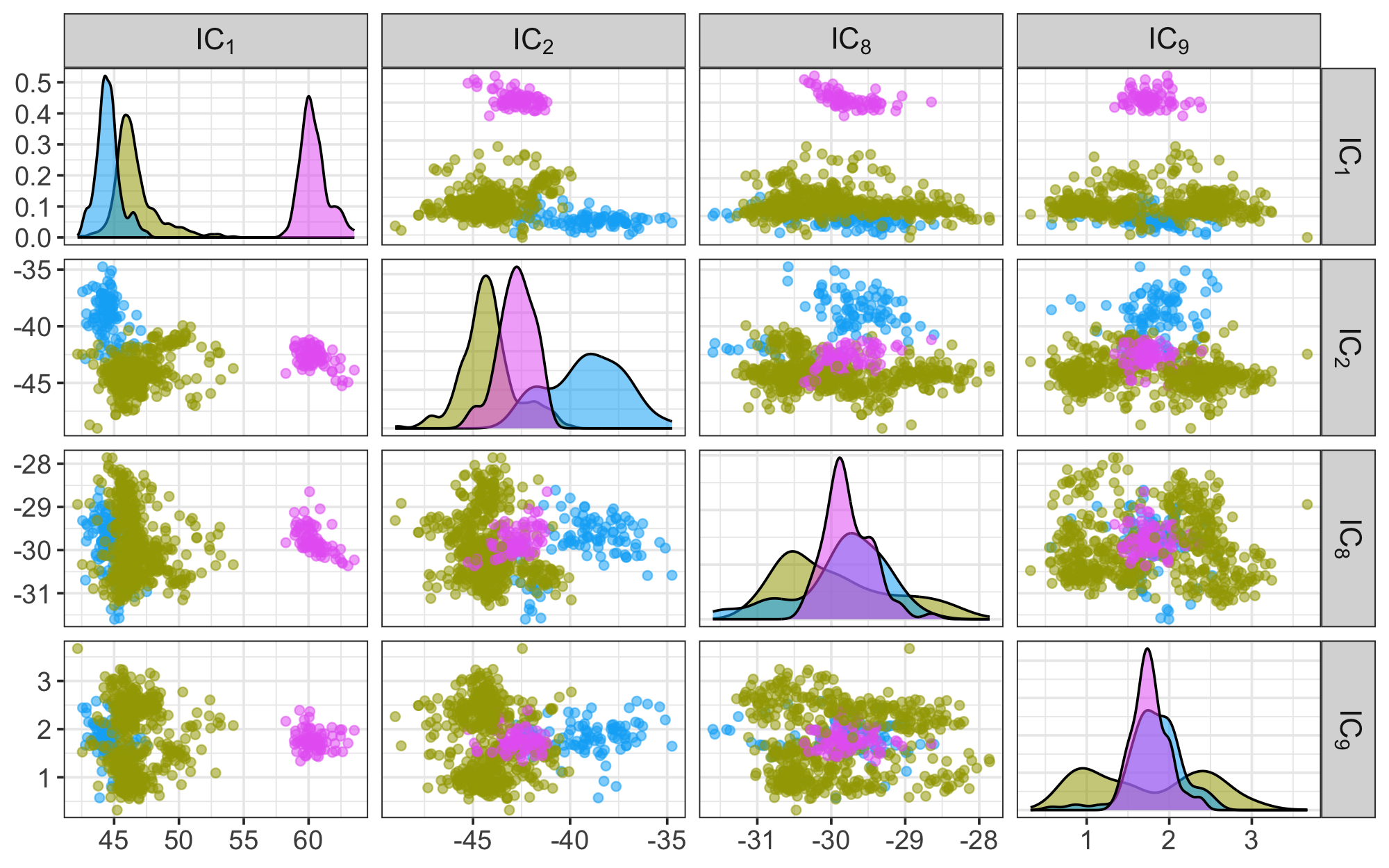}
\caption{First two and last two components from ICS with $\mcd_{0.50}-\cov$ on the Philips data set: scatterplot matrix with density estimates on the diagonal. The true clusters are indicated by different colors.}
\label{fig:ICS_philips_MCD50}
\end{figure}

For the Philips data set, the ARI on the initial data is rather low with a value of 0.26 for $k$means, but even lower for mclust (0.06). The most striking result in this example is that the ARI is by far the highest after ICS with $\mcd_{0.5}-\cov$ with a value of 0.89 for $k$means, while the other scatter pairs yield only a moderate improvement in the ARI (which is also the case for mclust). Furthermore, the med, var, and normal criteria yield similar results in terms of ARI for all considered scatter pairs. This is surprising as the med and var criteria always select the first two components, whereas the normal criterion selects many more (the first four in the case of $\mcd_{0.5}-\cov$). Figure~\ref{fig:ICS_philips_MCD50} indeed confirms that the first two components are necessary to reveal the clusters for the $\mcd_{0.5}-\cov$ scatter pair. Note that for this specific dataset, the clusters were labeled by Philips engineers, after investigation and interpretation, based on Mahalanobis distances calculated via the $\mcd$. This might explain why other scatter pairs do not succeed in the same way as $\mcd_{0.5}-\cov$ to reveal the structure.

Further results for the three considered data sets can be found in \ref{app:joint}, where we include methods that integrate dimension reduction into the $k$means algorithm \citep{soete1994k, vichi2001factorial} or into Gaussian mixture modeling \citep{mclachlan2003modelling, raftery2006variable}. None of those approaches attain the performance of tandem clustering with ICS.

\section{Conclusions and discussion} \label{sec:conclusion}

Overall, ICS outperforms PCA for tandem clustering. More specifically, the scatter pairs $\tcov-\cov$ and $\lcov-\cov$ yield the best performance, but also $\mcd_{\alpha}-\cov$ with a carefully selected parameter $\alpha \leq 0.5$ is promising. However, there is no scatter pair that is uniformly best in every situation. Regarding the component selection criteria, note that we restrict the med criterion to selecting $k-1$ components, where $k$ denotes the number of clusters to be specified in the clustering algorithm. The main reason for this restriction is to have a simple, automatic component selection rule for the simulations. In practice, an even smaller subspace may carry the relevant structure, hence the med criterion is not well adapted in this context and may select too many components. On the other hand, the normal criterion can estimate the dimension of the subspace of interest, but the D'Agostino test \citep{dagostino1970transformation} struggles when there are two balanced groups on one component, which can result in selecting none of the components, or in presence of outliers. Other normality tests may be more suitable. Further research on suitable component selection criteria is necessary, but in practice, researchers and practitioners may also select the components based on a scatterplot matrix of the components or a screeplot of the generalized eigenvalues. A more advanced procedure could be to consider the choice of the invariant coordinates as a tuning parameter of the subsequent clustering method. In this context, the best subspace could be chosen based on an internal validation index such as the silhouette coefficient \citep{kaufman2009finding} or those described in \cite{hennig2019cluster}.

In addition to improving the component selection criteria, the present work has many perspectives, including the treatment of mixed data \citep{caussinus2006projection,van2019distance} and functional data \citep{archimbaud2022ics}, but also the improvement of the ICS calculation algorithm by allowing to take into account (nearly) ill-conditioned data \citep{archimbaud2022numerical}, or by refining the research of the components using a projection pursuit algorithm as suggested in \citet{duembgen2021refining}. Generalizing ICS to high-dimensional data through some regularization approach is another interesting perspective in the context of clustering \citep[see][for a review for model-based clustering]{bouveyron2014model}. In the context of rating-scale data from surveys in the social and behavioral sciences, tandem clustering with ICS may further be useful for detecting different response styles among respondents \citep[e.g.][]{schoonees2015}.

\section*{Computational details}

All computations are performed with \proglang{R} version 4.3.1 \citep{R}. Our proposed approach for tandem clustering with ICS is implemented in package \pkg{ICSClust} \citep{ICSClust}, which is publicly available from \url{https://CRAN.R-project.org/package=ICSClust}. It builds upon packages \pkg{ICS} \citep{NordhausenOjaTyler:2008, ICS} for ICS, \pkg{rrcov} \citep{rrcov} for the MCD scatter matrix, \pkg{cluster} \citep{cluster} for PAM, \pkg{stats} \citep{R} for $k$means, \pkg{tclust} \citep{fritz2012} for t$k$means, and \pkg{mclust} \citep{scrucca2023} for mclust and rmclust. Furthermore, we use packages \pkg{EMMIXmfa} \citep{EMMIXmfa} for MFA, \pkg{clustvarsel} \citep{scrucca2018} for clustvarsel, and \pkg{clustrd} \citep{markos2019beyond} for reduced $k$means and factorial $k$means. Replication files for the simulations and the empirical applications are available from \url{https://github.com/aalfons/TandemICS-Replication}.

\section*{Acknowledgements}\label{sec:acknow}
We thank Raphaël Pinault for his preliminary work on the data as an intern.
This work was partly supported by a grant of the \emph{Dutch Research Council} (NWO, research program Vidi, project number VI.Vidi.195.141), by the Austrian Science Fund P31881-N32, by the HiTEc COST Action (CA21163), by the EUR CHESS at TSE, and by the French \emph{Agence Nationale de la Recherche} under grant ANR-17-EURE-0010 (Investissements d'Avenir program).

\bibliography{tandem_ics}

\begin{thebibliography}{94}
\expandafter\ifx\csname natexlab\endcsname\relax\def\natexlab#1{#1}\fi
\providecommand{\url}[1]{\texttt{#1}}
\providecommand{\href}[2]{#2}
\providecommand{\path}[1]{#1}
\providecommand{\DOIprefix}{doi:}
\providecommand{\ArXivprefix}{arXiv:}
\providecommand{\URLprefix}{URL: }
\providecommand{\Pubmedprefix}{pmid:}
\providecommand{\doi}[1]{\href{http://dx.doi.org/#1}{\path{#1}}}
\providecommand{\Pubmed}[1]{\href{pmid:#1}{\path{#1}}}
\providecommand{\bibinfo}[2]{#2}
\ifx\xfnm\relax \def\xfnm[#1]{\unskip,\space#1}\fi
\bibitem[{Aggarwal \& Reddy(2014)}]{aggarwal2013}
\bibinfo{editor}{Aggarwal, C.~C.}, \& \bibinfo{editor}{Reddy, C.~K.} (Eds.)
  (\bibinfo{year}{2014}).
\newblock {\it \bibinfo{title}{Data Clustering: Algorithms and
  Applications}\/}.
\newblock \bibinfo{publisher}{CRC Press}.
\newblock \URLprefix \url{http://www.charuaggarwal.net/clusterbook.pdf}.
\bibitem[{Alashwali \& Kent(2016)}]{AlashwaliKent2016}
\bibinfo{author}{Alashwali, F.}, \& \bibinfo{author}{Kent, J.~T.}
  (\bibinfo{year}{2016}).
\newblock \bibinfo{title}{The use of a common location measure in the invariant
  coordinate selection and projection pursuit}.
\newblock {\it \bibinfo{journal}{Journal of Multivariate Analysis}\/},  {\it
  \bibinfo{volume}{152}\/}, \bibinfo{pages}{145--161}.
  \DOIprefix\doi{10.1016/j.jmva.2016.08.007}.
\bibitem[{Anderson(1936)}]{anderson1936}
\bibinfo{author}{Anderson, E.} (\bibinfo{year}{1936}).
\newblock \bibinfo{title}{The species problem in iris}.
\newblock {\it \bibinfo{journal}{Annals of the Missouri Botanical Garden}\/},
  {\it \bibinfo{volume}{23}\/}, \bibinfo{pages}{457--509}.
  \DOIprefix\doi{10.2307/2394164}.
\bibitem[{Arabie \& Hubert(1994)}]{ArabieHubert1994}
\bibinfo{author}{Arabie, P.}, \& \bibinfo{author}{Hubert, L.}
  (\bibinfo{year}{1994}).
\newblock \bibinfo{title}{Cluster analysis in marketing research}.
\newblock In \bibinfo{editor}{R.~Bagozzi} (Ed.), {\it
  \bibinfo{booktitle}{Advanced methods in marketing research}\/} (pp.
  \bibinfo{pages}{160--189}).
\newblock \bibinfo{publisher}{Blackwell}.
\bibitem[{Archimbaud(2018)}]{archimbaud2018statistical}
\bibinfo{author}{Archimbaud, A.} (\bibinfo{year}{2018}).
\newblock {\it \bibinfo{title}{Statistical methods for outlier detection for
  high-dimensional data}\/}.
\newblock Ph.D. thesis Universit{\'e} Toulouse 1 Capitole.
\newblock \URLprefix \url{https://www.theses.fr/2018TOU10001}.
\bibitem[{Archimbaud et~al.(2023{\natexlab{a}})Archimbaud, Alfons, Nordhausen
  \& Ruiz-Gazen}]{ICSClust}
\bibinfo{author}{Archimbaud, A.}, \bibinfo{author}{Alfons, A.},
  \bibinfo{author}{Nordhausen, K.}, \& \bibinfo{author}{Ruiz-Gazen, A.}
  (\bibinfo{year}{2023}{\natexlab{a}}).
\newblock {\it \bibinfo{title}{ICSClust: Tandem Clustering with Invariant
  Coordinate Selection}\/}.
\newblock \URLprefix \url{https://CRAN.R-project.org/package=ICSClust}
  \bibinfo{note}{\proglang{R} package version~0.1.0}.
\bibitem[{Archimbaud et~al.(2022)Archimbaud, Boulfani, Gendre, Nordhausen,
  Ruiz-Gazen \& Virta}]{archimbaud2022ics}
\bibinfo{author}{Archimbaud, A.}, \bibinfo{author}{Boulfani, F.},
  \bibinfo{author}{Gendre, X.}, \bibinfo{author}{Nordhausen, K.},
  \bibinfo{author}{Ruiz-Gazen, A.}, \& \bibinfo{author}{Virta, J.}
  (\bibinfo{year}{2022}).
\newblock \bibinfo{title}{{ICS} for multivariate functional anomaly detection
  with applications to predictive maintenance and quality control}.
\newblock {\it \bibinfo{journal}{Econometrics and Statistics}\/},  {\it
  \bibinfo{volume}{in press}\/}. \DOIprefix\doi{10.1016/j.ecosta.2022.03.003}.
\bibitem[{Archimbaud et~al.(2023{\natexlab{b}})Archimbaud, Drma{\v{c}},
  Nordhausen, Radoji{\v{c}}i{\'c} \& Ruiz-Gazen}]{archimbaud2022numerical}
\bibinfo{author}{Archimbaud, A.}, \bibinfo{author}{Drma{\v{c}}, Z.},
  \bibinfo{author}{Nordhausen, K.}, \bibinfo{author}{Radoji{\v{c}}i{\'c}, U.},
  \& \bibinfo{author}{Ruiz-Gazen, A.} (\bibinfo{year}{2023}{\natexlab{b}}).
\newblock \bibinfo{title}{Numerical considerations and a new implementation for
  {ICS}}.
\newblock {\it \bibinfo{journal}{SIAM Journal on Mathematics of Data Science
  (SIMODS)}\/},  {\it \bibinfo{volume}{5}\/}, \bibinfo{pages}{97--121}.
  \DOIprefix\doi{10.1137/22M1498759}.
\bibitem[{Archimbaud et~al.(2018)Archimbaud, Nordhausen \&
  Ruiz-Gazen}]{archimbaud2018CSDA}
\bibinfo{author}{Archimbaud, A.}, \bibinfo{author}{Nordhausen, K.}, \&
  \bibinfo{author}{Ruiz-Gazen, A.} (\bibinfo{year}{2018}).
\newblock \bibinfo{title}{{ICS} for multivariate outlier detection with
  application to quality control}.
\newblock {\it \bibinfo{journal}{Computational Statistics \& Data Analysis}\/},
   {\it \bibinfo{volume}{128}\/}, \bibinfo{pages}{184--199}.
  \DOIprefix\doi{10.1016/j.csda.2018.06.011}.
\bibitem[{Art et~al.(1982)Art, Gnanadesikan \& Kettenring}]{art1982data}
\bibinfo{author}{Art, D.}, \bibinfo{author}{Gnanadesikan, R.}, \&
  \bibinfo{author}{Kettenring, J.} (\bibinfo{year}{1982}).
\newblock \bibinfo{title}{Data-based metrics for cluster analysis}.
\newblock {\it \bibinfo{journal}{Utilitas Mathematica A}\/},  {\it
  \bibinfo{volume}{21}\/}, \bibinfo{pages}{75--99}.
\bibitem[{Bouveyron \& Brunet-Saumard(2014)}]{bouveyron2014model}
\bibinfo{author}{Bouveyron, C.}, \& \bibinfo{author}{Brunet-Saumard, C.}
  (\bibinfo{year}{2014}).
\newblock \bibinfo{title}{Model-based clustering of high-dimensional data: A
  review}.
\newblock {\it \bibinfo{journal}{Computational Statistics \& Data Analysis}\/},
   {\it \bibinfo{volume}{71}\/}, \bibinfo{pages}{52--78}.
  \DOIprefix\doi{10.1016/j.csda.2012.12.008}.
\bibitem[{Bouveyron et~al.(2019)Bouveyron, Celeux, Murphy \&
  Raftery}]{BouveyronCeleuxMurphyRaftery2019}
\bibinfo{author}{Bouveyron, C.}, \bibinfo{author}{Celeux, G.},
  \bibinfo{author}{Murphy, T.~B.}, \& \bibinfo{author}{Raftery, A.~E.}
  (\bibinfo{year}{2019}).
\newblock {\it \bibinfo{title}{Model-Based Clustering and Classification for
  Data Science: With Applications in \proglang{R}}\/}.
\newblock \bibinfo{publisher}{Cambridge University Press}.
\newblock \DOIprefix\doi{10.1017/9781108644181}.
\bibitem[{Campbell \& Mahon(1974)}]{campbell1974multivariate}
\bibinfo{author}{Campbell, N.}, \& \bibinfo{author}{Mahon, R.}
  (\bibinfo{year}{1974}).
\newblock \bibinfo{title}{A multivariate study of variation in two species of
  rock crab of the genus leptograpsus}.
\newblock {\it \bibinfo{journal}{Australian Journal of Zoology}\/},  {\it
  \bibinfo{volume}{22}\/}, \bibinfo{pages}{417--425}.
  \DOIprefix\doi{10.1071/zo9740417}.
\bibitem[{Cardoso(1989)}]{Cardoso1989}
\bibinfo{author}{Cardoso, J.-F.} (\bibinfo{year}{1989}).
\newblock \bibinfo{title}{Source separation using higher order moments}.
\newblock In {\it \bibinfo{booktitle}{Proceedings of the IEEE International
  Conference on Acoustics, Speech and Signal Processing}\/} (pp.
  \bibinfo{pages}{2109--2112}).
\newblock \bibinfo{publisher}{IEEE}.
\newblock \DOIprefix\doi{10.1109/ICASSP.1989.266878}.
\bibitem[{Cator \& Lopuha{\"a}(2012)}]{CatorLopuhaa2012}
\bibinfo{author}{Cator, E.~A.}, \& \bibinfo{author}{Lopuha{\"a}, H.~P.}
  (\bibinfo{year}{2012}).
\newblock \bibinfo{title}{Central limit theorem and influence function for the
  {MCD} estimators at general multivariate distributions}.
\newblock {\it \bibinfo{journal}{Bernoulli}\/},  {\it \bibinfo{volume}{18}\/},
  \bibinfo{pages}{520 -- 551}. \DOIprefix\doi{10.3150/11-BEJ353}.
\bibitem[{Caussinus et~al.(2003)Caussinus, Fekri, Hakam \&
  Ruiz-Gazen}]{caussinus2003monitoring}
\bibinfo{author}{Caussinus, H.}, \bibinfo{author}{Fekri, M.},
  \bibinfo{author}{Hakam, S.}, \& \bibinfo{author}{Ruiz-Gazen, A.}
  (\bibinfo{year}{2003}).
\newblock \bibinfo{title}{A monitoring display of multivariate outliers}.
\newblock {\it \bibinfo{journal}{Computational Statistics \& Data Analysis}\/},
   {\it \bibinfo{volume}{44}\/}, \bibinfo{pages}{237--252}.
  \DOIprefix\doi{10.1016/S0167-9473(03)00059-8}.
\bibitem[{Caussinus \& Ruiz(1990)}]{caussinus1990interesting}
\bibinfo{author}{Caussinus, H.}, \& \bibinfo{author}{Ruiz, A.}
  (\bibinfo{year}{1990}).
\newblock \bibinfo{title}{Interesting projections of multidimensional data by
  means of generalized principal component analyses}.
\newblock In \bibinfo{editor}{K.~Momirovi{\'{c}}}, \&
  \bibinfo{editor}{V.~Mildner} (Eds.), {\it \bibinfo{booktitle}{Compstat}\/}
  (pp. \bibinfo{pages}{121--126}).
\newblock \bibinfo{publisher}{Physica-Verlag HD}.
\newblock \DOIprefix\doi{10.1007/978-3-642-50096-1_19}.
\bibitem[{Caussinus \& Ruiz-Gazen(1993)}]{caussinus1993ascona}
\bibinfo{author}{Caussinus, H.}, \& \bibinfo{author}{Ruiz-Gazen, A.}
  (\bibinfo{year}{1993}).
\newblock \bibinfo{title}{Projection pursuit and generalized principal
  component analysis}.
\newblock In \bibinfo{editor}{S.~Morgenthaler}, \bibinfo{editor}{E.~Ronchetti},
  \& \bibinfo{editor}{W.~A. Stahel} (Eds.), {\it \bibinfo{booktitle}{New
  Directions in Statistical Data Analysis and Robustness.}\/} Monte Verita,
  Proceedings of the Centro Stefano Franciscini Ascona Series.
\newblock \bibinfo{publisher}{Springer}.
\bibitem[{Caussinus \& Ruiz-Gazen(1995)}]{Caussinus1995metrics}
\bibinfo{author}{Caussinus, H.}, \& \bibinfo{author}{Ruiz-Gazen, A.}
  (\bibinfo{year}{1995}).
\newblock \bibinfo{title}{Metrics for finding typical structures by means of
  principal component analysis}.
\newblock In {\it \bibinfo{booktitle}{Data Science and its Applications}\/}
  (pp. \bibinfo{pages}{177--192}).
\newblock \bibinfo{publisher}{Harcourt Brace}.
\bibitem[{Caussinus \& Ruiz-Gazen(2006)}]{caussinus2006projection}
\bibinfo{author}{Caussinus, H.}, \& \bibinfo{author}{Ruiz-Gazen, A.}
  (\bibinfo{year}{2006}).
\newblock \bibinfo{title}{Projection-pursuit approach for categorical data}.
\newblock In \bibinfo{editor}{M.~Greenacre}, \& \bibinfo{editor}{J.~Blasius}
  (Eds.), {\it \bibinfo{booktitle}{Multiple Correspondence Analysis and Related
  Methods}\/} (pp. \bibinfo{pages}{405--418}).
\newblock \bibinfo{publisher}{Chapman \& Hall/CRC}.
\bibitem[{Caussinus \& Ruiz-Gazen(2007)}]{caussinus2007classification}
\bibinfo{author}{Caussinus, H.}, \& \bibinfo{author}{Ruiz-Gazen, A.}
  (\bibinfo{year}{2007}).
\newblock \bibinfo{title}{Classification and generalized principal component
  analysis}.
\newblock In {\it \bibinfo{booktitle}{Selected Contributions in Data Analysis
  and Classification}\/} (pp. \bibinfo{pages}{539--548}).
\newblock \bibinfo{publisher}{Springer}.
\newblock \DOIprefix\doi{10.1007/978-3-540-73560-1_50}.
\bibitem[{Cerioli(2005)}]{cerioli2005k}
\bibinfo{author}{Cerioli, A.} (\bibinfo{year}{2005}).
\newblock \bibinfo{title}{K-means cluster analysis and {Mahalanobis} metrics: A
  problematic match or an overlooked opportunity}.
\newblock {\it \bibinfo{journal}{Statistica Applicata}\/},  {\it
  \bibinfo{volume}{17}\/}, \bibinfo{pages}{61--73}.
\bibitem[{Chang(1983)}]{chang1983using}
\bibinfo{author}{Chang, W.-C.} (\bibinfo{year}{1983}).
\newblock \bibinfo{title}{On using principal components before separating a
  mixture of two multivariate normal distributions}.
\newblock {\it \bibinfo{journal}{Journal of the Royal Statistical Society,
  Series C}\/},  {\it \bibinfo{volume}{32}\/}, \bibinfo{pages}{267--275}.
  \DOIprefix\doi{10.2307/2347949}.
\bibitem[{Croux \& Haesbroeck(1999)}]{croux1999influence}
\bibinfo{author}{Croux, C.}, \& \bibinfo{author}{Haesbroeck, G.}
  (\bibinfo{year}{1999}).
\newblock \bibinfo{title}{Influence function and efficiency of the minimum
  covariance determinant scatter matrix estimator}.
\newblock {\it \bibinfo{journal}{Journal of Multivariate Analysis}\/},  {\it
  \bibinfo{volume}{71}\/}, \bibinfo{pages}{161--190}.
  \DOIprefix\doi{10.1006/jmva.1999.1839}.
\bibitem[{Cuesta-Albertos et~al.(1997)Cuesta-Albertos, Gordaliza \&
  Matr{\'a}n}]{cuesta1997trimmed}
\bibinfo{author}{Cuesta-Albertos, J.~A.}, \bibinfo{author}{Gordaliza, A.}, \&
  \bibinfo{author}{Matr{\'a}n, C.} (\bibinfo{year}{1997}).
\newblock \bibinfo{title}{Trimmed $ k $-means: An attempt to robustify
  quantizers}.
\newblock {\it \bibinfo{journal}{The Annals of Statistics}\/},  {\it
  \bibinfo{volume}{25}\/}, \bibinfo{pages}{553--576}.
  \DOIprefix\doi{10.1214/aos/1031833664}.
\bibitem[{D'Agostino(1970)}]{dagostino1970transformation}
\bibinfo{author}{D'Agostino, R.~B.} (\bibinfo{year}{1970}).
\newblock \bibinfo{title}{Transformation to normality of the null distribution
  of $g_1$}.
\newblock {\it \bibinfo{journal}{Biometrika}\/},  (pp.
  \bibinfo{pages}{679--681}). \DOIprefix\doi{10.2307/2334794}.
\bibitem[{Ding \& Li(2007)}]{ding2007adaptive}
\bibinfo{author}{Ding, C.}, \& \bibinfo{author}{Li, T.} (\bibinfo{year}{2007}).
\newblock \bibinfo{title}{Adaptive dimension reduction using discriminant
  analysis and k-means clustering}.
\newblock In {\it \bibinfo{booktitle}{Proceedings of the 24th International
  Conference on Machine learning}\/} (pp. \bibinfo{pages}{521--528}).
\newblock \DOIprefix\doi{10.1145/1273496.1273562}.
\bibitem[{D\"umbgen et~al.(2023)D\"umbgen, Gysel \&
  Perler}]{duembgen2021refining}
\bibinfo{author}{D\"umbgen, L.}, \bibinfo{author}{Gysel, K.}, \&
  \bibinfo{author}{Perler, F.} (\bibinfo{year}{2023}).
\newblock \bibinfo{title}{Refining invariant coordinate selection via local
  projection pursuit}.
\newblock In \bibinfo{editor}{M.~Yi}, \& \bibinfo{editor}{K.~Nordhausen}
  (Eds.), {\it \bibinfo{booktitle}{Robust and Multivariate Statistical Methods:
  Festschrift in Honor of David E. Tyler}\/} (pp. \bibinfo{pages}{121--136}).
\newblock \bibinfo{address}{Cham}: \bibinfo{publisher}{Springer}.
\newblock \DOIprefix\doi{10.1007/978-3-031-22687-8_6}.
\bibitem[{Fekri \& Ruiz-Gazen(2015)}]{Fekri2015}
\bibinfo{author}{Fekri, M.}, \& \bibinfo{author}{Ruiz-Gazen, A.}
  (\bibinfo{year}{2015}).
\newblock \bibinfo{title}{A {B}-robust non-iterative scatter matrix estimator:
  Asymptotics and application to cluster detection using invariant coordinate
  selection}.
\newblock In \bibinfo{editor}{K.~Nordhausen}, \& \bibinfo{editor}{S.~Taskinen}
  (Eds.), {\it \bibinfo{booktitle}{Modern Nonparametric, Robust and
  Multivariate Methods: Festschrift in Honour of Hannu Oja}\/} (pp.
  \bibinfo{pages}{395--423}).
\newblock \bibinfo{address}{Cham}: \bibinfo{publisher}{Springer}.
\newblock \DOIprefix\doi{10.1007/978-3-319-22404-6_22}.
\bibitem[{Fischer et~al.(2017)Fischer, Honkatukia, Tuiskula-Haavisto,
  Nordhausen, Cavero, Preisinger \&
  Vilkki}]{FischerHonkatukiaTuiskulaHaavistoNordhausenCaveroPreisingerVilkki:2017}
\bibinfo{author}{Fischer, D.}, \bibinfo{author}{Honkatukia, M.},
  \bibinfo{author}{Tuiskula-Haavisto, M.}, \bibinfo{author}{Nordhausen, K.},
  \bibinfo{author}{Cavero, D.}, \bibinfo{author}{Preisinger, R.}, \&
  \bibinfo{author}{Vilkki, J.} (\bibinfo{year}{2017}).
\newblock \bibinfo{title}{Subgroup detection in genotype data using invariant
  coordinate selection}.
\newblock {\it \bibinfo{journal}{BMC Bioinformatics}\/},  {\it
  \bibinfo{volume}{18}\/}, \bibinfo{pages}{173--181}.
  \DOIprefix\doi{10.1186/s12859-017-1589-9}.
\bibitem[{Fischer et~al.(2020)Fischer, Nordhausen \& Oja}]{FISCHER2020e05732}
\bibinfo{author}{Fischer, D.}, \bibinfo{author}{Nordhausen, K.}, \&
  \bibinfo{author}{Oja, H.} (\bibinfo{year}{2020}).
\newblock \bibinfo{title}{On linear dimension reduction based on
  diagonalization of scatter matrices for bioinformatics downstream analyses}.
\newblock {\it \bibinfo{journal}{Heliyon}\/},  {\it \bibinfo{volume}{6}\/},
  \bibinfo{pages}{e05732}. \DOIprefix\doi{10.1016/j.heliyon.2020.e05732}.
\bibitem[{Fisher(1936)}]{fisher1936}
\bibinfo{author}{Fisher, R.~A.} (\bibinfo{year}{1936}).
\newblock \bibinfo{title}{The use of multiple measurements in taxonomic
  problems}.
\newblock {\it \bibinfo{journal}{Annals of Eugenics}\/},  {\it
  \bibinfo{volume}{7}\/}, \bibinfo{pages}{179--188}.
  \DOIprefix\doi{10.1111/j.1469-1809.1936.tb02137.x}.
\bibitem[{Fraley \& Raftery(2002)}]{fraley2002model}
\bibinfo{author}{Fraley, C.}, \& \bibinfo{author}{Raftery, A.~E.}
  (\bibinfo{year}{2002}).
\newblock \bibinfo{title}{Model-based clustering, discriminant analysis, and
  density estimation}.
\newblock {\it \bibinfo{journal}{Journal of the American Statistical
  Association}\/},  {\it \bibinfo{volume}{97}\/}, \bibinfo{pages}{611--631}.
  \URLprefix \url{https://www.jstor.org/stable/3085676}.
\bibitem[{Fritz et~al.(2012)Fritz, Garc\'{i}a-Escudero \&
  Mayo-Iscar}]{fritz2012}
\bibinfo{author}{Fritz, H.}, \bibinfo{author}{Garc\'{i}a-Escudero, L.~A.}, \&
  \bibinfo{author}{Mayo-Iscar, A.} (\bibinfo{year}{2012}).
\newblock \bibinfo{title}{\pkg{tclust}: An \proglang{R} package for a trimming
  approach to cluster analysis}.
\newblock {\it \bibinfo{journal}{Journal of Statistical Software}\/},  {\it
  \bibinfo{volume}{47}\/}, \bibinfo{pages}{1--26}.
  \DOIprefix\doi{10.18637/jss.v047.i12}.
\bibitem[{Garc\'{i}a-Escudero et~al.(2011)Garc\'{i}a-Escudero, Gordaliza,
  Matr\'{a}n \& Mayo-Iscar}]{garcia2011}
\bibinfo{author}{Garc\'{i}a-Escudero, L.~A.}, \bibinfo{author}{Gordaliza, A.},
  \bibinfo{author}{Matr\'{a}n, C.}, \& \bibinfo{author}{Mayo-Iscar, A.}
  (\bibinfo{year}{2011}).
\newblock \bibinfo{title}{Exploring the number of groups in robust model-based
  clustering}.
\newblock {\it \bibinfo{journal}{Statistics and Computing}\/},  {\it
  \bibinfo{volume}{21}\/}, \bibinfo{pages}{585--599}.
  \DOIprefix\doi{10.1007/s11222-010-9194-z}.
\bibitem[{Gnanadesikan et~al.(1993)Gnanadesikan, Harvey \&
  Kettenring}]{gnanadesikan1993mahalanobis}
\bibinfo{author}{Gnanadesikan, R.}, \bibinfo{author}{Harvey, J.}, \&
  \bibinfo{author}{Kettenring, J.} (\bibinfo{year}{1993}).
\newblock \bibinfo{title}{Mahalanobis metrics for cluster analysis}.
\newblock {\it \bibinfo{journal}{Sankhy{\=a}: The Indian Journal of Statistics,
  Series A}\/},  (pp. \bibinfo{pages}{494--505}). \URLprefix
  \url{https://www.jstor.org/stable/25050956}.
\bibitem[{Hampel et~al.(1986)Hampel, Ronchetti, Rousseeuw \&
  Stahel}]{hampel1986robust}
\bibinfo{author}{Hampel, F.}, \bibinfo{author}{Ronchetti, E.},
  \bibinfo{author}{Rousseeuw, P.}, \& \bibinfo{author}{Stahel, W.}
  (\bibinfo{year}{1986}).
\newblock {\it \bibinfo{title}{Robust Statistics: The Approach Based on
  Influence Functions}\/}.
\newblock \bibinfo{address}{New York}: \bibinfo{publisher}{John Wiley \& Sons}.
\bibitem[{Hartigan \& Wong(1979)}]{hartigan1979algorithm}
\bibinfo{author}{Hartigan, J.~A.}, \& \bibinfo{author}{Wong, M.~A.}
  (\bibinfo{year}{1979}).
\newblock \bibinfo{title}{Algorithm {AS} 136: A k-means clustering algorithm}.
\newblock {\it \bibinfo{journal}{Journal of the Royal Statistical Society,
  Series C}\/},  {\it \bibinfo{volume}{28}\/}, \bibinfo{pages}{100--108}.
  \DOIprefix\doi{10.2307/2346830}.
\bibitem[{Hennig(2009)}]{Hennig2009}
\bibinfo{author}{Hennig, C.} (\bibinfo{year}{2009}).
\newblock \bibinfo{title}{Discussion of ``{Invariant} co-ordinate selection'',
  by {D. E. Tyler, F. Critchley, L. D\"umbgen, and H. Oja}}.
\newblock {\it \bibinfo{journal}{Journal of the Royal Statistical Society,
  Series B}\/},  {\it \bibinfo{volume}{71}\/}, \bibinfo{pages}{579--583}.
  \DOIprefix\doi{10.1111/j.1467-9868.2009.00706.x}.
\bibitem[{Hennig(2019)}]{hennig2019cluster}
\bibinfo{author}{Hennig, C.} (\bibinfo{year}{2019}).
\newblock \bibinfo{title}{Cluster validation by measurement of clustering
  characteristics relevant to the user}.
\newblock In {\it \bibinfo{booktitle}{Data Analysis and Applications 1}\/}
  chapter~\bibinfo{chapter}{1}. (pp. \bibinfo{pages}{1--24}).
\newblock \bibinfo{publisher}{John Wiley \& Sons}.
\newblock \DOIprefix\doi{10.1002/9781119597568.ch1}.
\bibitem[{Hennig et~al.(2015)Hennig, Meila, Murtagh \&
  Rocci}]{hennig2015handbook}
\bibinfo{author}{Hennig, C.}, \bibinfo{author}{Meila, M.},
  \bibinfo{author}{Murtagh, F.}, \& \bibinfo{author}{Rocci, R.}
  (\bibinfo{year}{2015}).
\newblock {\it \bibinfo{title}{Handbook of Cluster Analysis}\/}.
\newblock \bibinfo{publisher}{Chapman \& Hall/CRC}.
\newblock \DOIprefix\doi{10.1201/b19706}.
\bibitem[{Hubert \& Arabie(1985)}]{hubert1985comparing}
\bibinfo{author}{Hubert, L.}, \& \bibinfo{author}{Arabie, P.}
  (\bibinfo{year}{1985}).
\newblock \bibinfo{title}{Comparing partitions}.
\newblock {\it \bibinfo{journal}{Journal of Classification}\/},  {\it
  \bibinfo{volume}{2}\/}, \bibinfo{pages}{193--218}.
  \DOIprefix\doi{10.1007/BF01908075}.
\bibitem[{Hubert \& Debruyne(2010)}]{HubertDebruyne2010}
\bibinfo{author}{Hubert, M.}, \& \bibinfo{author}{Debruyne, M.}
  (\bibinfo{year}{2010}).
\newblock \bibinfo{title}{Minimum covariance determinant}.
\newblock {\it \bibinfo{journal}{Wiley Interdisciplinary Reviews: Computational
  Statistics}\/},  {\it \bibinfo{volume}{2}\/}, \bibinfo{pages}{36--43}.
  \DOIprefix\doi{https://doi.org/10.1002/wics.61}.
\bibitem[{Hubert et~al.(2018)Hubert, Debruyne \&
  Rousseeuw}]{HubertDebruyneRousseeuw2018}
\bibinfo{author}{Hubert, M.}, \bibinfo{author}{Debruyne, M.}, \&
  \bibinfo{author}{Rousseeuw, P.~J.} (\bibinfo{year}{2018}).
\newblock \bibinfo{title}{Minimum covariance determinant and extensions}.
\newblock {\it \bibinfo{journal}{Wiley Interdisciplinary Reviews: Computational
  Statistics}\/},  {\it \bibinfo{volume}{10}\/}, \bibinfo{pages}{e1421}.
  \DOIprefix\doi{https://doi.org/10.1002/wics.1421}.
\bibitem[{Hubert et~al.(2019)Hubert, Rousseeuw \& den Bossche}]{hubert2019}
\bibinfo{author}{Hubert, M.}, \bibinfo{author}{Rousseeuw, P.~J.}, \&
  \bibinfo{author}{den Bossche, W.~V.} (\bibinfo{year}{2019}).
\newblock \bibinfo{title}{{MacroPCA}: An all-in-one {PCA} method allowing for
  missing values as well as cellwise and rowwise outliers}.
\newblock {\it \bibinfo{journal}{Technometrics}\/},  {\it
  \bibinfo{volume}{61}\/}, \bibinfo{pages}{459--473}.
  \DOIprefix\doi{10.1080/00401706.2018.1562989}.
\bibitem[{Hubert et~al.(2005)Hubert, Rousseeuw \& Branden}]{hubert2005robpca}
\bibinfo{author}{Hubert, M.}, \bibinfo{author}{Rousseeuw, P.~J.}, \&
  \bibinfo{author}{Branden, K.~V.} (\bibinfo{year}{2005}).
\newblock \bibinfo{title}{{ROBPCA}: A new approach to robust principal
  component analysis}.
\newblock {\it \bibinfo{journal}{Technometrics}\/},  {\it
  \bibinfo{volume}{47}\/}, \bibinfo{pages}{64--79}.
  \DOIprefix\doi{10.1198/004017004000000563}.
\bibitem[{Hubert et~al.(2008)Hubert, Rousseeuw \& Van~Aelst}]{hubert2008high}
\bibinfo{author}{Hubert, M.}, \bibinfo{author}{Rousseeuw, P.~J.}, \&
  \bibinfo{author}{Van~Aelst, S.} (\bibinfo{year}{2008}).
\newblock \bibinfo{title}{High-breakdown robust multivariate methods}.
\newblock {\it \bibinfo{journal}{Statistical Science}\/},  {\it
  \bibinfo{volume}{23}\/}, \bibinfo{pages}{92--119}.
  \DOIprefix\doi{10.1214/088342307000000087}.
\bibitem[{Kaufman \& Rousseeuw(1990)}]{kaufman2009finding}
\bibinfo{author}{Kaufman, L.}, \& \bibinfo{author}{Rousseeuw, P.~J.}
  (\bibinfo{year}{1990}).
\newblock {\it \bibinfo{title}{Finding Groups in Data: An Introduction to
  Cluster Analysis}\/}.
\newblock \bibinfo{publisher}{John Wiley \& Sons}.
\newblock \DOIprefix\doi{10.1002/9780470316801}.
\bibitem[{Kent \& Tyler(1991)}]{KentTyler1991}
\bibinfo{author}{Kent, J.~T.}, \& \bibinfo{author}{Tyler, D.~E.}
  (\bibinfo{year}{1991}).
\newblock \bibinfo{title}{Redescending {M}-estimates of multivariate location
  and scatter}.
\newblock {\it \bibinfo{journal}{The Annals of Statistics}\/},  {\it
  \bibinfo{volume}{19}\/}, \bibinfo{pages}{2102 -- 2119}.
  \DOIprefix\doi{10.1214/aos/1176348388}.
\bibitem[{Kettenring(2006)}]{kettenring2006practice}
\bibinfo{author}{Kettenring, J.~R.} (\bibinfo{year}{2006}).
\newblock \bibinfo{title}{The practice of cluster analysis}.
\newblock {\it \bibinfo{journal}{Journal of Classification}\/},  {\it
  \bibinfo{volume}{23}\/}, \bibinfo{pages}{3--30}.
  \DOIprefix\doi{10.1007/s00357-006-0002-6}.
\bibitem[{Lapidot(2018)}]{lapidot2018convergence}
\bibinfo{author}{Lapidot, I.} (\bibinfo{year}{2018}).
\newblock \bibinfo{title}{Convergence problems of {Mahalanobis} distance-based
  k-means clustering}.
\newblock In {\it \bibinfo{booktitle}{2018 IEEE International Conference on the
  Science of Electrical Engineering in Israel (ICSEE)}\/} (pp.
  \bibinfo{pages}{1--5}).
\newblock \bibinfo{organization}{IEEE}.
\newblock \DOIprefix\doi{10.1109/ICSEE.2018.8646138}.
\bibitem[{Liski et~al.(2014)Liski, Nordhausen \& Oja}]{LiskiNordhausenOja2014}
\bibinfo{author}{Liski, E.}, \bibinfo{author}{Nordhausen, K.}, \&
  \bibinfo{author}{Oja, H.} (\bibinfo{year}{2014}).
\newblock \bibinfo{title}{Supervised invariant coordinate selection}.
\newblock {\it \bibinfo{journal}{Statistics: A Journal of Theoretical and
  Applied Statistics}\/},  {\it \bibinfo{volume}{4}\/},
  \bibinfo{pages}{711--731}. \DOIprefix\doi{10.1080/02331888.2013.800067}.
\bibitem[{Luo \& Li(2016)}]{LuoLi2016}
\bibinfo{author}{Luo, W.}, \& \bibinfo{author}{Li, B.} (\bibinfo{year}{2016}).
\newblock \bibinfo{title}{Combining eigenvalues and variation of eigenvectors
  for order determination}.
\newblock {\it \bibinfo{journal}{Biometrika}\/},  {\it
  \bibinfo{volume}{103}\/}, \bibinfo{pages}{875--887}.
  \DOIprefix\doi{10.1093/biomet/asw051}.
\bibitem[{Luo \& Li(2021)}]{luo2021order}
\bibinfo{author}{Luo, W.}, \& \bibinfo{author}{Li, B.} (\bibinfo{year}{2021}).
\newblock \bibinfo{title}{On order determination by predictor augmentation}.
\newblock {\it \bibinfo{journal}{Biometrika}\/},  {\it
  \bibinfo{volume}{108}\/}, \bibinfo{pages}{557--574}.
  \DOIprefix\doi{10.1093/biomet/asaa077}.
\bibitem[{Maechler et~al.(2022)Maechler, Rousseeuw, Struyf, Hubert \&
  Hornik}]{cluster}
\bibinfo{author}{Maechler, M.}, \bibinfo{author}{Rousseeuw, P.},
  \bibinfo{author}{Struyf, A.}, \bibinfo{author}{Hubert, M.}, \&
  \bibinfo{author}{Hornik, K.} (\bibinfo{year}{2022}).
\newblock {\it \bibinfo{title}{\pkg{cluster}: Cluster Analysis Basics and
  Extensions}\/}.
\newblock \URLprefix \url{https://CRAN.R-project.org/package=cluster}
  \bibinfo{note}{\proglang{R} package version~2.1.4}.
\bibitem[{Mardia et~al.(1982)Mardia, Bibby \& Kent}]{mardia1982multivariate}
\bibinfo{author}{Mardia, K.}, \bibinfo{author}{Bibby, J.}, \&
  \bibinfo{author}{Kent, J.} (\bibinfo{year}{1982}).
\newblock {\it \bibinfo{title}{Multivariate Analysis}\/}.
\newblock \bibinfo{publisher}{Academic Press}.
\bibitem[{Markos et~al.(2019)Markos, D'Enza \& van~de
  Velden}]{markos2019beyond}
\bibinfo{author}{Markos, A.}, \bibinfo{author}{D'Enza, A.~I.}, \&
  \bibinfo{author}{van~de Velden, M.} (\bibinfo{year}{2019}).
\newblock \bibinfo{title}{Beyond tandem analysis: Joint dimension reduction and
  clustering in \proglang{R}}.
\newblock {\it \bibinfo{journal}{Journal of Statistical Software}\/},  {\it
  \bibinfo{volume}{91}\/}, \bibinfo{pages}{1--24}.
\bibitem[{McLachlan(1992)}]{mclachlan1992discriminant}
\bibinfo{author}{McLachlan, G.} (\bibinfo{year}{1992}).
\newblock {\it \bibinfo{title}{Discriminant Analysis and Statistical Pattern
  Recognition}\/}.
\newblock \bibinfo{publisher}{John Wiley \& Sons}.
\bibitem[{McLachlan et~al.(2003)McLachlan, Peel \&
  Bean}]{mclachlan2003modelling}
\bibinfo{author}{McLachlan, G.~J.}, \bibinfo{author}{Peel, D.}, \&
  \bibinfo{author}{Bean, R.~W.} (\bibinfo{year}{2003}).
\newblock \bibinfo{title}{Modelling high-dimensional data by mixtures of factor
  analyzers}.
\newblock {\it \bibinfo{journal}{Computational Statistics \& Data Analysis}\/},
   {\it \bibinfo{volume}{41}\/}, \bibinfo{pages}{379--388}.
  \DOIprefix\doi{10.1016/S0167-9473(02)00183-4}.
\bibitem[{Nordhausen et~al.(2023)Nordhausen, Alfons, Archimbaud, Oja,
  Ruiz-Gazen \& Tyler}]{ICS}
\bibinfo{author}{Nordhausen, K.}, \bibinfo{author}{Alfons, A.},
  \bibinfo{author}{Archimbaud, A.}, \bibinfo{author}{Oja, H.},
  \bibinfo{author}{Ruiz-Gazen, A.}, \& \bibinfo{author}{Tyler, D.~E.}
  (\bibinfo{year}{2023}).
\newblock {\it \bibinfo{title}{\pkg{ICS}: Tools for Exploring Multivariate Data
  via ICS/ICA}\/}.
\newblock \URLprefix \url{https://CRAN.R-project.org/package=ICS}
  \bibinfo{note}{\proglang{R} package version~1.4-1}.
\bibitem[{Nordhausen et~al.(2011)Nordhausen, Oja \&
  Ollila}]{NordhausenOjaOllila2011b}
\bibinfo{author}{Nordhausen, K.}, \bibinfo{author}{Oja, H.}, \&
  \bibinfo{author}{Ollila, E.} (\bibinfo{year}{2011}).
\newblock \bibinfo{title}{Multivariate models and the first four moments}.
\newblock In {\it \bibinfo{booktitle}{Nonparametric Statistics and Mixture
  Models}\/} (pp. \bibinfo{pages}{267--287}).
\newblock \bibinfo{publisher}{World Scientific}.
\newblock \DOIprefix\doi{10.1142/9789814340564\_0016}.
\bibitem[{Nordhausen et~al.(2008)Nordhausen, Oja \&
  Tyler}]{NordhausenOjaTyler:2008}
\bibinfo{author}{Nordhausen, K.}, \bibinfo{author}{Oja, H.}, \&
  \bibinfo{author}{Tyler, D.~E.} (\bibinfo{year}{2008}).
\newblock \bibinfo{title}{Tools for exploring multivariate data: The package
  \pkg{ICS}}.
\newblock {\it \bibinfo{journal}{Journal of Statistical Software}\/},  {\it
  \bibinfo{volume}{28}\/}, \bibinfo{pages}{1--31}.
  \DOIprefix\doi{10.18637/jss.v028.i06}.
\bibitem[{Nordhausen et~al.(2022)Nordhausen, Oja \&
  Tyler}]{nordhausen2022asymptotic}
\bibinfo{author}{Nordhausen, K.}, \bibinfo{author}{Oja, H.}, \&
  \bibinfo{author}{Tyler, D.~E.} (\bibinfo{year}{2022}).
\newblock \bibinfo{title}{Asymptotic and bootstrap tests for subspace
  dimension}.
\newblock {\it \bibinfo{journal}{Journal of Multivariate Analysis}\/},  {\it
  \bibinfo{volume}{188}\/}, \bibinfo{pages}{104830}.
  \DOIprefix\doi{10.1016/j.jmva.2021.104830}.
\bibitem[{Nordhausen et~al.(2017)Nordhausen, Oja, Tyler \&
  Virta}]{nordhausen2017asymptotic}
\bibinfo{author}{Nordhausen, K.}, \bibinfo{author}{Oja, H.},
  \bibinfo{author}{Tyler, D.~E.}, \& \bibinfo{author}{Virta, J.}
  (\bibinfo{year}{2017}).
\newblock \bibinfo{title}{Asymptotic and bootstrap tests for the dimension of
  the non-{Gaussian} subspace}.
\newblock {\it \bibinfo{journal}{IEEE Signal Processing Letters}\/},  {\it
  \bibinfo{volume}{24}\/}, \bibinfo{pages}{887--891}.
  \DOIprefix\doi{10.1109/LSP.2017.2696880}.
\bibitem[{Nordhausen \& Ruiz-Gazen(2022)}]{nordhausen2022usage}
\bibinfo{author}{Nordhausen, K.}, \& \bibinfo{author}{Ruiz-Gazen, A.}
  (\bibinfo{year}{2022}).
\newblock \bibinfo{title}{On the usage of joint diagonalization in multivariate
  statistics}.
\newblock {\it \bibinfo{journal}{Journal of Multivariate Analysis}\/},  {\it
  \bibinfo{volume}{188}\/}, \bibinfo{pages}{104844}.
  \DOIprefix\doi{10.1016/j.jmva.2021.104844}.
\bibitem[{Nordhausen \& Tyler(2015)}]{NordhausenTyler2015}
\bibinfo{author}{Nordhausen, K.}, \& \bibinfo{author}{Tyler, D.~E.}
  (\bibinfo{year}{2015}).
\newblock \bibinfo{title}{{A cautionary note on robust covariance plug-in
  methods}}.
\newblock {\it \bibinfo{journal}{Biometrika}\/},  {\it
  \bibinfo{volume}{102}\/}, \bibinfo{pages}{573--588}.
  \DOIprefix\doi{10.1093/biomet/asv022}.
\bibitem[{Nordhausen \& Virta(2019)}]{NordhausenVirta2019}
\bibinfo{author}{Nordhausen, K.}, \& \bibinfo{author}{Virta, J.}
  (\bibinfo{year}{2019}).
\newblock \bibinfo{title}{An overview of properties and extensions of {FOBI}}.
\newblock {\it \bibinfo{journal}{Knowledge-Based Systems}\/},  {\it
  \bibinfo{volume}{173}\/}, \bibinfo{pages}{113--116}.
  \DOIprefix\doi{10.1016/j.knosys.2019.02.026}.
\bibitem[{Pe{\~n}a et~al.(2010)Pe{\~n}a, Prieto \&
  Viladomat}]{pena2010eigenvectors}
\bibinfo{author}{Pe{\~n}a, D.}, \bibinfo{author}{Prieto, F.~J.}, \&
  \bibinfo{author}{Viladomat, J.} (\bibinfo{year}{2010}).
\newblock \bibinfo{title}{Eigenvectors of a kurtosis matrix as interesting
  directions to reveal cluster structure}.
\newblock {\it \bibinfo{journal}{Journal of Multivariate Analysis}\/},  {\it
  \bibinfo{volume}{101}\/}, \bibinfo{pages}{1995--2007}.
  \DOIprefix\doi{10.1016/j.jmva.2010.04.014}.
\bibitem[{Radoji{\v{c}}i{\'c} \& Nordhausen(2020)}]{radojivcic2019non}
\bibinfo{author}{Radoji{\v{c}}i{\'c}, U.}, \& \bibinfo{author}{Nordhausen, K.}
  (\bibinfo{year}{2020}).
\newblock \bibinfo{title}{Non-{Gaussian} component analysis: Testing the
  dimension of the signal subspace}.
\newblock In {\it \bibinfo{booktitle}{Analytical Methods in Statistics: AMISTAT
  2019}\/} (pp. \bibinfo{pages}{101--123}).
\newblock \bibinfo{organization}{Springer}.
\newblock \DOIprefix\doi{10.1007/978-3-030-48814-7_6}.
\bibitem[{Radoji{\v{c}}i{\'c} et~al.(2021)Radoji{\v{c}}i{\'c}, Nordhausen \&
  Virta}]{radojivcic2021large}
\bibinfo{author}{Radoji{\v{c}}i{\'c}, U.}, \bibinfo{author}{Nordhausen, K.}, \&
  \bibinfo{author}{Virta, J.} (\bibinfo{year}{2021}).
\newblock \bibinfo{title}{Large-sample properties of unsupervised estimation of
  the linear discriminant using projection pursuit}.
\newblock {\it \bibinfo{journal}{Electronic Journal of Statistics}\/},  {\it
  \bibinfo{volume}{15}\/}, \bibinfo{pages}{6677--6739}.
  \DOIprefix\doi{10.1214/21-EJS1956}.
\bibitem[{Raftery \& Dean(2006)}]{raftery2006variable}
\bibinfo{author}{Raftery, A.~E.}, \& \bibinfo{author}{Dean, N.}
  (\bibinfo{year}{2006}).
\newblock \bibinfo{title}{Variable selection for model-based clustering}.
\newblock {\it \bibinfo{journal}{Journal of the American Statistical
  Association}\/},  {\it \bibinfo{volume}{101}\/}, \bibinfo{pages}{168--178}.
  \DOIprefix\doi{10.1198/016214506000000113}.
\bibitem[{Rathnayake et~al.(2019)Rathnayake, McLachlan, Peel, Baek \&
  {\proglang{R} Core Team}}]{EMMIXmfa}
\bibinfo{author}{Rathnayake, S.}, \bibinfo{author}{McLachlan, G.},
  \bibinfo{author}{Peel, D.}, \bibinfo{author}{Baek, J.}, \&
  \bibinfo{author}{{\proglang{R} Core Team}} (\bibinfo{year}{2019}).
\newblock {\it \bibinfo{title}{\pkg{EMMIXmfa}: Mixture Models with
  Component-Wise Factor Analyzers}\/}.
\newblock \URLprefix \url{https://CRAN.R-project.org/package=EMMIXmfa}
  \bibinfo{note}{\proglang{R} package version~2.0.11}.
\bibitem[{Raymaekers \& Rousseeuw(2023)}]{cellWise}
\bibinfo{author}{Raymaekers, J.}, \& \bibinfo{author}{Rousseeuw, P.}
  (\bibinfo{year}{2023}).
\newblock {\it \bibinfo{title}{\pkg{cellWise}: Analyzing Data with Cellwise
  Outliers}\/}.
\newblock \URLprefix \url{https://CRAN.R-project.org/package=cellWise}
  \bibinfo{note}{\proglang{R} package version~2.5.3}.
\bibitem[{{\proglang{R} Core Team}(2023)}]{R}
\bibinfo{author}{{\proglang{R} Core Team}} (\bibinfo{year}{2023}).
\newblock {\it \bibinfo{title}{\proglang{R}: A Language and Environment for
  Statistical Computing}\/}.
\newblock \bibinfo{organization}{\proglang{R} Foundation for Statistical
  Computing} \bibinfo{address}{Vienna, Austria}.
\newblock \URLprefix \url{https://www.R-project.org/}.
\bibitem[{Rousseeuw(1985)}]{rousseeuw1985multivariate}
\bibinfo{author}{Rousseeuw, P.} (\bibinfo{year}{1985}).
\newblock \bibinfo{title}{Multivariate estimation with high breakdown point}.
\newblock In {\it \bibinfo{booktitle}{Mathematical statistics and applications,
  {V}ol. {B} ({B}ad {T}atzmannsdorf, 1983)}\/} (pp. \bibinfo{pages}{283--297}).
\newblock \bibinfo{publisher}{Reidel, Dordrecht}.
\bibitem[{Rousseeuw \& Van~Driessen(1999)}]{rousseeuw1999fast}
\bibinfo{author}{Rousseeuw, P.~J.}, \& \bibinfo{author}{Van~Driessen, K.}
  (\bibinfo{year}{1999}).
\newblock \bibinfo{title}{A fast algorithm for the minimum covariance
  determinant estimator}.
\newblock {\it \bibinfo{journal}{Technometrics}\/},  {\it
  \bibinfo{volume}{41}\/}, \bibinfo{pages}{212--223}.
  \DOIprefix\doi{10.1080/00401706.1999.10485670}.
\bibitem[{Ruiz-Gazen(1996)}]{RuizGazen1996}
\bibinfo{author}{Ruiz-Gazen, A.} (\bibinfo{year}{1996}).
\newblock \bibinfo{title}{A very simple robust estimator of a dispersion
  matrix}.
\newblock {\it \bibinfo{journal}{Computational Statistics \& Data Analysis}\/},
   {\it \bibinfo{volume}{21}\/}, \bibinfo{pages}{149--162}.
  \DOIprefix\doi{https://doi.org/10.1016/0167-9473(95)00009-7}.
\bibitem[{Schoonees et~al.(2015)Schoonees, van~de Velden \&
  Groenen}]{schoonees2015}
\bibinfo{author}{Schoonees, P.~C.}, \bibinfo{author}{van~de Velden, M.}, \&
  \bibinfo{author}{Groenen, P. J.~F.} (\bibinfo{year}{2015}).
\newblock \bibinfo{title}{Constrained dual scaling for detecting response
  styles in categorical data}.
\newblock {\it \bibinfo{journal}{Psychometrika}\/},  {\it
  \bibinfo{volume}{80}\/}, \bibinfo{pages}{968--994}.
  \DOIprefix\doi{10.1007/s11336-015-9458-9}.
\bibitem[{Scrucca(2010)}]{scrucca2010}
\bibinfo{author}{Scrucca, L.} (\bibinfo{year}{2010}).
\newblock \bibinfo{title}{Dimension reduction for model-based clustering}.
\newblock {\it \bibinfo{journal}{Statistics and Computing}\/},  {\it
  \bibinfo{volume}{20}\/}, \bibinfo{pages}{471--484}.
  \DOIprefix\doi{10.1007/s11222-009-9138-7}.
\bibitem[{Scrucca et~al.(2023)Scrucca, Fraley, Murphy \& Raftery}]{scrucca2023}
\bibinfo{author}{Scrucca, L.}, \bibinfo{author}{Fraley, C.},
  \bibinfo{author}{Murphy, T.~B.}, \& \bibinfo{author}{Raftery, A.~E.}
  (\bibinfo{year}{2023}).
\newblock {\it \bibinfo{title}{Model-Based Clustering, Classification, and
  Density Estimation Using \pkg{mclust} in \proglang{R}}\/}.
\newblock \bibinfo{publisher}{Chapman \& Hall/CRC}.
\newblock \URLprefix \url{https://mclust-org.github.io/book/}.
  \DOIprefix\doi{10.1201/9781003277965}.
\bibitem[{Scrucca \& Raftery(2018)}]{scrucca2018}
\bibinfo{author}{Scrucca, L.}, \& \bibinfo{author}{Raftery, A.~E.}
  (\bibinfo{year}{2018}).
\newblock \bibinfo{title}{\pkg{clustvarsel}: A package implementing variable
  selection for gaussian model-based clustering in \proglang{R}}.
\newblock {\it \bibinfo{journal}{Journal of Statistical Software}\/},  {\it
  \bibinfo{volume}{84}\/}, \bibinfo{pages}{1--28}.
  \DOIprefix\doi{10.18637/jss.v084.i01}.
\bibitem[{Soete \& Carroll(1994)}]{soete1994k}
\bibinfo{author}{Soete, G.~D.}, \& \bibinfo{author}{Carroll, J.~D.}
  (\bibinfo{year}{1994}).
\newblock \bibinfo{title}{K-means clustering in a low-dimensional {Euclidean}
  space}.
\newblock In {\it \bibinfo{booktitle}{New Approaches in Classification and Data
  Analysis}\/} (pp. \bibinfo{pages}{212--219}).
\newblock \bibinfo{publisher}{Springer}.
\newblock \DOIprefix\doi{10.1007/978-3-642-51175-2_24}.
\bibitem[{Stahel \& M{\"a}chler(2009)}]{stahel2009comment}
\bibinfo{author}{Stahel, W.}, \& \bibinfo{author}{M{\"a}chler, M.}
  (\bibinfo{year}{2009}).
\newblock \bibinfo{title}{Comment on ``{Invariant} co-ordinate selection''}.
\newblock {\it \bibinfo{journal}{Journal of the Royal Statistical Society,
  Series B}\/},  {\it \bibinfo{volume}{71}\/}.
\bibitem[{Timmerman et~al.(2013)Timmerman, Ceulemans, De~Roover \&
  Van~Leeuwen}]{timmerman2013subspace}
\bibinfo{author}{Timmerman, M.~E.}, \bibinfo{author}{Ceulemans, E.},
  \bibinfo{author}{De~Roover, K.}, \& \bibinfo{author}{Van~Leeuwen, K.}
  (\bibinfo{year}{2013}).
\newblock \bibinfo{title}{Subspace k-means clustering}.
\newblock {\it \bibinfo{journal}{Behavior Research Methods}\/},  {\it
  \bibinfo{volume}{45}\/}, \bibinfo{pages}{1011--1023}.
  \DOIprefix\doi{10.3758/s13428-013-0329-y}.
\bibitem[{Todorov(2007)}]{todorov2007robust}
\bibinfo{author}{Todorov, V.} (\bibinfo{year}{2007}).
\newblock \bibinfo{title}{Robust selection of variables in linear discriminant
  analysis}.
\newblock {\it \bibinfo{journal}{Statistical Methods and Applications}\/},
  {\it \bibinfo{volume}{15}\/}, \bibinfo{pages}{395--407}.
  \DOIprefix\doi{10.1007/s10260-006-0032-6}.
\bibitem[{Todorov \& Filzmoser(2009)}]{rrcov}
\bibinfo{author}{Todorov, V.}, \& \bibinfo{author}{Filzmoser, P.}
  (\bibinfo{year}{2009}).
\newblock \bibinfo{title}{An object-oriented framework for robust multivariate
  analysis}.
\newblock {\it \bibinfo{journal}{Journal of Statistical Software}\/},  {\it
  \bibinfo{volume}{32}\/}, \bibinfo{pages}{1--47}.
  \DOIprefix\doi{10.18637/jss.v032.i03}.
\bibitem[{Tyler(2010)}]{Tyler2010}
\bibinfo{author}{Tyler, D.~E.} (\bibinfo{year}{2010}).
\newblock \bibinfo{title}{A note on multivariate location and scatter
  statistics for sparse data sets}.
\newblock {\it \bibinfo{journal}{Statistics \& Probability Letters}\/},  {\it
  \bibinfo{volume}{80}\/}, \bibinfo{pages}{1409--1413}.
  \DOIprefix\doi{10.1016/j.spl.2010.05.006}.
\bibitem[{Tyler et~al.(2009)Tyler, Critchley, D{\"u}mbgen \&
  Oja}]{tyler2009invariant}
\bibinfo{author}{Tyler, D.~E.}, \bibinfo{author}{Critchley, F.},
  \bibinfo{author}{D{\"u}mbgen, L.}, \& \bibinfo{author}{Oja, H.}
  (\bibinfo{year}{2009}).
\newblock \bibinfo{title}{Invariant co-ordinate selection}.
\newblock {\it \bibinfo{journal}{Journal of the Royal Statistical Society,
  Series B}\/},  {\it \bibinfo{volume}{71}\/}, \bibinfo{pages}{549--592}.
  \DOIprefix\doi{10.1111/j.1467-9868.2009.00706.x}.
\bibitem[{van~de Velden et~al.(2019{\natexlab{a}})van~de Velden, D’Enza \&
  Yamamoto}]{van2019special}
\bibinfo{author}{van~de Velden, M.}, \bibinfo{author}{D’Enza, A.~I.}, \&
  \bibinfo{author}{Yamamoto, M.} (\bibinfo{year}{2019}{\natexlab{a}}).
\newblock \bibinfo{title}{Special feature: Dimension reduction and cluster
  analysis}.
\newblock {\it \bibinfo{journal}{Behaviormetrika}\/},  {\it
  \bibinfo{volume}{46}\/}, \bibinfo{pages}{239--241}.
  \DOIprefix\doi{10.1007/s41237-019-00092-6}.
\bibitem[{van~de Velden et~al.(2019{\natexlab{b}})van~de Velden, Iodice~D'Enza
  \& Markos}]{van2019distance}
\bibinfo{author}{van~de Velden, M.}, \bibinfo{author}{Iodice~D'Enza, A.}, \&
  \bibinfo{author}{Markos, A.} (\bibinfo{year}{2019}{\natexlab{b}}).
\newblock \bibinfo{title}{Distance-based clustering of mixed data}.
\newblock {\it \bibinfo{journal}{Wiley Interdisciplinary Reviews: Computational
  Statistics}\/},  {\it \bibinfo{volume}{11}\/}, \bibinfo{pages}{e1456}.
  \DOIprefix\doi{10.1002/wics.1456}.
\bibitem[{Venables \& Ripley(2002)}]{mass}
\bibinfo{author}{Venables, W.~N.}, \& \bibinfo{author}{Ripley, B.~D.}
  (\bibinfo{year}{2002}).
\newblock {\it \bibinfo{title}{Modern Applied Statistics with \proglang{S}}\/}.
\newblock (\bibinfo{edition}{4th} ed.).
\newblock \bibinfo{publisher}{Springer}.
\newblock \URLprefix \url{https://www.stats.ox.ac.uk/pub/MASS4/}.
\bibitem[{Vichi \& Kiers(2001)}]{vichi2001factorial}
\bibinfo{author}{Vichi, M.}, \& \bibinfo{author}{Kiers, H.~A.}
  (\bibinfo{year}{2001}).
\newblock \bibinfo{title}{Factorial k-means analysis for two-way data}.
\newblock {\it \bibinfo{journal}{Computational Statistics \& Data Analysis}\/},
   {\it \bibinfo{volume}{37}\/}, \bibinfo{pages}{49--64}.
  \DOIprefix\doi{10.1016/S0167-9473(00)00064-5}.
\bibitem[{Vichi \& Saporta(2009)}]{vichi2009clustering}
\bibinfo{author}{Vichi, M.}, \& \bibinfo{author}{Saporta, G.}
  (\bibinfo{year}{2009}).
\newblock \bibinfo{title}{Clustering and disjoint principal component
  analysis}.
\newblock {\it \bibinfo{journal}{Computational Statistics \& Data Analysis}\/},
   {\it \bibinfo{volume}{53}\/}, \bibinfo{pages}{3194--3208}.
  \DOIprefix\doi{10.1016/j.csda.2008.05.028}.
\bibitem[{Xu \& Wunsch(2005)}]{xu2005survey}
\bibinfo{author}{Xu, R.}, \& \bibinfo{author}{Wunsch, D.}
  (\bibinfo{year}{2005}).
\newblock \bibinfo{title}{Survey of clustering algorithms}.
\newblock {\it \bibinfo{journal}{IEEE Transactions on Neural Networks}\/},
  {\it \bibinfo{volume}{16}\/}, \bibinfo{pages}{645--678}.
  \DOIprefix\doi{10.1109/TNN.2005.845141}.

\end{thebibliography}

\appendix

\section{Cluster settings in the simulation study}
\label{app:sim-details}

We consider the following 22 combinations of the mixture weights $\epsilon_1, \dots, \epsilon_q$ in the simulation study from Section~\ref{sec:sim}. We refer to the different cluster settings using the notation $100\epsilon_{1}$--\dots--$100\epsilon_{q}$:
\begin{itemize}
    \item $q=2$ clusters: 50--50, 55--45, 60--40, 65--35, 70--30, 75--35, 80--20, 85--15, 90--10, 95--5,
    \item $q=3$ clusters: 33--33--33, 30--40--30, 20--50--30, 10--50--40, 10--60--30, 10--70--20,  10--80--10,
    \item $q=5$ clusters: 20--20--20--20--20, 10--20--20--20--30, 10--10--20--20--40, 10--10--10--30--40, 10--10--20--30--30.
\end{itemize}
The dimension reduction strategies, which are the main focus of this paper, are evaluated in all 22 settings in terms of the discriminatory power of the selected components. 

However, to keep computation time low, we apply the clustering methods on the obtained components only for the following subset of the cluster settings:
\begin{itemize}
    \item $q=2$ clusters: 50--50, 70--30, 80--20, 90--10, 95--5,
    \item $q=3$ clusters: 33--33--33, 20--50--30, 10--80--10,
    \item $q=5$ clusters: 20--20--20--20--20, 10--10--20--20--40.
\end{itemize}
Hence, results for the clustering performance (see Section~\ref{sec:best-ARI} and \ref{app:ARI-clusters}) are based on those 10 cluster settings.

\section{Additional simulation results}

\subsection{Overall comparison of methods in setting without outliers}
\label{app:selection}

\begin{figure}[t!]
\includegraphics[width=\textwidth]{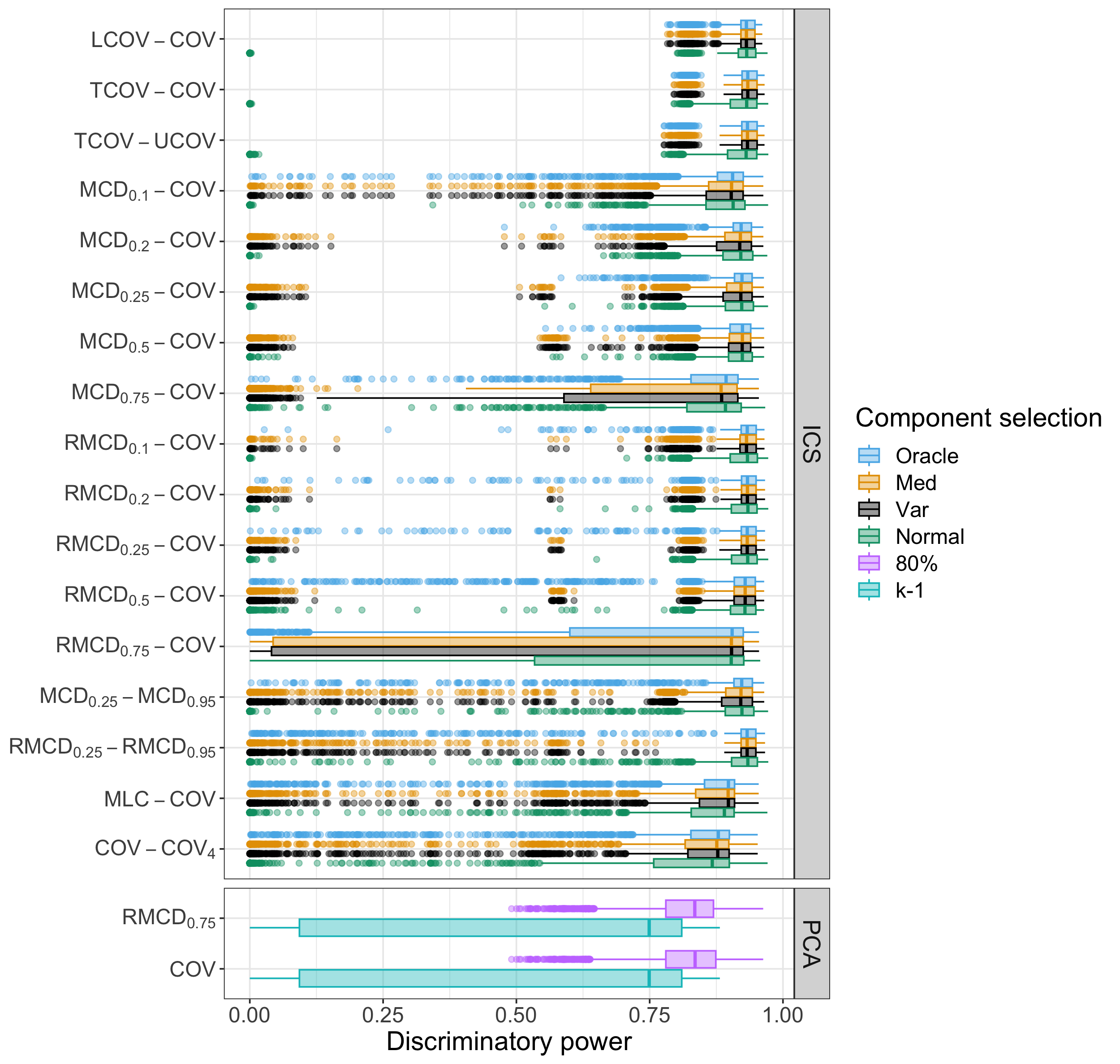}
\caption{Discriminatory power as measured by the $\eta^{2}$ of the investigated dimension reduction methods with different component selection criteria. Boxplots display the results across the different cluster settings with $100$ simulation runs each.} 
\label{fig:eta2}
\end{figure}

To establish which methods should be discussed in more detail, we take a look at their performance in the baseline scenario without outliers. For all investigated dimension reduction methods and component selection criteria, Figure~\ref{fig:eta2} provides boxplots of the discriminatory power of the selected components across the different cluster settings and simulation runs. The discriminatory power is thereby measured by the $\eta^{2}$.

\begin{figure}[t!]
\centering
\includegraphics[width=0.46\textwidth]{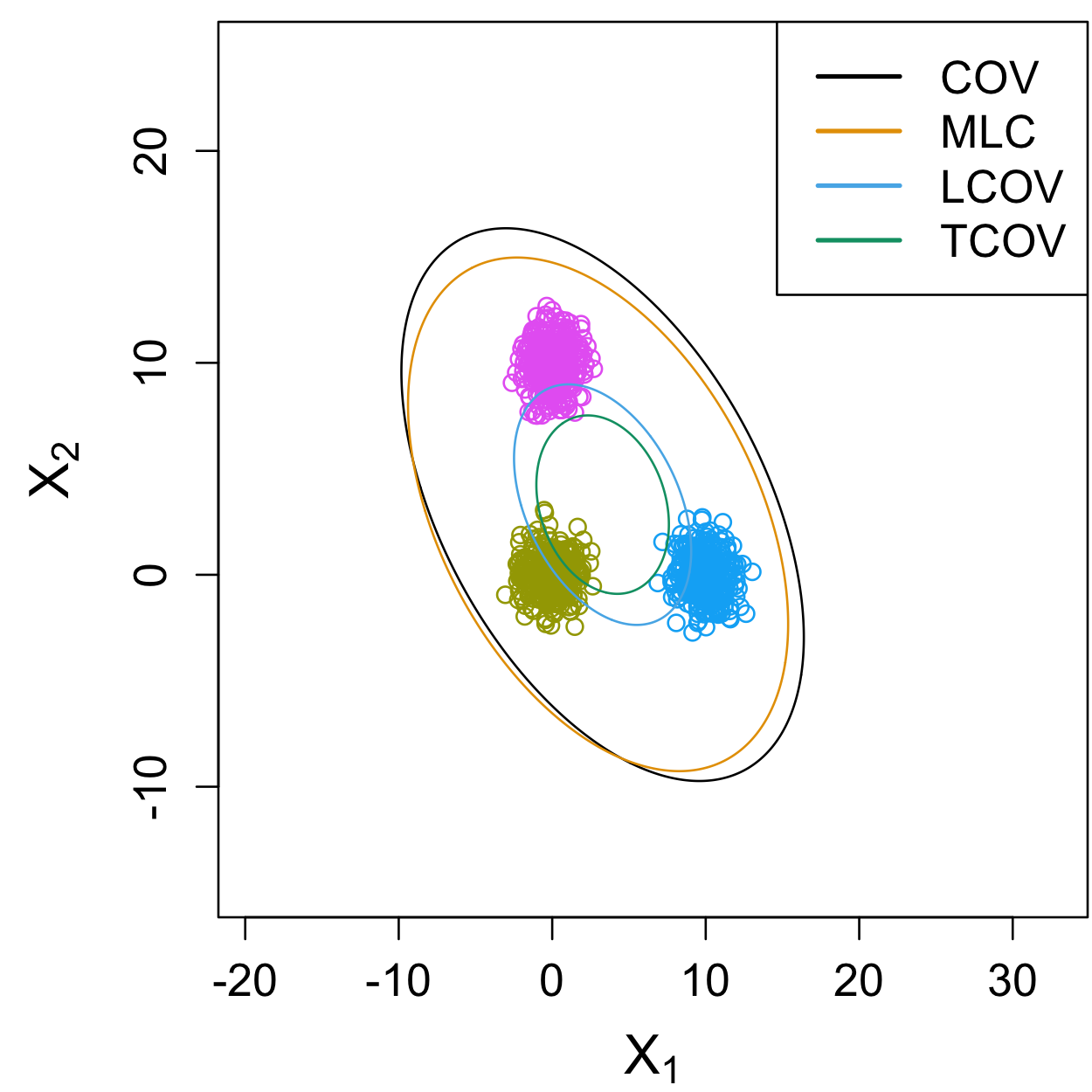} 
\includegraphics[width=0.46\textwidth]{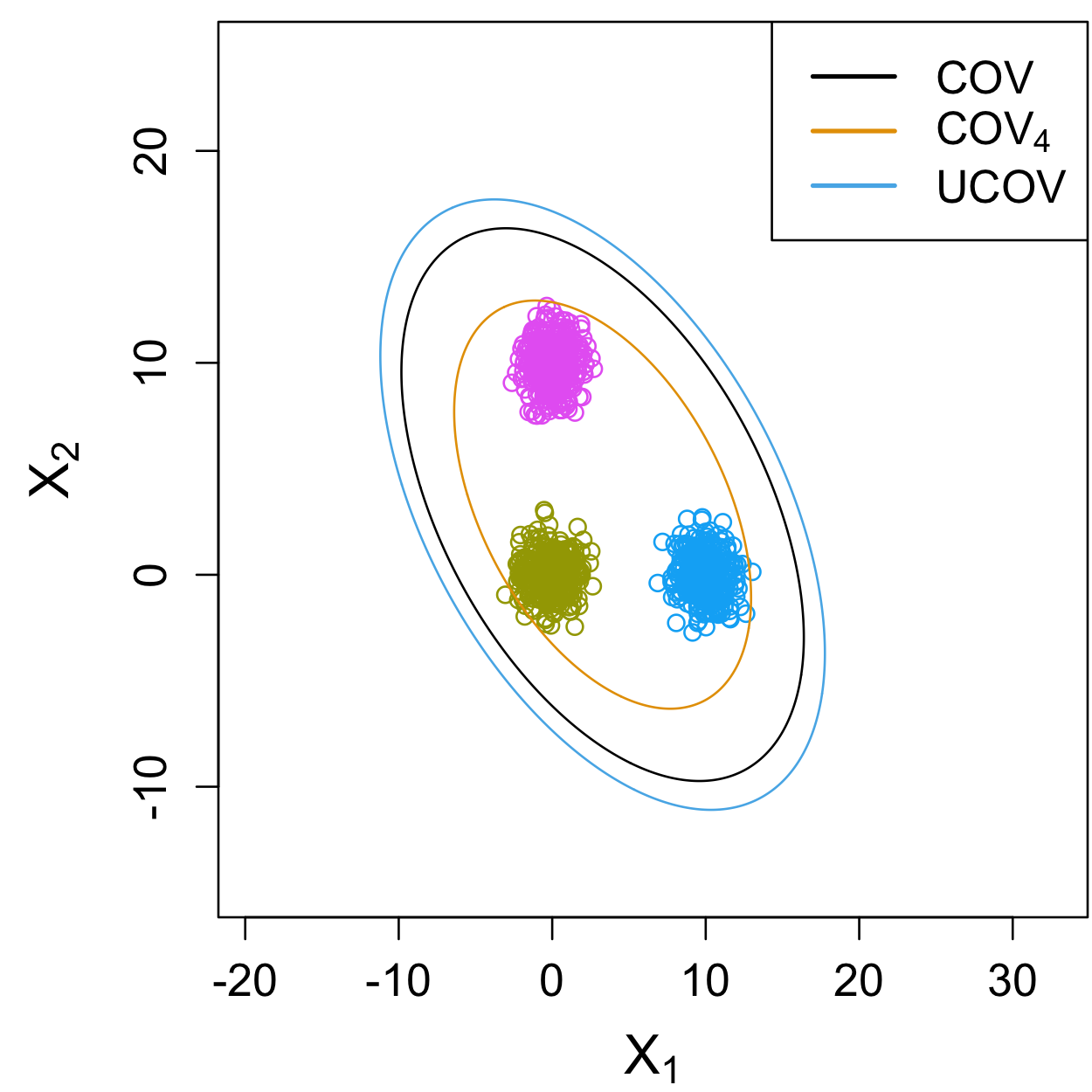}
\includegraphics[width=0.46\textwidth]{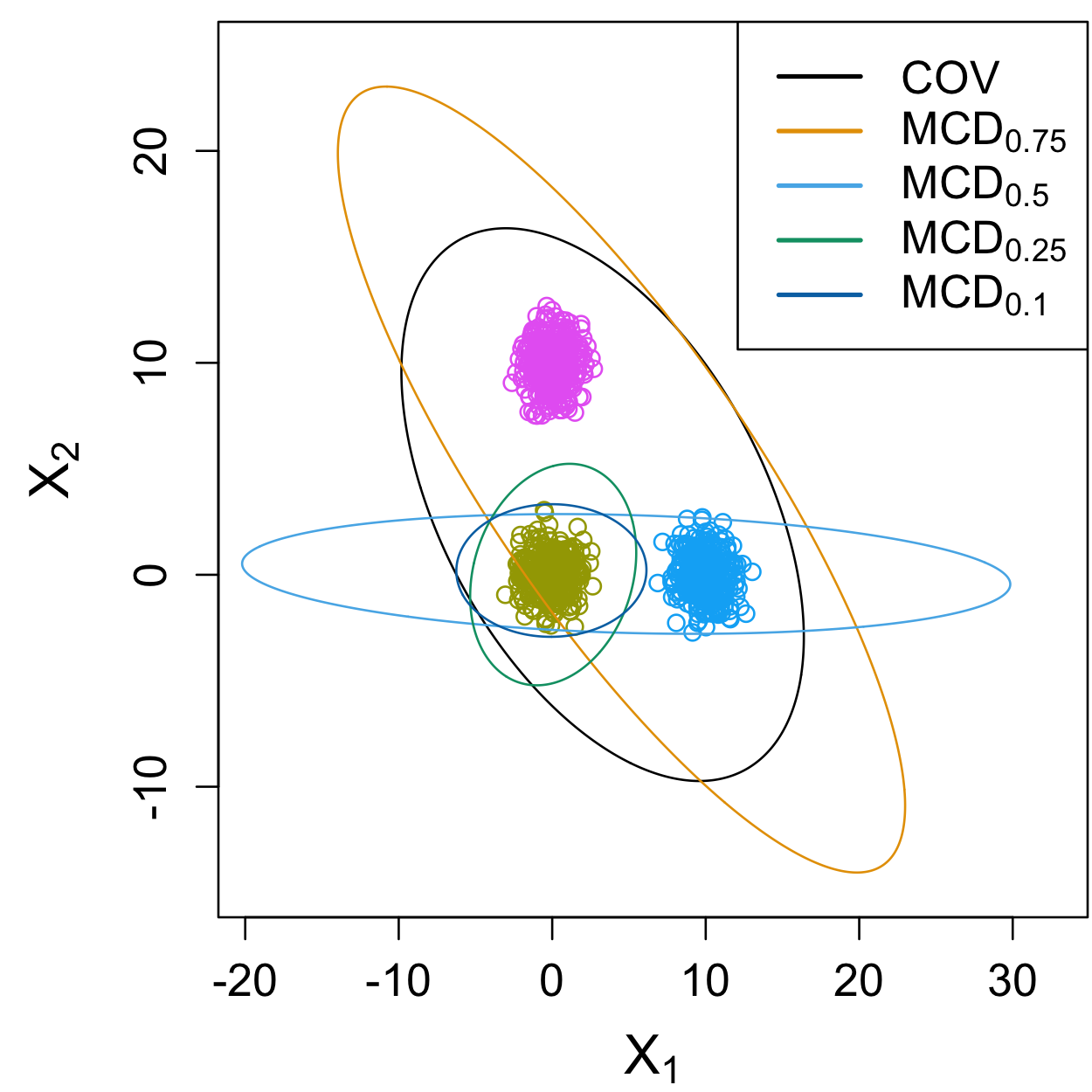}
\includegraphics[width=0.46\textwidth]{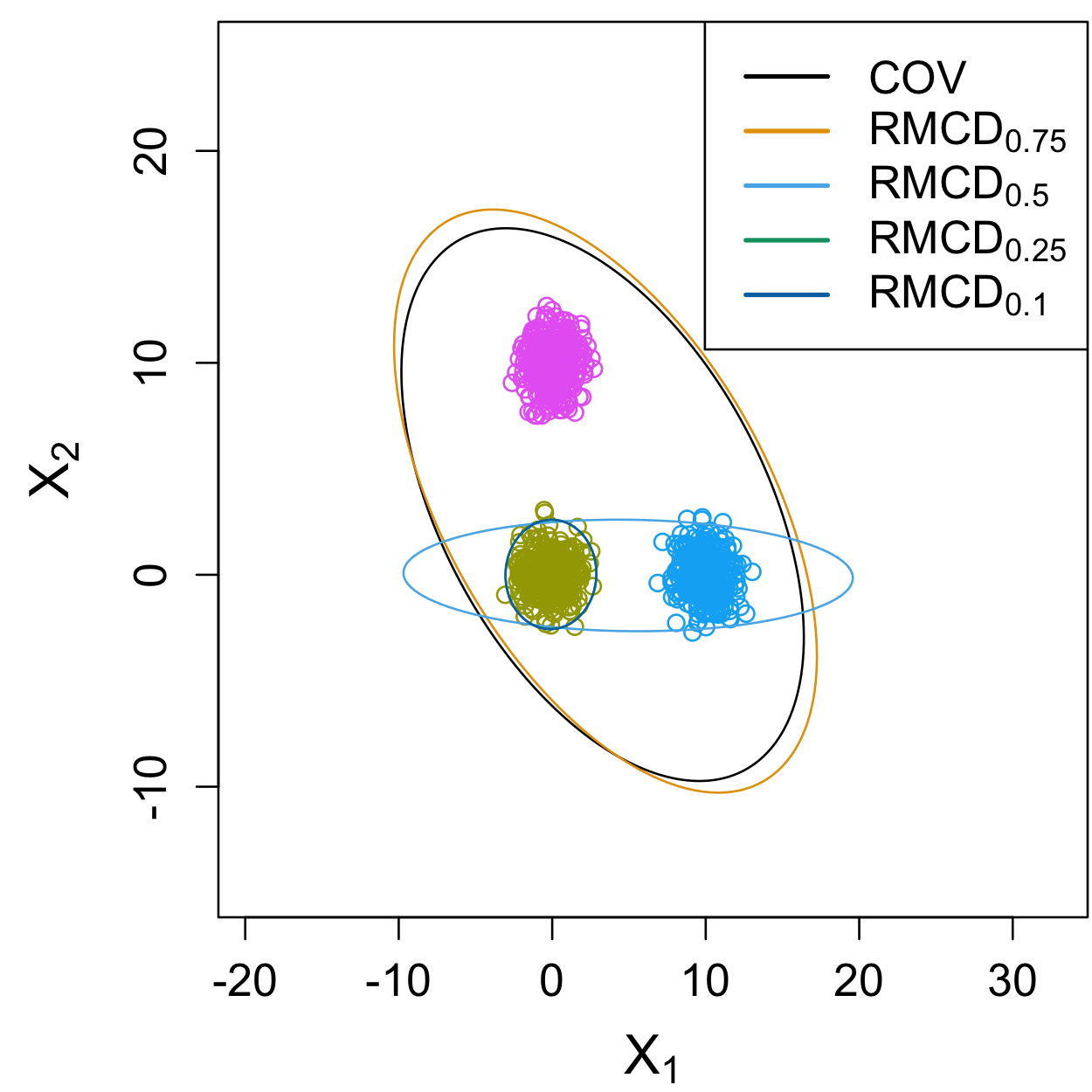}
\caption{Shapes of different scatter matrices for a sample from a bivariate Gaussian mixture distribution with three balanced components. In the first row, the local scatters (together with $\cov$) are shown in the left panel and the global scatters in the right panel. In the second row, $\mcd_{\alpha}$ is shown in the left panel and $\rmcd_{\alpha}$ in the right panel for different values of $\alpha$ (together with $\cov$).} \label{fig:scatter_shapes}
\end{figure}

We clearly observe that both the highest median and the lowest variability in discriminatory power are obtained with ICS using the scatter pairs $\lcov-\cov$ and $\tcov-\cov$, followed by the robust scatter pair $\tcov-\ucov$. For those scatter pairs, there is not much difference in the performance of different component selection criteria, which are all close to the oracle criterion. For illustration purposes, the first row of Figure~\ref{fig:scatter_shapes} visualizes the shapes of those scatters as ellipses based on a sample from a bivariate Gaussian mixture distribution with three balanced components. $\lcov$ and $\tcov$ nicely capture the local structure while $\cov$ and $\ucov$ recover the global structure. Suitable combinations therefore lead to highlighting the structure of interest. 

Going back to the results in Figure~\ref{fig:eta2}, we find for PCA that the $80\%$ criterion yields components with a higher median and lower variability in discriminatory power than the $k-1$ criterion. Using the $80\%$ criterion, $\rmcd_{0.75}$ clearly outperforms the sample covariance matrix $\cov$, as the median discriminatory power is higher while variability is lower. However, ICS with the aforementioned three scatter pairs clearly outperforms PCA with $\rmcd_{0.75}$.

Overall, ICS with a scatter pair involving the $\mcd$ performs surprisingly well, but the results depend strongly on the parameter $\alpha$. When paired with $\cov$, somewhat smaller values of $\alpha$ tend to perform better in general. This is expected since a smaller subset size increases the chances that the shape of the MCD estimate corresponds to the local covariance structure within one of the clusters versus the global covariance structure captured by $\cov$ \citep[cf.\ the discussion in][]{Hennig2009}. Another important question for $\mcd$ scatters is whether to use the raw estimator $\mcd_{\alpha}$ or the reweighted estimator $\rmcd_{\alpha}$. Using $\mcd_{\alpha}$ gives (almost) full control over the subset size, which can be desirable in a clustering context to capture the local covariance structure. While scatter pairs with $\rmcd_{\alpha}$ in general yield a higher median $\eta^{2}$ than scatter pairs with $\mcd_{\alpha}$, small differences in the median are not that relevant as long as the $\eta^{2}$ is large enough. On the other hand, scatter pairs with $\rmcd_{\alpha}$ fail to pick up the relevant structure in more instances compared to their counterparts with $\mcd_{\alpha}$, i.e., there are more points with small $\eta^{2}$ in Figure~\ref{fig:eta2}, which is most visible for the oracle criterion. For a rather large subset size (e.g., $\alpha = 0.75$), using $\rmcd_{\alpha}$ can be considered detrimental to performance according to our results. Nevertheless, for a small subset size (e.g., $\alpha = 0.1$), using $\rmcd_{\alpha}$ rather than $\mcd_{\alpha}$ can be beneficial to prevent that the estimate is based on too few observations. This could be of particular relevance when $n$ is small, although we emphasize that we did not study such a setting. The second row of Figure~\ref{fig:scatter_shapes}, which visualizes the shapes of $\mcd_{\alpha}$ and $\rmcd_{\alpha}$ in our illustrative example of a bivariate Gaussian mixture, provides intuition for the above discussion by exposing the dramatic impact of the parameter $\alpha$ and the reweighting step on finding the local structure.

In Figure~\ref{fig:eta2}, the scatter pairs $\mlt-\cov$ and $\cov-\covF$ clearly perform worse than the other scatter pairs for ICS in terms of discriminatory power, as they have a lower median $\eta^{2}$ and fail in many instances. The first row of Figure~\ref{fig:scatter_shapes} further illustrates that those scatter pairs are not the best choices for finding the local structure. Nevertheless, since $\cov-\covF$ is likely the most widely used scatter pair for ICS, we find it relevant to discuss its behavior in the main text. 

Regarding the component selection criteria for ICS, keep in mind that in this paper, we only consider clustering methods that take the number of clusters $k$ as an input. The med and var criteria, which are based on selecting \mbox{$k-1$} components, are therefore conceptually appealing and, across the different scatter pairs, lead to the best overall performance in discriminatory power.

\subsection{Clustering performance in selected cluster settings}
\label{app:ARI-clusters}

\begin{figure}[b!]
\includegraphics[width=\textwidth]{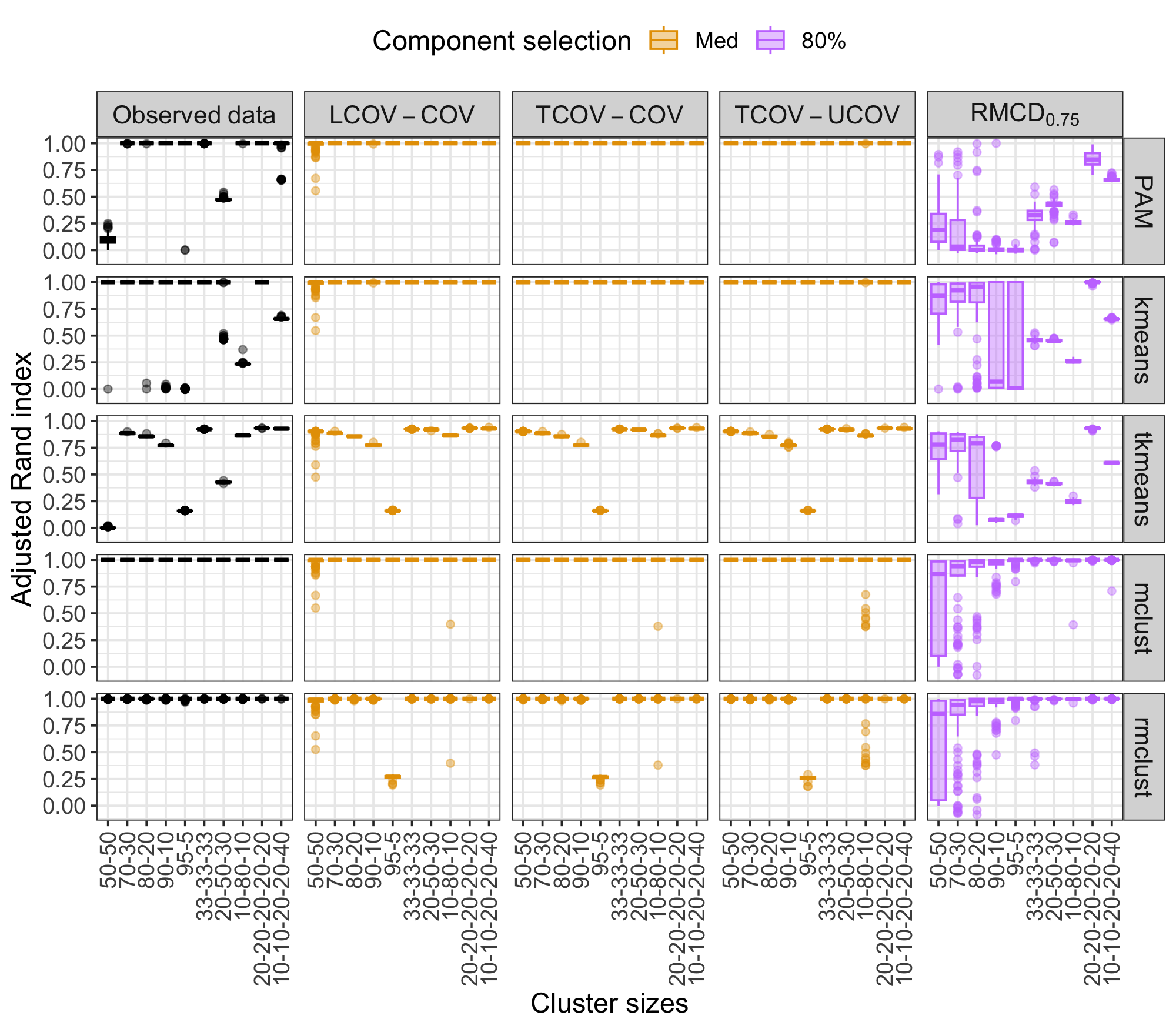}
\caption{Clustering performance as measured by the ARI after the best-performing dimension reduction methods for ICS and PCA, respectively, using the corresponding best component selection criteria. Boxplots display the results for the different cluster settings over $100$ simulation runs in the setting without outliers. Results for different clustering methods are shown in separate rows, and results for different dimension reduction methods are shown in separate columns.}
\label{fig:best-ARI-no}
\end{figure}

\begin{figure}[t!]
\includegraphics[width=\textwidth]{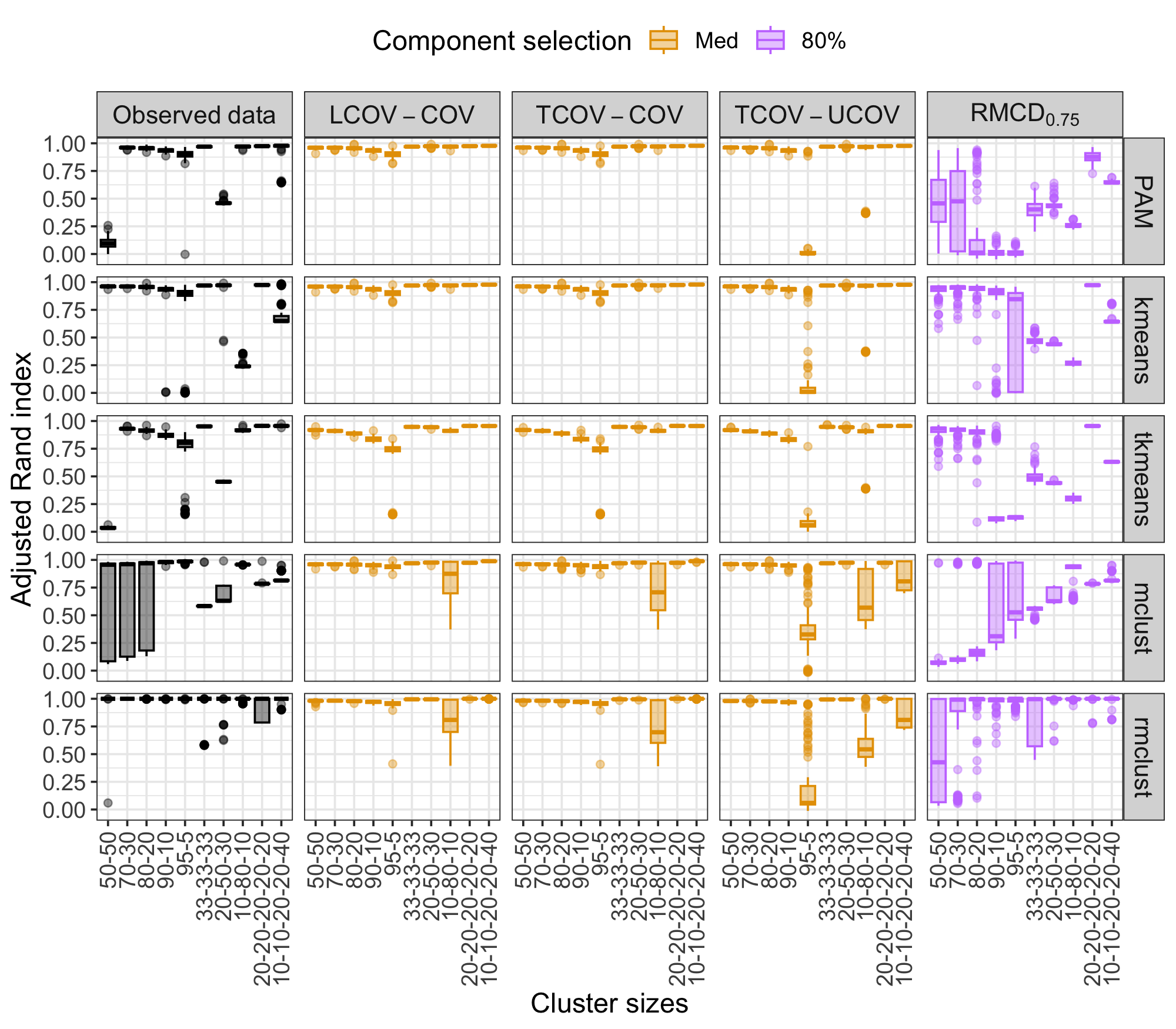}
\caption{Clustering performance as measured by the ARI after the best-performing dimension reduction methods for ICS and PCA, respectively, using the corresponding best component selection criteria. Boxplots display the results for the different cluster settings over $100$ simulation runs in the setting with 2\% outliers. Results for different clustering methods are shown in separate rows, and results for different dimension reduction methods are shown in separate columns.}
\label{fig:best-ARI-0.02}
\end{figure}

\begin{figure}[t!]
\includegraphics[width=\textwidth]{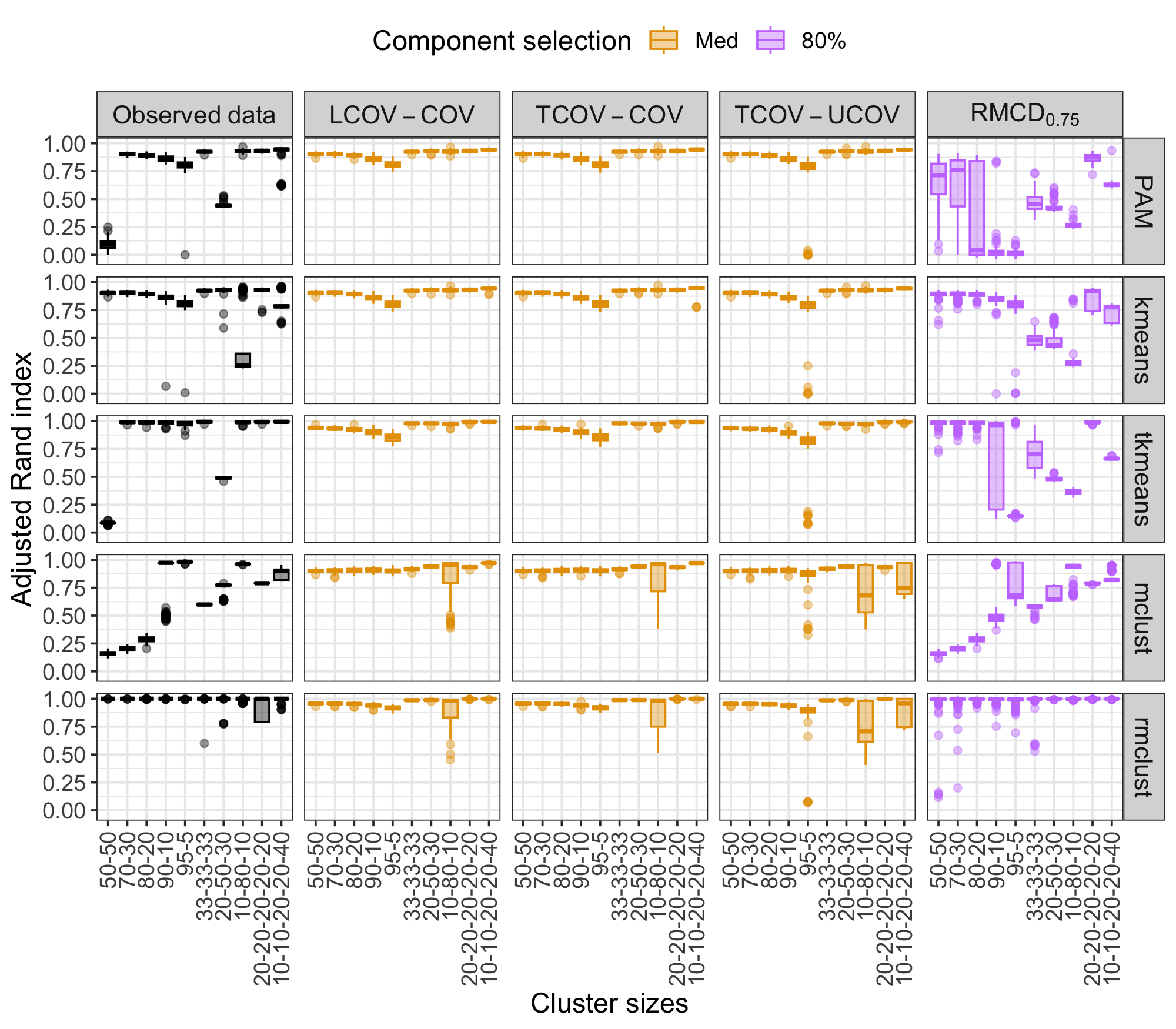}
\caption{Clustering performance as measured by the ARI after the best-performing dimension reduction methods for ICS and PCA, respectively, using the corresponding best component selection criteria. Boxplots display the results for the different cluster settings over $100$ simulation runs in the setting with 5\% outliers. Results for different clustering methods are shown in separate rows, and results for different dimension reduction methods are shown in separate columns.}
\label{fig:best-ARI-0.05}
\end{figure}

Figures~\ref{fig:best-ARI-no},~\ref{fig:best-ARI-0.02}, and~\ref{fig:best-ARI-0.05} display the clustering results for the setting without outliers, with 2\% outliers, and with 5\% outliers, respectively. The figures show boxplots of the ARI of the five clustering methods in the selected cluster settings, applied after the best-performing dimension reduction methods for ICS and PCA.

In Figure~\ref{fig:best-ARI-no} for the setting without outliers, we observe a clear improvement in clustering performance with ICS for PAM and $k$means. Specifically, $\tcov-\cov$ and $\tcov-\ucov$ yield a nearly perfect ARI in all cluster settings. $\lcov-\cov$ performs similarly in most cluster settings, but fails on occasion for perfectly balanced clusters (50-50). For t$k$means, we find a substantial improvement only for certain cluster settings (50-50, 20-50-30). On the other hand, mclust and rmclust perform almost perfectly on the observed data. While ICS keeps this near-perfect performance in most settings, it can increase failures in some settings (in particular for rmclust). Finally, PCA with $\rmcd_{0.75}$ is detrimental to clustering performance, yielding a lower ARI or larger variability than on the observed data in almost every setting.

From Figures~\ref{fig:best-ARI-0.02} and~\ref{fig:best-ARI-0.05}, we see that the main findings remain valid in the settings with outliers. For all clustering methods, ICS with $\tcov-\cov$ and $\lcov-\cov$ clearly yields the best performance except for rmclust, where performance improves for some settings but deteriorates in others. The simple clustering methods $k$means and PAM after ICS even outperform rmclust on observed data in the settings with $q=5$ clusters. Finally, PCA with $\rmcd_{0.75}$ still has a detrimental effect on clustering performance.

\subsection{Illustration for the barrow wheel distribution} \label{app:barrowwheel}

In the simulation study of Section~\ref{sec:sim}, we generate the data from a Gaussian mixture model due to its simplicity and popularity. To provide an illustration of tandem clustering using ICS in a non-elliptical framework, we simulate a data set from the barrow wheel distribution, which was introduced in \citet{hampel1986robust} and used by \citet{stahel2009comment} as an example in the discussion of \citet{tyler2009invariant}. A $d$-variate random vector $\bo X$ follows a  barrow wheel distribution if
$$
{\bo X} \sim \bo O ((1-\epsilon) {\cal N}(\bo 0,\diag(\sigma_1^2, 1, \dots, 1)) + \epsilon H),
$$
with $H$ so that $h = (h_1, \bo h_2')' \sim H$ implies $h_1 \sim {\chi}_{d-1}$ and $h_2 \sim {\cal N}(0, \sigma_2^2I_{d-1})$. The $d \times d$ matrix $\bo O$ can be used to rotate and rescale the vector.

\begin{figure}[t!]
\centering
\includegraphics[width=\textwidth]{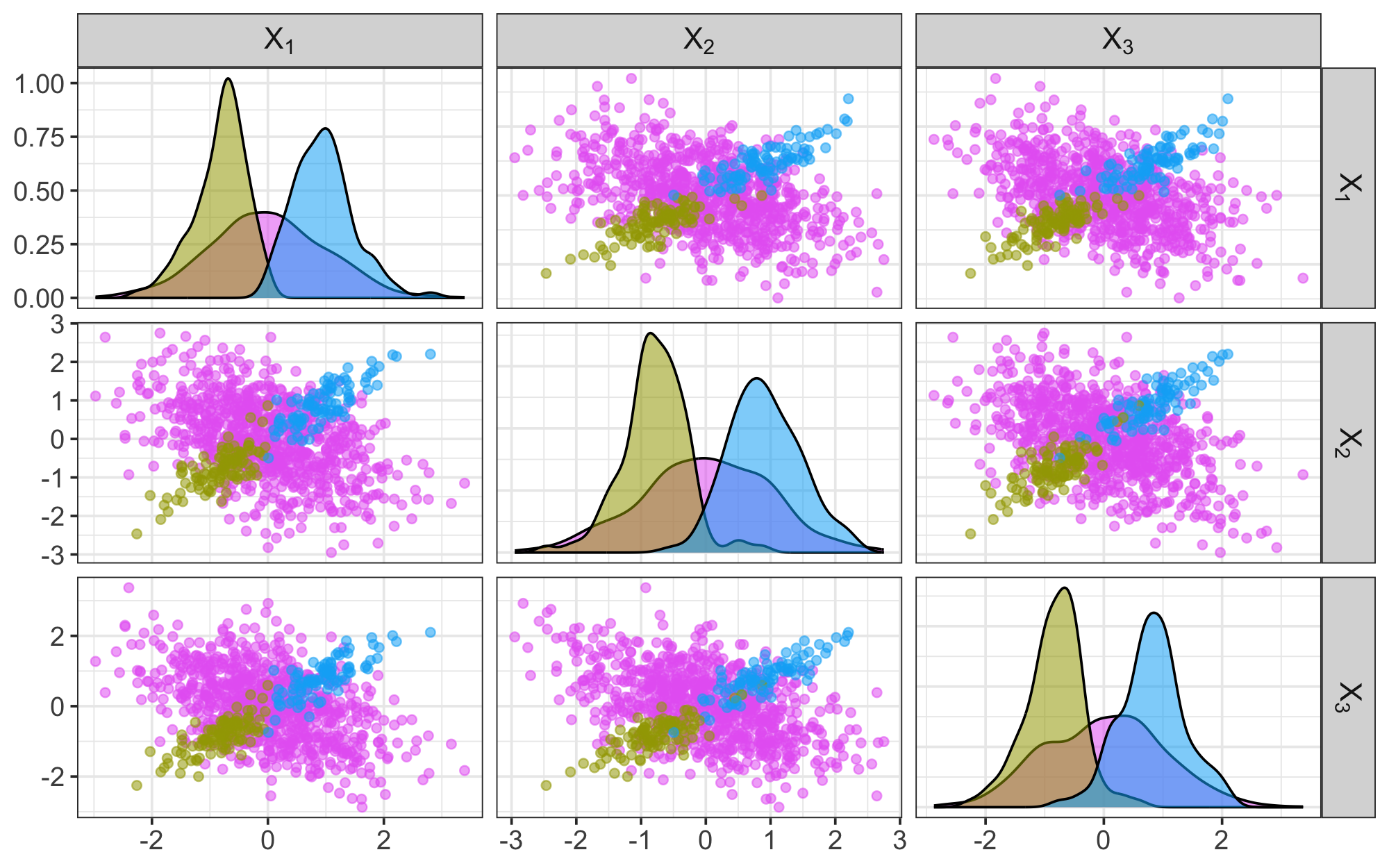}
\caption{Scatter plot matrix of a simulated data set from the barrow wheel distribution.}
\label{fig:rbwheel_data}
\end{figure}

\begin{table}[t!]
\begin{center}
\begin{tabular}{ c c c c c c c c}
          &        &         &  ICS +   & ICS +  & ICS +   &     & \\
 $k$means & mclust & rmclust & $k$means & mclust & rmclust & MFA & clustvarsel \\
\hline
 0.030 & 0.346 & 0.339 & 0.835 & 0.958  & 0.955 &  0.102 & 0.346 \\ 
\end{tabular}
\end{center}
\caption{\label{tab:rbwheel}Adjusted Rand Index for different clustering techniques on the simulated data from the barrow wheel distribution. ICS is applied using the scatter pair $\tcov-\cov$.}
\end{table}

\begin{figure}[t!]
\centering
\includegraphics[width=\textwidth]{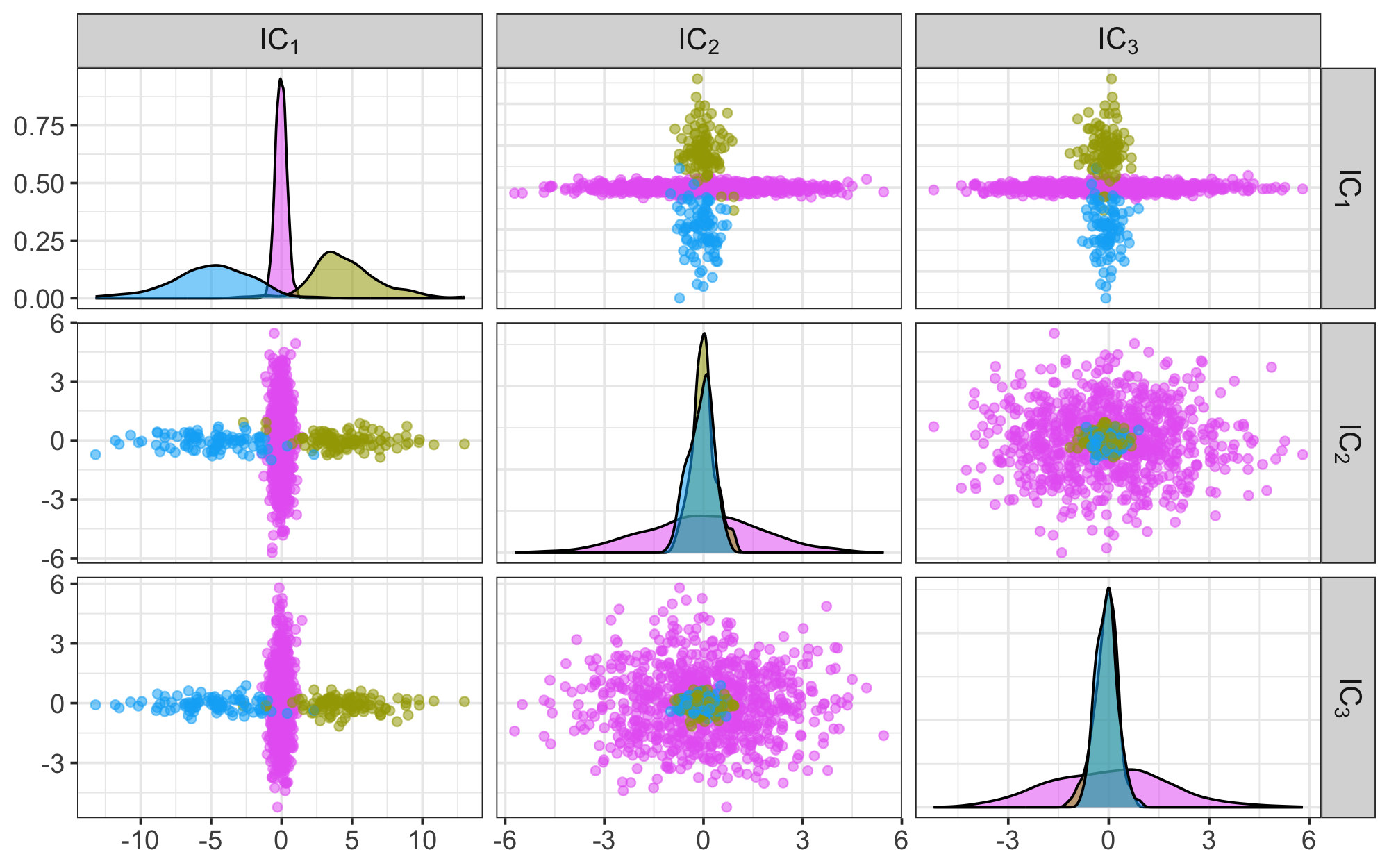}
\caption{Scatter plot matrices of the invariant coordinates obtained by ICS with $\tcov-\cov$ on the simulated data set from the barrow wheel distribution.}
\label{fig:rbwheel_IC}
\end{figure}

Figure~\ref{fig:rbwheel_data} shows a simulated data set of size $n=1000$ from the barrow wheel distribution with $d=3$, $\sigma_1=0.1$, $\sigma_2=0.2$ and $\epsilon = 0.2$. Table~\ref{tab:rbwheel} reports the ARI values of $k$means, mclust, and rmclust (applied directly to this data set and after ICS with $\tcov-\cov$), as well as MFA and clustvarsel. The approaches based on a Gaussian mixture model (mclust, rmclust, MFA, and clustvarsel) fail to detect the clusters in the simulated data set with ARI values between 0.102 and 0.346, while $k$means performs even worse with an ARI close to 0. However, using tandem clustering with ICS and TCOV-COV recovers the cluster structure in the first coordinate, as illustrated in Figure~\ref{fig:rbwheel_IC}. Applying the clustering methods to the first invariant coordinate greatly improves the identification of the three clusters and yields ARI values ranging from 0.835 for $k$means to about 0.96 for mclust and rmclust.

\section{Additional results for the empirical applications} \label{app:joint}

\begin{figure}[t!]
\centering
\includegraphics[width=\textwidth]{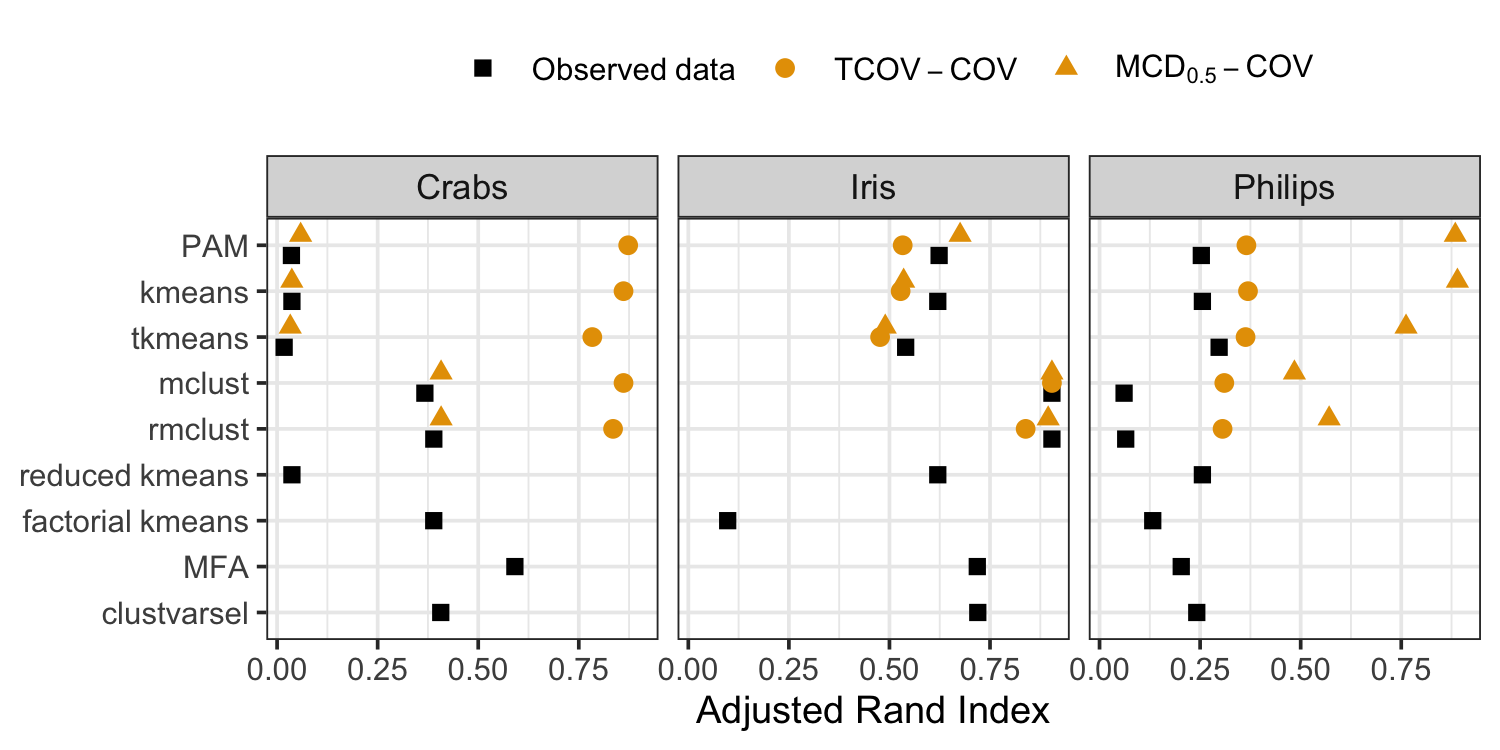}
\caption{Adjusted Rand Index (ARI) in the empirical applications for different clustering methods on the observed data, as well as after ICS with the med criterion for $\tcov-\cov$ and $\mcd_{0.5}-\cov$. Results for different data sets are shown in separate columns.}
\label{fig:emp_applications}
\end{figure}

For the three empirical applications discussed in Section~\ref{sec:applications}, Figure~\ref{fig:emp_applications} compares the five considered clustering methods (applied to the observed data, as well as after ICS with the med criterion for $\tcov-\cov$ and $\mcd_{0.5}-\cov$) to two variants of $k$means with a built-in approach for subspace estimation, namely reduced $k$means \citep{soete1994k} and factorial $k$means \citep{vichi2001factorial}, and the two previously considered dimension reduction methods for Gaussian mixture modeling, MFA \citep{mclachlan2003modelling} and clustvarsel \citep{raftery2006variable}. Clearly, these four additional methods do not improve the results and are always outperformed by at least one version of tandem clustering with ICS. 

\end{document}